\documentclass[12pt,a4paper,uplatex]{revtex4-1}
\usepackage[hidelinks,linktoc=all]{hyperref} 
\usepackage{here}
\usepackage{siunitx}
\usepackage{braket}
\usepackage[dvipdfmx]{graphicx}
\usepackage[version=4]{mhchem}
\begin{document}
\title{Recent Progress in Ultrafast Dynamics of Transition-Metal Compounds Studied by Time-Resolved X-ray Techniques}
\author{Hiroki Wadati}
\email{wadati@sci.u-hyogo.ac.jp}
\affiliation{Department of Material Science, Graduate School of Science, University of Hyogo, Ako, Hyogo 678-1297, Japan}
\affiliation{Institute of Laser Engineering, Osaka University, Suita, Osaka 565-0871, Japan}
\author{Kohei Yamamoto}
\affiliation{NanoTerasu Center, National Institutes for Quantum Science and Technology, Sendai, Miyagi 980-8572, Japan}
\author{Kohei Yamagami}
\affiliation{Japan Synchrotron Radiation Research Institute, Sayo, Hyogo 679-5198, Japan}

\maketitle
\section*{abstract}
X-ray absorption spectroscopy and X-ray magnetic circular dichroism have long served as indispensable tools for probing the electronic and magnetic properties of transition-metal compounds with elemental selectivity. In recent years, the emergence of femtosecond lasers has opened a new avenue for studying nonequilibrium dynamics in condensed matter. However, conventional optical techniques lack elemental and orbital specificity, making it difficult to disentangle the coupled charge, spin, and lattice responses in complex materials. The development of X-ray free-electron lasers (XFEL) and laboratory high-harmonic generation (HHG) sources has enabled the extension of X-ray absorption and scattering techniques into the femtosecond time domain. Time-resolved X-ray absorption spectroscopy, X-ray magnetic circular dichroism, and resonant soft X-ray scattering now provide direct, complementary access to element- and momentum-resolved ultrafast dynamics. This review summarizes recent progress in these techniques, focusing on pump-probe measurements of laser-induced demagnetization, spin-state transitions, and valence and structural changes in transition-metal compounds. We also discuss advances in tabletop HHG-based X-ray spectroscopy and its integration with large-scale XFEL facilities. These developments provide powerful routes for visualizing the nonequilibrium evolution of charge, spin, orbital, and lattice degrees of freedom, offering new insights into the ultrafast control of quantum materials.

\section{Introduction}
X-ray spectroscopy at elemental absorption edges has long been an essential experimental technique for probing the ground- and excited-state properties of transition-metal compounds on an element-specific basis \cite{Ueda2023}. 
By exploiting resonance phenomena and the polarization of X-rays, one can obtain detailed information on the local valence, spin, and orbital states of atoms in solids. 
In particular, X-ray absorption spectroscopy (XAS) measures the absorption coefficient as a function of photon energy near a core-level edge \cite{XASrev}.
The resulting fine structure reflects electronic transitions from well-defined core levels (such as $1s$ or $2p$) to unoccupied states above the Fermi level, providing element-specific information on the local electronic configuration, oxidation state, and coordination environment.
Since XAS directly probes the density of unoccupied states, it has become a powerful tool for examining both the electronic structure and chemical bonding in solids. The advent of circularly polarized X-rays enabled the development of X-ray magnetic circular dichroism (XMCD), which provides an element-specific magnetic signal through the difference in absorption between left- and right-circularly polarized light \cite{Vaz2025}.
XMCD has since become a standard and powerful probe for the study of magnetic transition-metal compounds.

With the emergence of femtosecond lasers, ultrafast optical spectroscopy opened the way to direct observation of nonequilibrium phenomena in condensed matter.
Techniques such as transient reflectivity, time-resolved magneto-optical Kerr effect (MOKE), and ultrafast photoemission have revealed that electronic, lattice, and spin degrees of freedom can respond on femtosecond timescales \cite{store,RevModPhys.82.2731}.
For example, the pioneering work by Beaurepaire \textit{et al.} demonstrated that the magnetization of Ni collapses within a picosecond after optical excitation. This observation launched the field of ultrafast magnetism \cite{PhysRevLett.76.4250}.
However, these optical techniques, while offering excellent temporal resolution, are inherently limited in their ability to distinguish contributions from individual elements and electronic orbitals. 
Elemental and orbital selectivity is crucial for disentangling the coupled charge, spin, and lattice dynamics in complex materials.

A major breakthrough occurred with the invention of X-ray free-electron lasers (XFELs) \cite{Schoenlein2019}. XFELs deliver intense, coherent, and ultrashort X-ray pulses typically tens of femtoseconds in duration, with peak brightness several orders of magnitude higher than that of synchrotron sources. 
Since the first operation of the Linac Coherent Light Source (LCLS) \cite{Emma2010} and SACLA \cite{Ishikawa2012}, XFELs have provided a new experimental platform that extends ultrafast spectroscopy from the optical to the X-ray regime.
This technological innovation has enabled femtosecond time-resolved X-ray absorption spectroscopy (trXAS) and time-resolved XMCD (trXMCD) measurements, which directly probe the element-specific electronic and spin dynamics following optical excitation \cite{Uemura2022}. 
As a result, it has become possible to visualize the nonequilibrium evolution of charge, spin, and lattice degrees of freedom in quantum materials with simultaneous femtosecond temporal and elemental selectivity.

Here, we provide an overview of the use of these advanced X-ray pump-probe spectroscopy techniques in the study of transition-metal compounds.
The ultrafast laser-induced magnetization dynamics discussed here are investigated using pump-probe methods. In these methods, the magnetization dynamics are excited by an ultrashort infrared or visible laser pulse, and the subsequent change in the magnetic state is monitored using a delayed probe pulse.
In the following section, we describe the experimental techniques employed, namely the pump-probe scheme and XFEL-based setups.
We then present representative examples of ultrafast magnetic phenomena, including demagnetization, magnetic phase transitions, and magnetization reversal induced by laser excitation.
Finally, we summarize the current achievements and discuss future prospects, including the potential of laboratory-based HHG sources for tabletop femtosecond X-ray spectroscopy.

\section{Experimental techniques}
\subsection{Magnetic materials}
The magnetic properties of solids arise from the spin and orbital angular momenta of electrons, particularly those in partially filled $3d$, $4f$, and $5f$ shells. The collective alignment of these atomic magnetic moments is determined by the exchange interaction, which originates from the Pauli exclusion principle and the Coulomb interaction between electrons. 
Depending on the strength, sign, and spatial dependence of the exchange coupling, various types of magnetic order emerge.

Ferromagnetism is characterized by a spontaneous parallel alignment of magnetic moments below the Curie temperature $T_{\mathrm{C}}$, resulting in a finite macroscopic magnetization even in the absence of an external magnetic field. Typical examples include Fe, Co, Ni, and their alloys, where strong exchange interactions favor parallel spin alignment. In itinerant ferromagnets, the exchange splitting of spin-up and spin-down bands can be described by the Stoner model.
In contrast, antiferromagnetism arises when neighboring magnetic moments are aligned antiparallel, leading to zero net magnetization despite long-range spin order below the Néel temperature $T_{\mathrm{N}}$. 
This order is stabilized by superexchange interactions, as found in transition-metal oxides such as NiO and MnO. Antiferromagnets exhibit characteristic magnetic excitations such as magnons with opposite spin sublattices, and their dynamics can be significantly faster than those of ferromagnets due to the absence of net magnetization.
A related but distinct type of magnetic order is ferrimagnetism, in which magnetic moments on different sublattices are antiparallel but have unequal magnitudes, producing a finite net magnetization. 
This behavior is typical of oxide spinels such as NiCo$_2$O$_4$ \cite{Shen}, as well as rare-earth-transition-metal alloys such as GdFeCo, which play a central role in all-optical magnetization switching.
Ferromagnetic, antiferromagnetic, and ferrimagnetic states are schematically shown in Fig.~1. 

In some systems, competing exchange interactions or Dzyaloshinskii-Moriya interactions (DMI) give rise to non-collinear magnetic structures such as helical or spiral magnetism. Helical spin order, as seen in materials like MnSi or TbMnO$_3$, is characterized by a spatial rotation of the spin direction with a well-defined propagation vector \cite{Tokura}. Such chiral spin textures, including skyrmions, have attracted considerable attention as candidates for ultrafast spintronic devices.


Understanding these various magnetic orders and anisotropies is fundamental for interpreting time-resolved X-ray spectroscopic experiments. Techniques such as XAS, XMCD, and resonant soft X-ray scattering provide element-specific information on the spin and orbital contributions to magnetism, making them indispensable tools for studying ultrafast magnetic phenomena in complex materials.

\begin{figure}[H]
\centerline{\includegraphics[width=6cm]{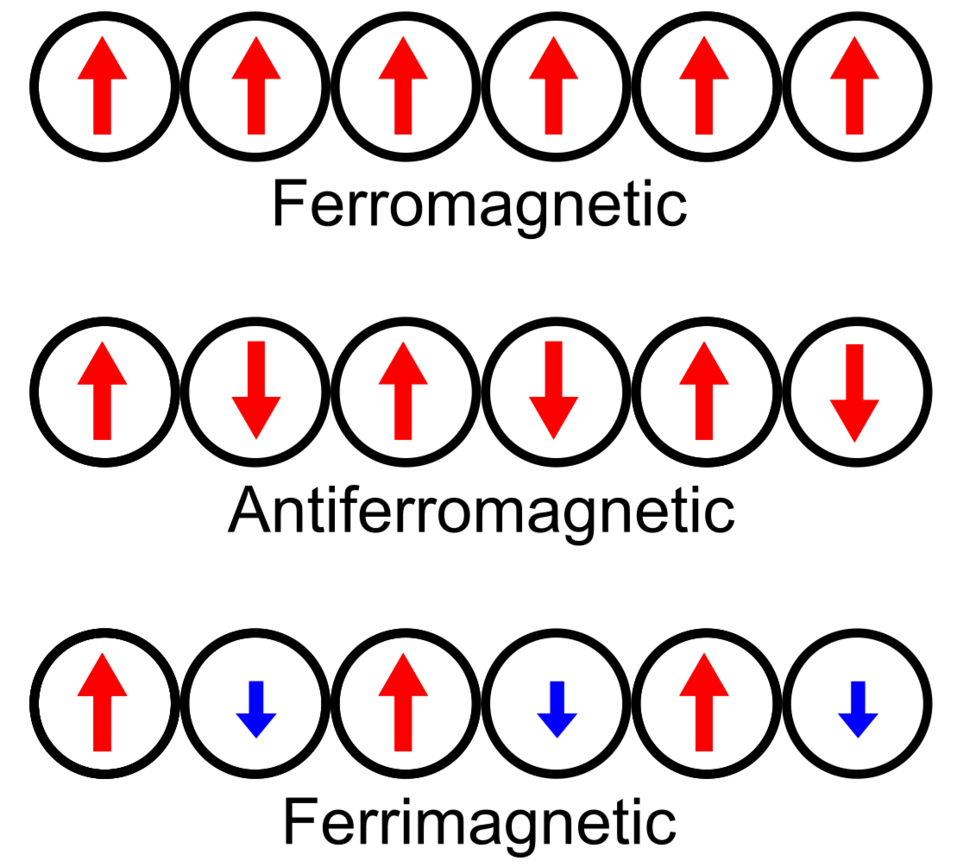}}
\caption{Schematic illustration of ferromagnetic, antiferromagnetic, and ferrimagnetic configurations. Ferromagnets show parallel spin alignment producing a net moment, antiferromagnets have antiparallel sublattices canceling the moment, and ferrimagnets possess unequal sublattices leading to finite magnetization.}
\label{magfig}
\end{figure}

\subsection{Magneto-optical effects}
A robust experimental technique for measuring magnetization is the use of superconducting quantum interference devices (SQUIDs). However, SQUIDs cannot be used to observe magnetization at the femtosecond scale under laser irradiation, and spin dynamics is investigated using magneto-optical effects. The magneto-optical effect is the interaction of light with a magnetic material. The most easily understood magneto-optical effects are the Faraday effect and MOKE. When polarized light passes through a material, the plane of polarization rotates or becomes elliptically polarized depending on the direction of magnetization of the sample. This phenomenon is called the Faraday effect. A similar effect on reflected light is called the Kerr effect.

In the visible-to-infrared range, these effects are often used because the rotation of the plane of polarization can be detected relatively easily. For example, in the case of iron, when the Faraday effect is observed by irradiating it with 578 nm visible light, the angle of rotation is $3.825 \times 10^5$ (deg/cm). In other words, it is about $380,000^{\circ}$ per cm of thickness, but in reality, visible light does not pass through iron with a thickness of 1 cm. For example, the transmittance of a 30 nm thick iron sample is about 70 \%, resulting in a rotation angle of about $1^{\circ}$. On the other hand, when the same iron sample is observed with infrared light of 0.75 eV in MOKE, the rotation angle is 0.87$^{\circ}$. These values are taken from \cite{Satokatsu}.

\begin{figure}[H]
\centerline{\includegraphics[width=12cm]{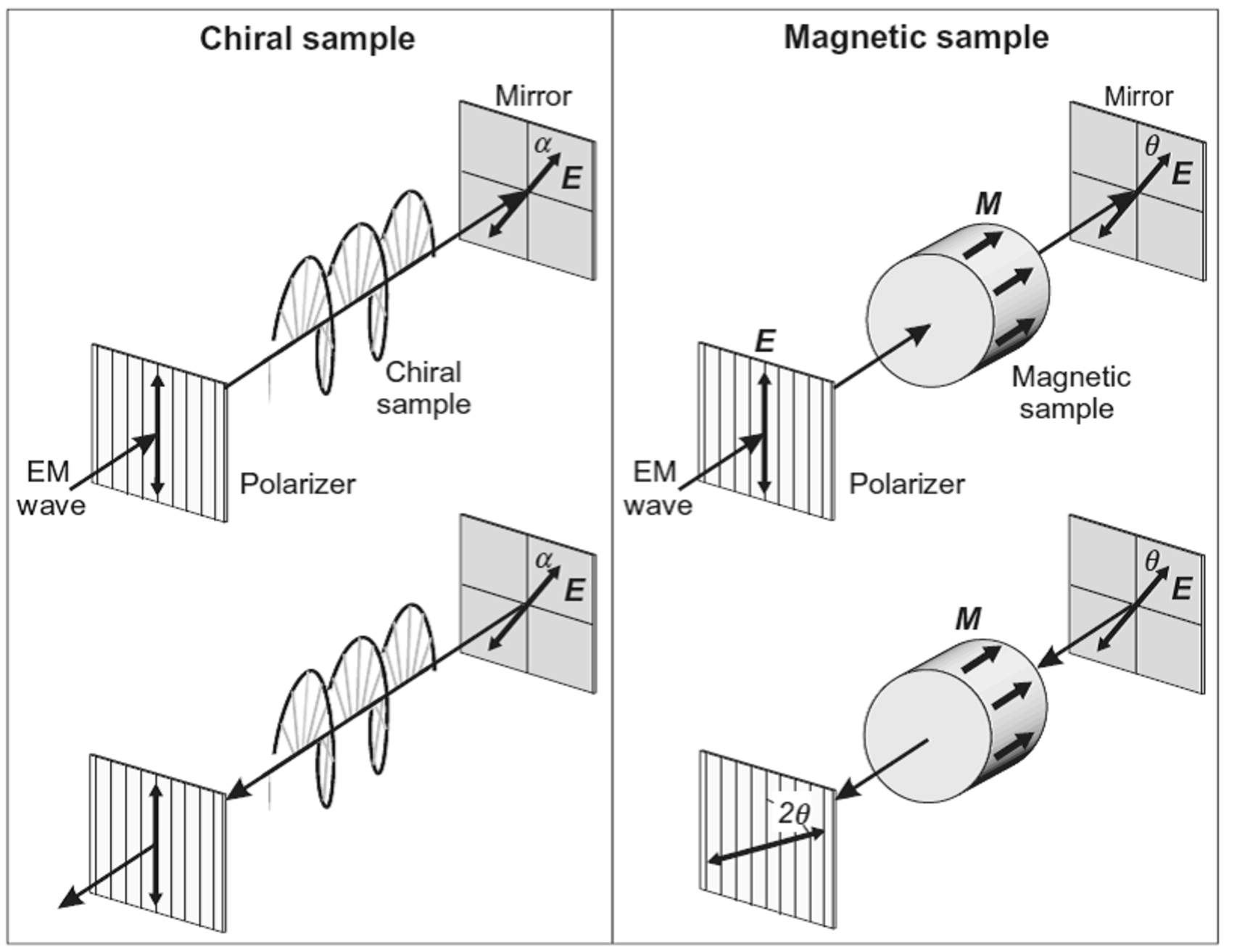}}
\caption{Comparison between the transmission of linearly polarized light through a chiral or magnetic sample (upper panel) and the transmission of the same light after reflection from the sample (lower panel) \cite{store}. The difference in polarization state provides the basis of magneto-optical contrast.}
\label{fig4}
\end{figure}

Here, we compare the transmission of light through a magnetic material and a chiral sample. Examples of chiral samples include an aqueous glucose solution and a single crystal composed of chiral molecules. Figure \ref{fig4} shows a comparison of linearly polarized light passing through a chiral sample and a magnetic material (top) and the same sample after reflection (bottom) \cite{store}. For example, consider a right-handed helix as a chiral sample. To the light, it appears to be a right-handed helix in both directions. Therefore, the rotation of the polarization plane follows the helicity of the helix, and the rotation direction relative to the direction of light propagation is the same. Because the direction of light propagation is reversed on the way there and back, the reflected wave returns to its original polarization. 

A magnetic material interacts differently with light on the forward and return paths. Since magnetization MM is a vector, it is crucial whether this vector is aligned with or opposed to the direction of light propagation. If the polarization vector rotates clockwise by $\theta$ during forward propagation, it rotates counterclockwise by $\theta$ during the return. When the electric field passes back and forth through the sample, these two rotations add, resulting in a total rotation angle of $2\theta$. Therefore, the Faraday effect in magnetic materials is fundamentally different from the optical rotation observed in chiral samples.

In the X-ray region, detecting the polarization angle is difficult, so XMCD using circularly polarized light is often employed. XMCD is the difference in response (typically absorption) when a magnetic material with magnetization aligned in one direction is irradiated with right-handed and left-handed circularly polarized light parallel to the magnetization direction.

\begin{figure}[H]
\centerline{\includegraphics[width=12cm]{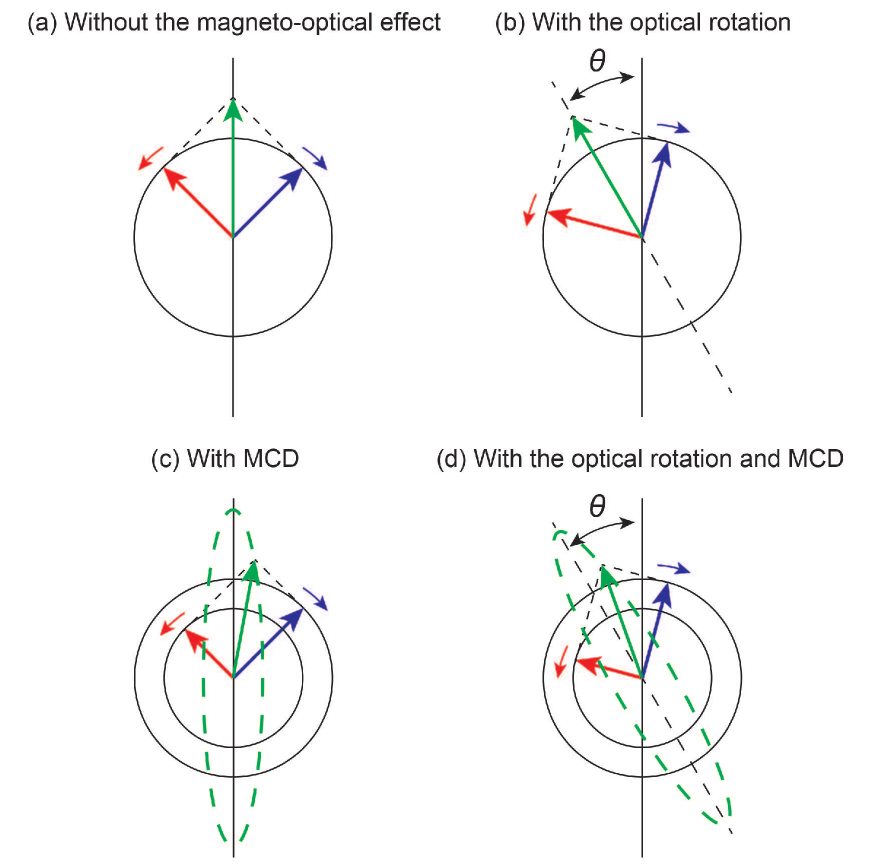}}
\caption{Schematic illustration of the origin of the magneto-optical effect. 
Rotation of the polarization angle $\theta$ arises from the interaction of circularly polarized light with spin-polarized electronic states. This interaction enables probing magnetization. \cite{phdthesis}}
\label{fig2}
\end{figure}

To summarize magneto-optical effects, Figure \ref{fig2} illustrates the relationship between the Faraday effect, which occurs with linearly polarized light, and MCD, which arises with circularly polarized light \cite{phdthesis}. The trajectory of the electric field of linearly polarized light is represented by a superposition of LCP and RCP with the same amplitude and rotation speed, as shown in Figure \ref{fig2} (a). When a phase difference occurs between the LCP and RCP after interacting with a magnetic material, the polarization plane of the linearly polarized light rotates as shown in Figure \ref{fig2} (b). This phenomenon is called optical rotation. When a difference occurs in the amplitude of the electric field of the LCP and RCP, as shown in Figure \ref{fig2} (c), it is called MCD, and the linearly polarized light changes to elliptically polarized light. When both optical rotation and MCD are observed, the light interacting with a magnetic material becomes elliptically polarized and its plane rotates (see Figure \ref{fig2} (d)).

\newpage
\subsection{Interaction of X-rays with matter}
\subsubsection{Lorentz model}
The interaction of X-rays with matter can, in some cases, be explained based on classical theory \cite{Atto}. In the classical model, a multi-electron atom is treated as a collection of harmonic oscillators, each with its characteristic resonance frequency $\omega_1, \omega_2, \cdots$. These resonances can be associated with transitions between stationary states of the atom. In this model, each bound electron is forced into simple harmonic motion by the applied electric field, while under the central restoring force of the massive, positively charged nucleus ($+Ze$). The equation of motion can then be written as follows: 
\begin{equation}
m\frac{d^2\boldsymbol{x}}{dt^2}+m\gamma \frac{d\boldsymbol{x}}{dt}
+m\omega_s^2\boldsymbol{x}=-e(\boldsymbol{E}_i+\boldsymbol{v}\times\boldsymbol{B}_i)
\end{equation}
Here, the electron charge is taken to be $-e$ ($e>0$). The first term is the acceleration, the second term is a dissipative force term with $\gamma \ll \omega_s$, and the third term is the restoring force for an oscillator of resonant frequency $\omega_s$. $-e(\boldsymbol{E}+\boldsymbol{v}\times\boldsymbol{B})$ is the Lorentz force and the term $\boldsymbol{v}\times\boldsymbol{B}$ can be neglected in the relativistic case of $v/c \ll 1$. 

With an incident electric field 
\begin{equation}
\boldsymbol{E}=\boldsymbol{E}_ie^{-i\omega t}, 
\end{equation}
the displacement $\boldsymbol{x}$, velocity, and acceleration 
all have the same $e^{-i\omega t}$ time dependence, 
and the time derivative can be replaced by $-i\omega t$. 
By substituting $\boldsymbol{x}(t)=\boldsymbol{x}_0e^{-i\omega t}$, the equation of motion becomes 
\begin{equation}
m(-i\omega)^2\boldsymbol{x}_0+m\gamma(-i\omega)\boldsymbol{x}_0+
m\omega_s^2\boldsymbol{x}_0=-e\boldsymbol{E}_i. 
\end{equation}
The harmonic displacement is given by 
\begin{equation}
\boldsymbol{x}_0=\frac{1}{\omega^2-\omega_s^2+i\gamma\omega}\frac{e\boldsymbol{E}_i}{m}, 
\end{equation}
and the acceleration is
\begin{equation}
\boldsymbol{a}=\frac{-\omega^2}{\omega^2-\omega_s^2+i\gamma\omega}\frac{e\boldsymbol{E}_i}{m}. 
\end{equation}

The radiated field $\boldsymbol{E}_{rad}$ at distance $r$ and time $t$ 
is given by the acceleration ${\boldsymbol{a}}(t-R/c)$ at 
the retarded time $t-R/c$ 
\begin{equation}
\boldsymbol{E}_{rad}(\boldsymbol{r},\ t)=
\frac{e}{4\pi\epsilon_0c^2r}{\boldsymbol{a}}
\left(t-r/c\right) 
\end{equation}
By inserting 
${\boldsymbol{a}}\left(t-r/c\right)
=-\omega^2\boldsymbol{x}_0e^{-i\omega t}e^{i\omega r/c},$
\begin{equation}
\boldsymbol{E}_{rad}(\boldsymbol{r},\ t)=
\frac{-\omega^2}{\omega^2-\omega_s^2+i\gamma\omega}
\frac{e^2}{4\pi\epsilon_0mc^2}\frac{e^{ikr}}{r}\boldsymbol{E}
\end{equation}
The atomic scattering factor $f_s$ is given in units of the electron classical radius 
$r_0=e^2/4\pi\epsilon_0mc^2=2.82\times 10^{-5}$ $\mbox{\AA}$, 
\begin{equation}
f_s=\frac{\omega^2}{\omega^2-\omega_s^2+i\gamma\omega}. 
\end{equation}
The cross section $\sigma$ is given by 
\begin{equation}
\sigma=\frac{8\pi}{3}r_e^2\frac{\omega^4}{(\omega^2-\omega_s^2)^2+(\gamma\omega)^2}. 
\end{equation}
This value shows a strong resonance at resonance $\omega=\omega_s$. 

In the limit $\omega^2 \ll {\omega_s}^2$ and $\gamma \ll \omega_s$, 
\begin{equation}
\sigma=\frac{8\pi}{3}r_e^2\left(\frac{\omega}{\omega_s}\right)^4
=\frac{8\pi}{3}r_e^2\left(\frac{\lambda_s}{\lambda}\right)^4, 
\end{equation}
which has a $\lambda^{-4}$ wavelength dependence, 
known as Rayleigh's law. 
Rayleigh first used this result to explain the blue color of the sky.
\begin{figure}[H]
\centering
\includegraphics[width=0.5\linewidth]{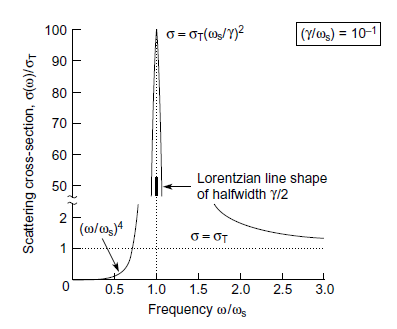}
\caption{Semi-classical scattering cross-section \cite{Atto} with definitions of the relevant physical quantities, providing the basis for interpreting resonant scattering intensities. The formulation links electronic structure to experimentally measured scattering profiles.}
\label{fig:placeholder}
\end{figure}

In the Lorentz model, physical quantities are calculated by assuming multiple oscillators. However, the strength $f_j$, the natural frequency $\omega_j$, and the damping rate $\gamma_j$ of each oscillator cannot be derived from the Lorentz model itself, and must instead be obtained by fitting to experimental data. To quantitatively discuss the characteristics of each oscillator, it is necessary to treat the electronic system quantum-mechanically.

\subsubsection{Quantum mechanical treatment}
As in the previous chapter, explanations based on classical theory have their limitations, and a rigorous analysis using quantum theory is often required. Quantum theory helps to understand the interaction between X-rays and matter. In the semi-classical theory, the electromagnetic field is treated classically, while the electronic system is treated quantum mechanically. On the other hand, in the fully quantum theory, both the electromagnetic field and the electronic system are treated quantum mechanically.

The Hamiltonian of the electronic system within an atom is given by 
\begin{equation}
H=\sum_j\frac{1}{2m}\boldsymbol{p}_j^2+H_e    
\end{equation}
where $p_j$ is the momentum of the j-th electron.
The first and second terms represent the kinetic and potential energies of the electronic system, respectively.
When an electromagnetic field is present, the vector potential $\mathbf{A}$ of the field modifies the first term by replacing $p$ with $p+e\mathbf{A}$, 
so that the Hamiltonian becomes 
\begin{equation}
H=\sum_j\frac{1}{2m}\boldsymbol{p}_j^2+\sum_j\frac{e}{m}\boldsymbol{p}_j\cdot \boldsymbol{A}+\sum_j\frac{e^2}{2m}\boldsymbol{A}^2+H_e
\end{equation}
Thus, the Hamiltonian for the interaction between the electronic system and the electromagnetic field is given as
\begin{equation}
H^{\prime}=\sum_j\frac{e}{m}\boldsymbol{p}_j\cdot \boldsymbol{A}
\end{equation}
by neglecting the term of $\boldsymbol{A}^2$. 
The time-dependent electromagnetic field induces transitions 
between an initial state $\ket{i}$ and final state $\ket{f}$. 

\begin{figure}[H]
\centering
\includegraphics[width=0.5\linewidth]{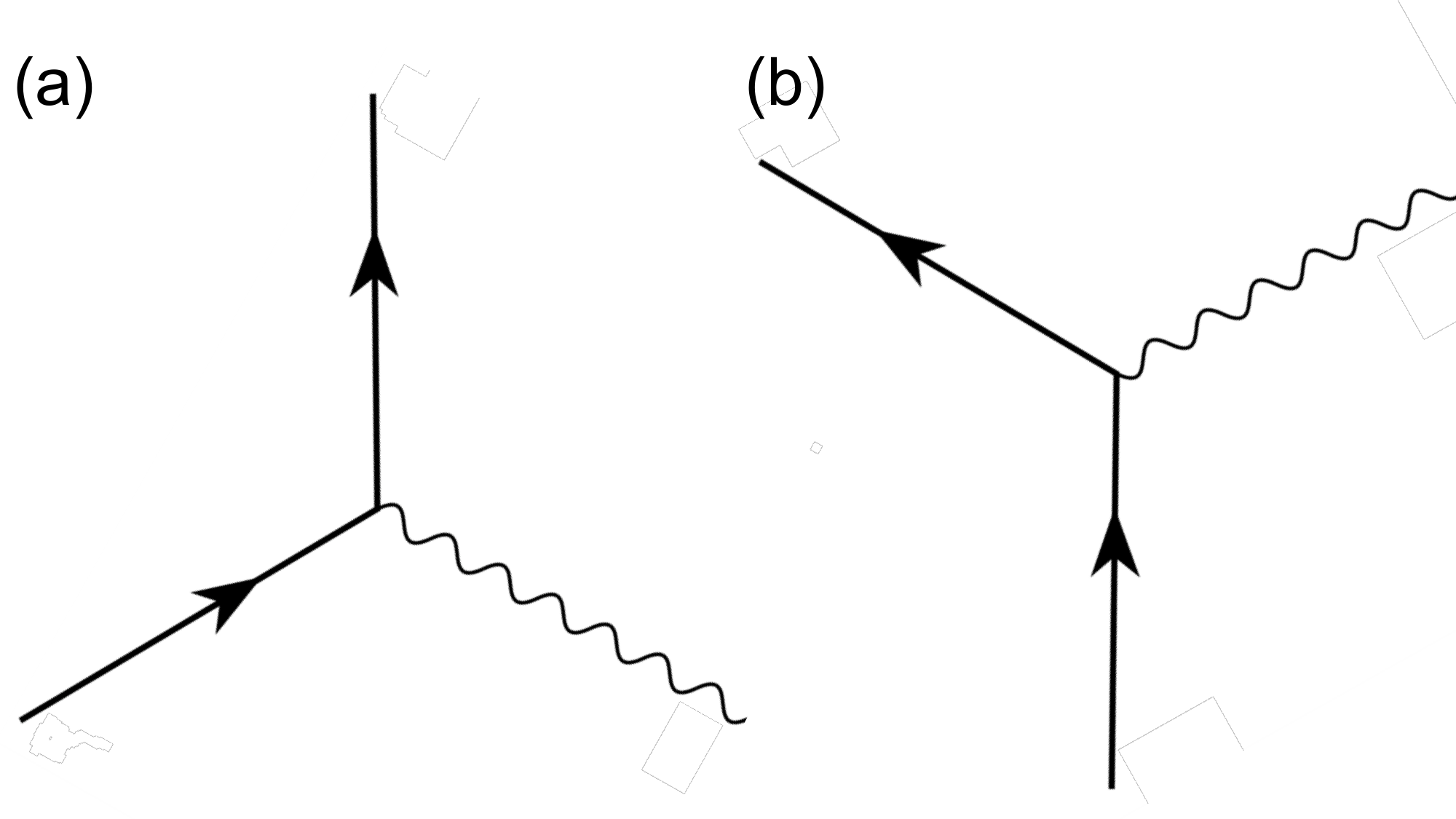}
\caption{The diagram of the first-order perturbation of $\boldsymbol{p}\cdot \boldsymbol{A}$, illustrating the interaction between the electromagnetic field and momentum operator within perturbation theory. These directly contribute to the transition probability in absorption and resonant scattering.}
\label{place}
\end{figure}

The $\boldsymbol{p}\cdot \boldsymbol{A}$ term contains one factor of $A$, 
thus its first-order perturbation corresponds to the first-order optical process involving the absorption or emission of a single photon. Figure \ref{place} shows the diagram of the first-order perturbation of $\boldsymbol{p}\cdot \boldsymbol{A}$, where panels (a) and (b) show X-ray absorption and fluorescence X-ray emission, respectively. 

\begin{figure}[H]
\centering
\includegraphics[width=0.45\linewidth]{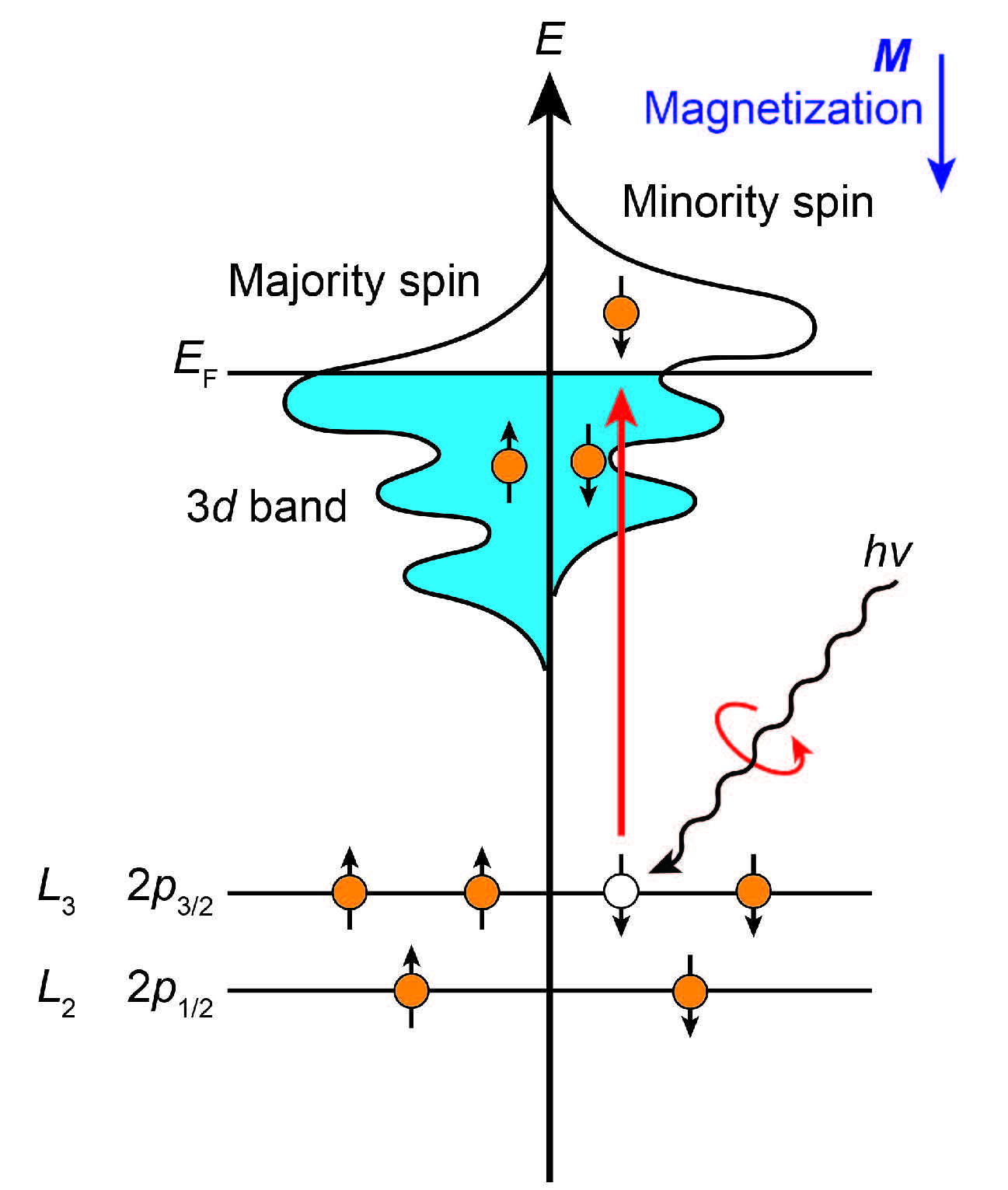}
\caption{Energy diagram of the $2p \rightarrow 3d$ transition in a ferromagnetic material \cite{phdthesis}. $E_F$ marks the Fermi level, and the spin-dependent splitting of the $3d$ states explains XMCD contrast. The diagram shows that dichroism arises from unequal spin occupation of the final states.}
\label{figX}
\end{figure}

The transition probability $T_{if}$ from $\ket{i}$ to $\ket{f}$ 
is given by Fermi's golden rule as: 
\begin{equation}
T_{if} \propto |\bra{f}{\boldsymbol{p}_j\cdot \boldsymbol{A}}\ket{i}|^2 \delta(E_f-E_i-\hbar\omega)
\end{equation}
Here, the first factor gives the modulus square over the matrix element. 
The second factor is the delta distribution, which ensures energy conservation during excitation. In the following, $\epsilon$ and $k$ denote the unit polarization vector and the wave vector, respectively, of the incident photon with energy $\hbar\omega$. The vector potential can be written as $\boldsymbol{A}=\boldsymbol{\epsilon}e^{i\boldsymbol{k}\cdot\boldsymbol{r}}$, and for $\boldsymbol{k}\cdot\boldsymbol{r}\ll 1$, it can be expanded as 
\begin{equation}
e^{i\boldsymbol{k}\cdot\boldsymbol{r}}=1+i\boldsymbol{k}\cdot\boldsymbol{r}+ \cdots 
\end{equation}
The first term yields the E1 (electric-dipole) term, and the second one the E2 (electric-quadrupole) term. In the soft X-ray region, the E1 approximation is sufficient, as will be described in the next paragraph. The relevant matrix element becomes 
\begin{equation}
\bra{f}{\boldsymbol{p}\cdot \boldsymbol{A}}\ket{i}=\bra{f}{\boldsymbol{p}\cdot \boldsymbol{\epsilon}e^{i\boldsymbol{k}\cdot\boldsymbol{r}}}\ket{i}\simeq
\bra{f}{\boldsymbol{p}\cdot \boldsymbol{\epsilon}}\ket{i}=
im\omega\bra{f}{\boldsymbol{r}\cdot \boldsymbol{\epsilon}}\ket{i},
\end{equation}
Here, $\omega = \omega_f-\omega_i$ is the photon frequency associated with the transition from the initial state $\ket{i}$ to the final state $\ket{f}$. This results in the transition operator $\boldsymbol{\epsilon}\cdot\boldsymbol{r}$, which induces transitions with $\Delta l =\pm 1$. Therefore, when ferromagnetism causes a difference in the occupation numbers of spin-up and down $3d$ electrons in the ground state, this can be observed as circular dichroism in the $2p\rightarrow 3d$ absorption shown in Fig.~\ref{figX} \cite{PhysRevLett.75.152}. 

The dipole approximation is based on the assumption that the size of the absorbing atomic shell is much smaller than the X-ray wavelength, i.e., $|\boldsymbol{r}| \ll 1/|\boldsymbol{k}| = \lambda/2\pi$, so that the electric field which causes the electronic transition is constant over the atomic volume. In the soft X-ray regime, the photon energy range around 1 keV corresponds to a wavelength $\lambda \sim 1.2$ nm and transitions from the $2p$ core shell of radius $|\boldsymbol{r}| \sim 0.01$ nm; thus, we have $|\boldsymbol{r}| \sim 0.01$ nm $\ll \lambda/2\pi \sim 0.2$ nm, which make it reasonable to use the dipole approximation. 

On the other hand, in the hard X-ray region around 10 keV,  $|\boldsymbol{r}| \sim 0.01$ nm becomes comparable to $\lambda/2\pi \sim 0.02$ nm.  Consequently, the E2 term corresponding to transitions with $\Delta l =\pm 2$ can be observed. For $3d$ electrons, this is the $1s\rightarrow 3d$ absorption. More detailed theoretical treatment is written in Refs .~\cite{VANDERLAAN2014}.

\subsection{Resonant soft X-ray scattering}
\begin{figure}[H]
\centering
\centerline{\includegraphics[width=9 cm]{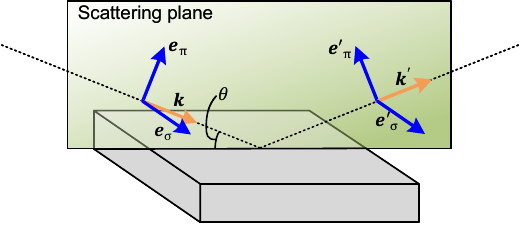}}
\caption{Experimental geometry for RSXS, indicating incidence and scattering angles, sample orientation, and the configuration used to detect magnetic Bragg peaks. This geometry is key to achieving momentum-resolved and element-specific magnetic contrast.}
\label{RSXSRSXS}
\end{figure}

Resonant soft X-ray scattering (RSXS) has recently been established as a powerful experimental method for elucidating the electric charge, orbital, and magnetic ordered structures \cite{Fink}. Conventionally, X-ray diffraction has been used to determine the crystal structures and lattice constants of materials because X-rays are primarily scattered by the charge of electrons in materials. On the other hand, RSXS is a diffraction method that exploits resonance scattering induced by inner-shell absorption. This measurement method uses the element absorption edge to enhance the interaction between X-rays and specific elements. 
As shown in Fig.~\ref{figX} \cite{phdthesis}, X-ray absorption occurs when irradiated using X-rays with energy at the absorption edge from $2p$ to 3d, and the $2p$ electrons transit to the 3d unoccupied state (intermediate state). Then, the excited $3d$ electrons recombine with the remaining holes in the 2p orbitals, emitting X-rays. In this elastic-scattering process, the initial and final states coincide. Here, spin-orbit interaction causes the inner shell of $2p$ to have an energy split of approximately 10 eV in $2p_{3/2}$ and $2p_{1/2}$, resulting in the acquisition of a magnetic signal. Figure \ref {RSXSRSXS} shows the experimental geometry for RSXS. Bragg peaks, such as those for antiferromagnetism, can be observed by a detector at a scattering angle of $2\theta$. In addition to observing antiferromagnetic peaks, RSXS can also be used to observe ferromagnetism through X-ray magnetic circular dichroism in reflectivity (XMCDR). In XMCDR, a ferromagnetic signal can be obtained by taking the difference in the reflectance values of X-rays with left and right circularly polarized light. 
\subsection{Pump-probe method}
\begin{figure}[H]
\centering
\includegraphics[width=10cm]{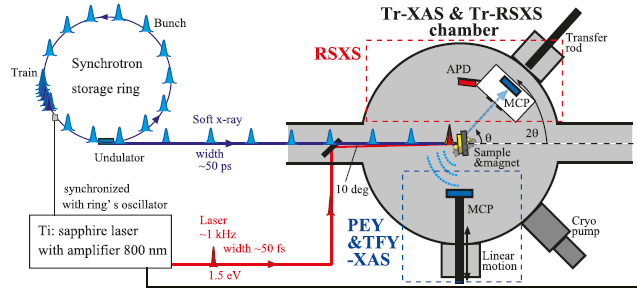}
\caption{Overview of the experimental setup for trXAS, trXMCD, and trRSXS measurements at BL07LSU of SPring-8. The Ti: sapphire laser, with a $\sim 1$ kHz repetition rate, is synchronized with the synchrotron bunches and delayed electronically before being introduced into the experimental chamber. \cite{TakuboAPL}.}
\label{BL07setup}
\end{figure}

First, we describe the pump–probe experimental setup at BL07LSU of SPring-8. Figure~\ref{BL07setup} shows an overview of the experimental setup for trXAS, trXMCD, and time-resolved RSXS (trRSXS) measurements in the soft X-ray region \cite{BL07LSU,TakuboAPL}. The trRSXS measurements are performed on the $\theta < 90^{\circ}$ side of the experimental chamber, where $\theta$ denotes the incident angle between the X-ray beam and the sample surface, as defined in conventional X-ray diffraction geometry. Scattered X-rays are detected using either a microchannel plate (MCP) or an avalanche photodiode (APD) mounted on the $2\theta$ arm of the diffractometer.
 
On the opposite side ($\theta > 90^{\circ}$), XAS and XMCD measurements are carried out in partial electron yield (PEY) and total fluorescence yield modes. Photoelectrons or fluorescence photons emitted from the sample are collected by a chevron-type (dual-plate) MCP mounted on a linear motion stage, which can be positioned as close as $\sim2$~cm from the sample surface. The MCP detects both photoelectrons and X-ray fluorescence; when a positive bias is applied to the MCP-in terminal, the signal predominantly originates from photoelectrons, whose larger current contribution allows spectra to be regarded as PEY-derived.

A femtosecond Ti: sapphire laser with a wavelength of $\sim 800$ nm, installed at the BL07LSU laser station \cite{Ogawa}, is introduced into the XAS and RSXS chamber. The laser beam is incident on the sample at an angle of $\sim 10^{\circ}$ below the X-ray beam. Photo-induced dynamics of the electronic and structural states are examined using the pump-probe technique, in which the laser pulses (repetition rate $\sim$1~kHz) are electronically synchronized with specific synchrotron bunches and delayed with respect to the X-ray probe. The laser pulse duration is $\sim 50$ fs, whereas the single-bunch width of the synchrotron X-rays in the H- and F-modes at SPring-8 is $\sim 50$ ps, which limits the overall temporal resolution.

The temporal width of synchrotron radiation X-ray pulses is on the order of several tens of picoseconds, which is much longer than that of a Ti: sapphire laser. Consequently, when synchrotron radiation X-rays are used as a probe, the overall time resolution of the measurement is limited to several tens of picoseconds. To investigate electron dynamics on the femtosecond timescale, it is therefore necessary to employ advanced X-ray sources \cite{Stamm2007,Higley}. 
Such X-ray sources are 
XFELs, synchrotron radiation with X-ray laser slicing, or X-rays generated via high-harmonic generation (HHG) from a laboratory laser \cite{Schoenlein2019}, which will be described in detail in the following section.  

\subsection{Ultrafast X-ray sources}
\subsubsection{X-ray free-electron lasers}
One of the most significant recent advances in ultrafast X-ray science is the advent of XFELs. XFELs deliver coherent X-ray pulses with durations of 10-100 fs and peak intensities orders of magnitude higher than those of conventional synchrotron radiation, enabled by advances in electron accelerator technology. Whereas earlier X-ray sources, such as synchrotrons, were primarily used for static measurements, XFELs are well-suited for probing materials dynamically on timescales of lattice vibrations and spin dynamics. Table 1 summarizes the key parameters of major XFEL facilities worldwide, including photon energy range, pulse energy and duration, repetition rate, and the corresponding country and beamline name \cite{Schoenlein2019}. Typical pulse widths are on the order of 10 - 100 fs, with pulse energies of approximately 1 mJ.

Next-generation XFELs represent a significant advancement in X-ray science and are currently under development. These facilities employ superconducting RF accelerator technology to deliver ultrafast X-ray pulses at high repetition rates. The upgrade of the U.S. XFEL facility, LCLS (LCLS-II), will be the first of these next-generation systems capable of producing soft X-rays at repetition rates of up to 1 MHz, a breakthrough that could transform the capabilities of XFEL-based research. 

\begin{table}[H]
\begin{center}
\caption{Characteristic parameters of the various XFEL facilities worldwide adapted from Ref.~\cite{Schoenlein2019}. These include the supporting country, the facility name, photon energy (keV), pulse energy (mJ), pulse duration (fs), and repetition rate (Hz).}
\begin{tabular}{cccccc}\hline
 & & Photon & X-ray pulse & X-ray pulse & Repetition\\
Country & Name & energy (keV) & energy (mJ) & duration (fs) & rate (Hz)\\ \hline
Japan & SACLA BL2, 3 & 4-20 & 0.1-1 & 2-10 & 60 \\ 
 & SACLA BL1 & 0.04-0.15 & 0.1 & 60 & 60 \\ \hline
Italy & FERMI-FEL-1 & 0.01-0.06 & 0.08-0.2 & 40-90 & 10 (50) \\
 & FERMI-FEL-2 & 0.06-0.3 & 0.01-0.1 & 20-50 & 10 (50) \\ \hline
Germany & FLASH1 & 0.02-0.3 & 0.01-0.5 & 30-200 & (1-800)$\times$10 \\
 & FLASH2 & 0.01-0.3 & 0.01-1 & 10-200 & (1-800)$\times$10 \\ \hline
Korea & PAL-XFEL & 2.5-15 & 0.8-1.5 & 5-50 & 60 \\
 &  & 0.25-1.2 & 0.2 & 5-50 & 60 \\ \hline
Switzerland & SwissFEL & 1.8-12.4 & 1 & 10-70 & 100 \\ 
 &  & 0.2–2 & 1 & 10-70 & 100 \\\hline
Europe & XFEL-SASE1,2 & 3-25 & 2 & 10-100 & 2700$\times$10 \\ 
 & XFEL-SASE3 & 0.2-3 & 2 & 10-100 & 2700$\times$10 \\ \hline
USA & LCLS & 0.3-12 & 2-4 & 2-500 & 120 \\
 & LCLS-II-HE & 0.2-13 & 0.02-1 & 10-200 & $10^6$ \\ \hline
\end{tabular}
\end{center}
\end{table}

\subsubsection{Synchrotron-based X-ray}
Synchrotron radiation X-ray pulses, owing to the intrinsic stability of an electron beam circulating in a closed orbit, can be used effectively for time-resolved X-ray measurements. However, in electron storage rings, accelerator physics imposes fundamental limits on the achievable electron bunch duration and, consequently, on the time resolution. Modern synchrotron facilities are designed to optimize both the average beam brightness and the electron-bunch storage lifetime, typically resulting in bunch lengths of several tens of picoseconds. Consequently, synchrotron sources are well suited for time-resolved X-ray measurements on timescales longer than tens of picoseconds. Nevertheless, innovative approaches are required to push the time resolution into the ultrafast femtosecond regime.

\begin{figure}[H]
\centerline{\includegraphics[width=10cm]{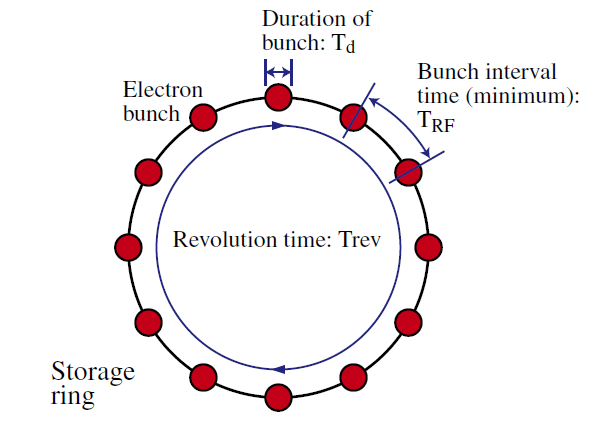}}
\caption{Time structure of electron storage rings \cite{JPSJ.82.021003}, showing the bunch timing pattern relevant for synchronization in ultrafast X-ray experiments. The pulse spacing defines the accessible time resolution and repetition rate in pump–probe studies.}
\label{figsusumu}
\end{figure}

Figure \ref{figsusumu} summarizes the time structures of electron storage rings at various synchrotron radiation facilities \cite{JPSJ.82.021003}. The time structure is characterized by three main time constants: the round-trip time ($T_{rev}$), the bunch interval ($T_{RF}$), and the bunch length or duration ($T_d$). The round-trip time, $T_{rev}$, for an electron bunch traveling at nearly the speed of light, is on the order of microseconds. The bunch interval, $T_{RF}$, determined by the electron filling pattern in the storage ring, is typically on the order of nanoseconds. The bunch length, $T_d$ - which corresponds to the synchrotron radiation pulse width - is on the order of several tens of picoseconds. These values reveal a hierarchical time structure, with each timescale differing by roughly three orders of magnitude.
The pump-probe method, in which a synchrotron radiation pulse is used as the probe and a femtosecond laser pulse as the pump, has been implemented. The time resolution in this case is on the order of several tens of picoseconds, corresponding to the temporal width of the synchrotron radiation pulse. To achieve ultrafast femtosecond time resolution with synchrotron radiation, a technique known as laser slicing has been developed to produce ultrashort X-ray pulses. In this method, a high-intensity ultrashort laser pulse is incident on an electron bunch. Owing to the strong electric field of the laser pulse ($10^9$ V/m), a very short electron bunch is extracted from the longer bunch. The intense femtosecond optical pulse is resonantly matched to the periodic motion of electrons in an undulator. The excited electrons are then spatially separated from their neighbors—this process is called laser slicing. The spike-like portion of the bunch emits ultrashort X-ray pulses, while the remainder generates normal synchrotron radiation. This laser slicing technique has been implemented at several facilities, including ALS \cite{Schoenlein}, BESSY \cite{PhysRevLett.97.074801,Hol}, SLS \cite{PhysRevLett.99.174801}, and SOLEIL \cite{Labat}. Although laser slicing can reliably generate femtosecond optical pulses, the photon flux is very low, making practical measurements challenging. This can be understood from the following simple estimate: the repetition rate of a high-intensity laser is typically on the order of kilohertz, i.e., about $10^{-5}$ of the synchrotron repetition frequency (typically around 100 MHz). In addition, the ratio of the spike duration to the total electron bunch duration is 100 fs/50 ps $= 1/500 \sim 10^{-3}$. Therefore, the photon flux of a laser-sliced femtosecond optical pulse is roughly $10^{-8}$ of that of a standard synchrotron radiation pulse.

\begin{figure}[H]
\centerline{\includegraphics[width=10cm]{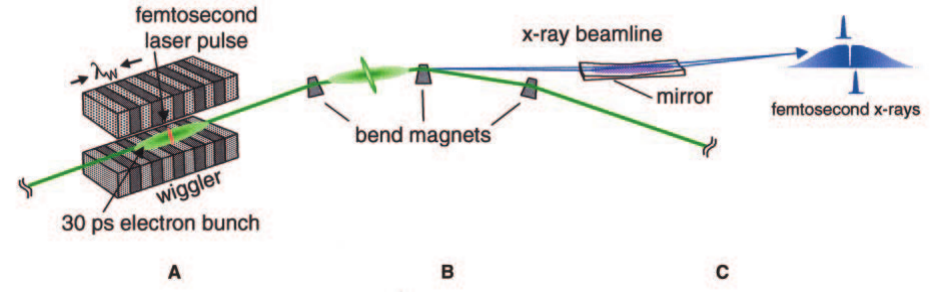}}
\caption{Schematic of the laser-slicing method for generating femtosecond synchrotron pulses \cite{Schoenlein}, where energy-modulated electrons emit ultrashort bursts of X-rays. This method expands synchrotron capabilities toward femtosecond-resolved magnetic measurements.}
\label{fig5}
\end{figure}

\subsubsection{High-harmonic generation}
In parallel with the development of XFELs, ultrafast X-ray sources 
have also made great progress due to dramatic advances in tabletop laser technology. 
The most successful approach is HHG \cite{PhysRevA.39.5751,Paul2001,Hentschel2001}. 
HHG is a nonlinear optical phenomenon that generates light at a high frequency, an integer multiple of the original laser frequency, when a medium, such as a gas or a solid, is irradiated with very intense laser light. By irradiating rare gases such as argon or neon with a femtosecond-pulse laser, one can generate harmonics extending into the extreme ultraviolet (XUV) and soft X-ray regions. A so-called three-step model can explain this phenomenon. In this model, electrons in atoms are extracted by tunneling induced by a strong laser electric field. The electrons are then accelerated by the laser field and emit light upon recombining with the parent ion. The emitted light has a much shorter wavelength than the original laser, and it exhibits unique coherence and ultrashort-pulse characteristics. These unique characteristics make HHG a powerful tool in optics and laser physics. High-order harmonics have attracted much attention because they can produce ultrashort X-ray pulses with tabletop-sized equipment, a capability previously reserved for large-scale facilities such as XFEL.

\begin{figure}[H]
\centerline{\includegraphics[width=7.5cm]{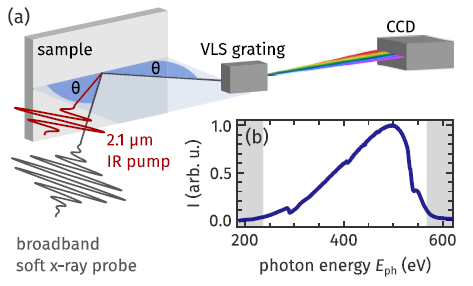}}
\caption{
(a) Setup for time-resolved energy-dispersive soft-X-ray diffraction using an HHG source. (b) Typical HHG soft-X-ray emission spectrum spanning 200–600 eV \cite{RSXSWater}. The broadband emission enables simultaneous probing of multiple resonances in a single measurement.}
\label{figRR}
\end{figure}

The key to obtaining shorter-wavelength photons via HHG is to increase the driving laser wavelength ($\lambda$). However, it is essential to note that the HHG efficiency drops significantly in proportion to $\lambda^{-(5-6)}$. Using an ultrashort laser with a wavelength of $\sim 2000$ nm, HHG can be generated over a wide spectral range, up to the water window at about 500 eV. 
This versatile broadband emission makes HHG sources in the soft X-ray region intriguing for a wide range of applications, including absorption spectroscopy and even diffraction experiments. 
Recently, a soft X-ray diffraction setup based on HHG in the water-window region has been developed, enabling time-resolved experiments as shown in Fig.~\ref{figRR} \cite{RSXSWater}.

\subsubsection{Complementarity with neutron scattering}
Recent advancements in neutron techniques provide information that is highly complementary to X-ray–based approaches. Unlike X-rays, neutron scattering cross-sections do not scale monotonically with atomic number, and neutrons couple directly to magnetic moments. Owing to this sensitivity, neutron scattering has long been a central tool for determining magnetic structures and collective spin behaviors in solids. In addition, neutrons offer outstanding contrast for light elements—particularly hydrogen—enabling quantitative studies of water adsorption, proton transport, and hydrogen-bonding networks that are extremely difficult to access using X-rays alone \cite{Chakraborty2025}. 
However, time-resolved neutron experiments are fundamentally limited by neutron flux and pulse duration, which restricts the achievable temporal resolution to the millisecond–microsecond regime. Consequently, ultrafast phenomena on femtosecond–picosecond timescales have been almost exclusively investigated using X-rays. Taken together, neutrons and ultrafast X-rays should be viewed as strongly complementary probes: neutrons map equilibrium magnetic correlations and low-energy collective excitations with high sensitivity, whereas X-rays reveal nonequilibrium electronic, spin, and structural dynamics that emerge immediately after photoexcitation.
The comparison is summarized in Table 2. 

\begin{table}[H]
\centering
\caption{Comparison of major probes for ultrafast and magnetic dynamics.}
\begin{tabular}{lcccc}
\hline
Technique &
Timescale &
Element selectivity &
Magnetic sensitivity &
Limitations \\
\hline
Synchrotron X-ray &
50 ps -- ns &
Yes (XAS/XMCD) &
Yes (XMCD/RSXS) &
Temporal resolution limited \\
 &
 &
 &
 &
to $>$ 50 ps (unless slicing) \\
\hline
XFEL &
10 -- 100 fs &
Yes (XAS/XMCD) &
Yes (XMCD/RSXS) &
Number of facilities limited \\
 &
 &
 &
 &
Sample damage at high fluence \\
\hline
HHG (tabletop) &
10 -- 100 fs &
Yes (XAS/XMCD) &
Yes (XMCD/RSXS) &
Flux limited\\
 &
 &
 &
 &
S/N challenging for weak signals \\
\hline
Neutrons &
ms -- $\mu$s  &
No  &
Excellent &
Time resolution limited \\
 &
 &
 &
 &
Weak signals from thin films \\
\hline
\end{tabular}
\label{tab:probe_comparison}
\end{table}

\newpage
\section{Ultrafast demagnetization}
\subsection{Background}
\begin{figure}[H]
\centerline{\includegraphics[width=10cm]{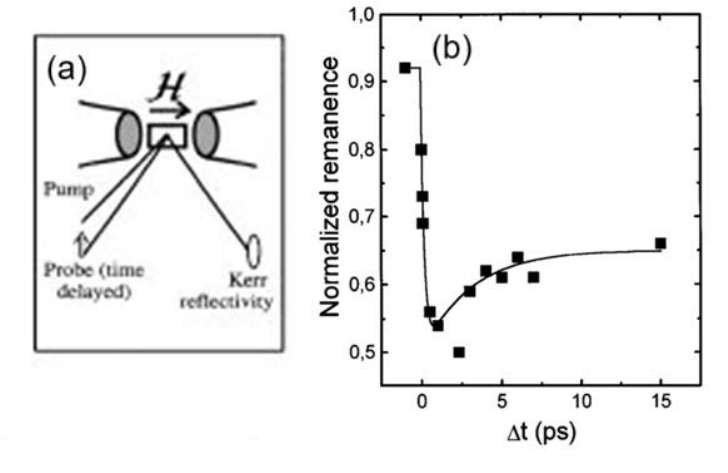}}
\caption{(a) Experimental pump-probe setup allowing dynamic longitudinal MOKE and transient transmissivity or reflectivity measurements. (b) Transient remanent longitudinal MOKE signal of a Ni(20 nm)/MgF$_2$ (100 nm) film for 7 mJ/cm$^2$ pump fluence \cite{PhysRevLett.76.4250}. This rapid decrease is the ultrafast demagnetization.}
\label{fig55}
\end{figure}

The magneto-optical effect was first used to study ferromagnetic spin dynamics. A strong pump pulse with duration $\Delta t_1$ is used to excite electron-hole pairs. Then, a weak probe pulse with duration $\Delta t_2$ is applied at varying delays relative to the pump to obtain information on the magnetization dynamics using MOKE. The first such trMOKE measurements with a time resolution below 1 ps were reported in Ref.~\cite{PhysRevLett.76.4250}. Beaurepaire \textit{et al.} performed measurements using laser pulses with durations of $\Delta t_1 = \Delta t_2 = 60$ fs, a pump light with an energy of 2 eV, and an intensity of 7 mJ/$\mathrm{cm}^2$. Figure \ref{fig55} (a) shows the results for a polycrystalline Ni film. To measure the remanent magnetization of the Ni film relative to the remanent magnetization in the absence of a pump pulse, they rotated the polarization plane of the linearly polarized explanatory pulse in MOKE. By varying the delay between the pump and probe pulses, the magnetization exhibits a rapid decrease on the first timescale of approximately 1 ps, followed by a slow recovery that levels off to a plateau of approximately two-thirds for delays up to 15 ps.
This work attracted considerable attention because it demonstrated ultrafast demagnetization on a timescale of less than 1 ps, which contradicted predictions based on spin-lattice relaxation times of several hundred picoseconds. It is considered the pioneering paper in ultrafast spin dynamics. After the first observation of ultrafast demagnetization on the femtosecond scale, it became clear that in metals, there is a much stronger coupling between the spin and the other two structures, the electron and the lattice. The effective electron-spin or phonon-spin coupling mechanism responsible for demagnetization timescales shorter than 100 fs has been the subject of active debate to date, with no clear conclusion yet reached \cite{CHEN2025}.

\subsection{Phenomenological three-temperature model}
\begin{figure}[H]
\centerline{\includegraphics[width=15cm]{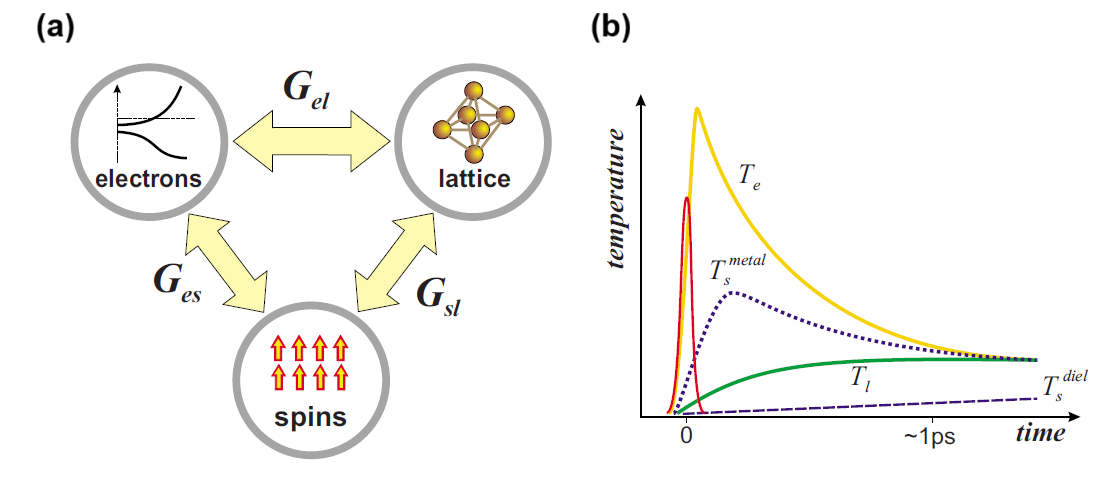}}
\caption{Three-temperature model \cite{RevModPhys.82.2731}.
(a) Interacting electrons, spins, and lattice.
(b) Temporal evolution of the electron, spin, and lattice temperatures after laser-pulse excitation. The pulse centered at $t=0$ represents the incident laser intensity. This framework offers a physical interpretation of ultrafast demagnetization.}
\label{fig23}
\end{figure}

The description of the laser-excited state can be qualitatively explained by a model involving three interacting degrees of freedom, namely, electron, lattice, and spin, as shown in Fig.~\ref{fig23} (a) \cite{RevModPhys.82.2731}. Interactions of different origins link these three degrees of freedom. Each of these degrees of freedom can be assigned a specific effective temperature, denoted as the electron temperature $T_e$, the lattice temperature $T_l$, and the spin temperature $T_s$. Then, the time evolution of $T_e$, $T_l$, and $T_s$ can be described by the following phenomenological three-temperature model:
\begin{eqnarray}
C_e\frac{dT_e}{dt} & = & -G_{el}(T_e-T_l)-G_{es}(T_e-T_s)+P(t)\\
C_s\frac{dT_s}{dt} & = & -G_{es}(T_s-T_e)-G_{sl}(T_s-T_l)\\
C_l\frac{dT_l}{dt} & = &  -G_{el}(T_l-T_e)-G_{sl}(T_l-T_s).
\end{eqnarray}
Here, $G_{el}$ denotes the interaction between electrons and the lattice, $G_{es}$ between electrons and the spin, and $G_{sl}$ between spin and the lattice. These are phenomenological parameters, and even if the details of the interaction mechanism remain unknown, one can still compare and discuss the magnitudes of the interactions. $C_e$, $C_s$, and $C_l$ are the heat capacities of the electrons, spins, and lattices, respectively. $P(t)$ denotes the increase in electron temperature due to laser irradiation.

\begin{figure}[H]
\centerline{\includegraphics[width=8cm]{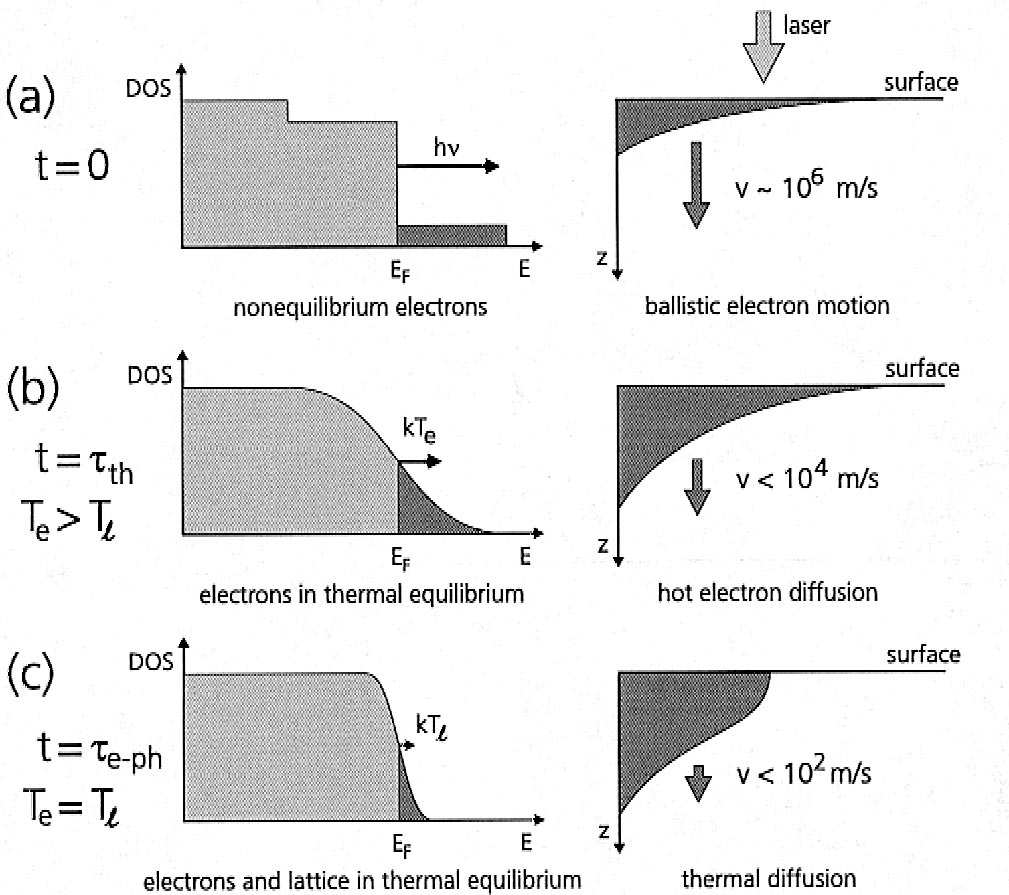}}
\caption{Three characteristic relaxation phases of optically excited electrons in metals \cite{feld}. (a) Photon absorption generates nonequilibrium electrons, which move with ballistic velocities. (b) At $t=\tau_{th}$, electrons have equilibrated by electron-electron collisions, forming a Fermi distribution with a well-defined electron temperature $T_e$. (c) Via electron–phonon coupling, the electrons come into equilibrium with the lattice at $t = \tau_{e-ph}$.}
\label{feldfig}
\end{figure}

The magnitude of the heat capacity determines the extent to which the effective temperature rises. For example, in ordinary materials, $C_e$ is 1-2 orders of magnitude smaller than $C_l$, and $T_e$ reaches several thousand K within the first tens of femtoseconds after laser excitation, while $T_l$ can remain relatively low. Figure \ref{fig23} (b) shows the behavior of $T_e$, $T_s$, and $T_l$ after femtosecond laser excitation \cite{RevModPhys.82.2731}. It has been observed that $T_s$ behaves differently in metals and dielectrics (insulators). The process that occurs after laser excitation is typically described as follows. 
\begin{enumerate}
\item In the visible to infrared laser range, only electrons whose energy satisfies $E_F-h\nu<E<E_F$ respond to laser irradiation and absorb photons almost instantly. Therefore, the laser first interacts with the sample, generating electron-hole pairs on a time scale of about 1 fs. 
\item The electron system equilibrates to high-temperature $T_e$ by electron-electron interactions within about 50-500 fs. 
\item The equilibrated electronic excitations heat the lattice and increase $T_l$ on a time scale determined by the electron-lattice interaction time, which is about 100 fs - 1 ps. \end{enumerate}
Thus, after about 1 ps, the electron system and the lattice system are in thermal equilibrium with each other, i.e., $T_e = T_l$. These three processes are summarized as panels (a)-(c) in Fig.~\ref{feldfig} \cite{feld}. 

The next question is, "How does the spin temperature $T_s$ behave?" Since the essence of magnetization is angular momentum, it is necessary to consider the conservation of angular momentum separately from energy. The demagnetization process, which removes magnetization by irradiating the spin system with a laser, requires removing a certain amount of angular momentum from the spin system. Generally speaking, both the electron system and the lattice system can absorb this angular momentum, even if only temporarily. The spin-lattice interaction is usually considered to be on the same order as the magnetocrystalline anisotropy of 100 $\mu$eV. Therefore, it was theoretically expected that the corresponding spin-lattice interaction time would be very long, on the order of several hundred ps. On the other hand, the question arises, "How does it behave when observed experimentally?" To do this, time-resolved magneto-optical measurements must be performed, as described in the next section.

\subsection{Synchrotron time-resolved measurements}
\begin{figure}[H]
\centering
\includegraphics[width=10cm]{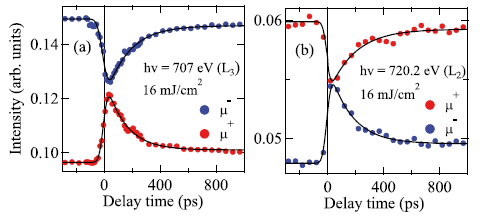}
\caption{
Time evolutions of XMCD in the PEY mode at the $L_3$ ($h\nu = 707.0$ eV) and $L_2$ ($h\nu=720.2$ eV) edges for the FePt thin film at room temperature \cite{TakuboAPL}, demonstrating ultrafast demagnetization broadened by the 50 ps X-ray pulse width of SPring-8.}
\label{BL07setup2}
\end{figure}

In SPring-8 BL07LSU, the photo-induced dynamics of the FePt thin film was examined in the PEY mode. The temporal evolutions of the FePt film excited by a laser fluence of 16~mJ/cm$^{2}$ are shown in Fig.~\ref{BL07setup2} (a) at $h\nu = 707.0$ eV (Fe $L_{3}$ edge) and Fig.~\ref{BL07setup2} (b) at 720.2 eV (Fe $L_{2}$ edge), respectively. As shown, the time-dependent behaviors at the Fe $L_{2,3}$ edges are nearly identical. The XMCD, defined as the difference in absorption intensities between right- and left-circularly polarized light, decreases immediately after the pump-pulse excitation at both Fe $L_{2,3}$ edges. The XMCD intensity is reduced by approximately 90\% of its initial value within about 30 ps after the pump pulse, which corresponds to the temporal resolution limited by the 50 ps X-ray pulse width of SPring-8. 

These results clearly indicate that synchrotron-based measurements are inherently limited by their temporal resolution, and that femtosecond-scale dynamics can only be captured with XFELs.
\subsection{Effect of time resolution}
\begin{figure}[H]
\centerline{\includegraphics[width=8cm]{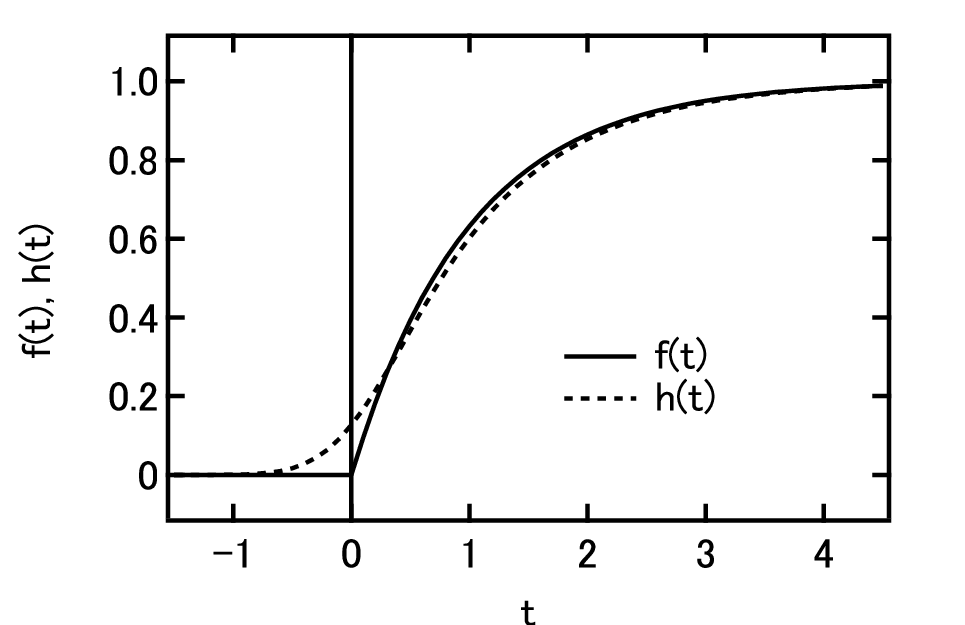}}
\caption{$f(t)$ and $h(t)$ with finite temporal width used to model the instrumental response and convolution in pump–probe measurements. These functions are essential for quantitative fitting of ultrafast magnetization dynamics.}
\label{fig6}
\end{figure}

Here, we demonstrate how the dynamics following laser excitation are observed with a finite time resolution. First, the dynamics induced by laser irradiation at $t=0$ are described by an exponential function $f(t)$ as shown below.
\begin{align}
f(t)=
\begin{cases}
0 & (t<0) \\
a(1-\exp(-\lambda t)) & (t\geq0)
\end{cases}
\end{align}
Here, the probe delay time is set to zero, and the blurring effect caused by the finite time resolution is represented by a Gaussian function $g(t)$ as shown below.
\begin{equation}
g(t)=\frac{1}{\sqrt{2\pi}\sigma}\exp(-t^2/2\sigma^2)
\end{equation}
Then, the actually measured signal corresponds to the convolution of 
$f(t)$ and $g(t)$, denoted as $h(t)$.
\begin{eqnarray}
 h(t) & = & (f*g)(t)\\
& = & \int^{\infty}_{-\infty}f(t-y)g(y)dy\\
& = & \frac{a}{2}\left(\mbox{erfc}\left(\frac{-t}{\sqrt{2}\sigma}\right)-\exp\left(\frac{\lambda^2\sigma^2}{2}\right)
\mbox{erfc}\left(\frac{\lambda\sigma^2-t}{\sqrt{2}\sigma}\right)\exp(-\lambda t)\right)
\end{eqnarray}
Here, erfc(x) is called the complementary error function.
\begin{equation}
\mbox{erfc}(x)=\frac{2}{\sqrt{\pi}}\int^{\infty}_{x}\exp(-y^2)dy
\end{equation}
Figure \ref{fig6} shows $f(t)$ and $h(t)$ for $\lambda=1$ and $\sigma=0.4$. 
In fitting the experimental results, Eq.~(25) is used.

The ultrafast laser-induced demagnetization discussed in this chapter represents the fundamental form of spin dynamics in systems with a single magnetic element or a single sublattice. In the next chapter, we extend this framework to multi-element and multi-sublattice systems, enabling investigation of element-selective spin dynamics.

\newpage
\section{Element-resolved ultrafast demagnetization}
\subsection{Background}
The control of magnetism using light has recently attracted significant attention, both for its potential applications and for the fundamental understanding of its underlying mechanisms. About three decades ago, trMOKE measurements revealed an unexpected phenomenon: the ferromagnetism in Ni foil disappears within 1 picosecond, indicating ultrafast demagnetization \cite{PhysRevLett.76.4250}. Since this discovery, many studies have sought to uncover new light-induced magnetic phenomena and clarify their mechanisms.

One of the systems studied intensively is the element-dependent spin dynamics observed in ferrimagnetic alloys such as GdFeCo, which contain multiple magnetic elements. Another notable phenomenon, first observed in GdFeCo, is light-induced helicity-dependent magnetization switching. More recent research has shown that such optical control of magnetization, depending on the light helicity, is also possible in ferromagnets such as Co/Pt multilayers and granular L1${}_0$-FePt ~\cite{Lambert2014}.

These light-induced magnetization switching phenomena typically occur in materials composed of multiple magnetic elements. Therefore, it is essential to measure spin dynamics with element specificity. This can be achieved through magneto-optic measurements that exploit the tunability of synchrotron radiation and XFELs, enabling experiments at the absorption edges of specific elements.

\subsection{Pt magnetic moment in L1${}_0$-FePt thin films}
\begin{figure}[H]
\centerline{\includegraphics[width=10cm]{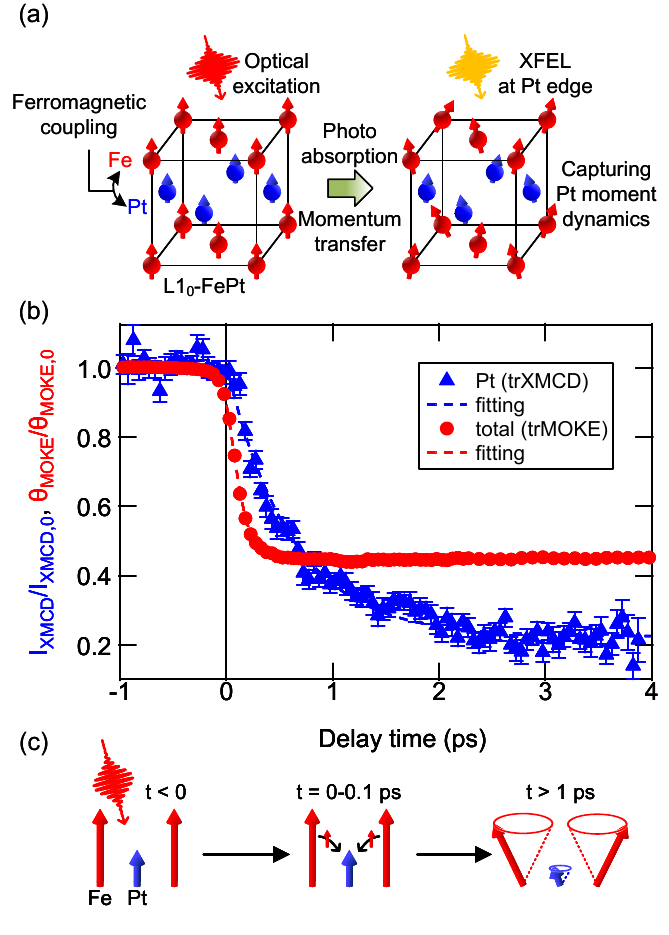}}
\caption{
(a) Schematics of capturing the ultrafast transient dynamics of the Pt moments. 
(b) Normalized trXMCD and trMOKE intensities tracing the demagnetization on Pt and on the whole FePt. 
(c) Schematic diagram of the laser-induced magnetization dynamics in FePt.
\cite{Yamamotonjp}}
\label{sacla_bl3}
\end{figure}

Yamamoto \textit{et al.} performed trXMCD measurements at the Pt $L_3$-edge using circularly polarized hard X-rays at SACLA BL3 \cite{SuzukiMCD,KubotaMCD}, as schematically shown in Fig.~\ref{sacla_bl3} (a) \cite{Yamamotonjp}. The sample was a 20-nm-thick L1${}_0$-FePt thin film grown on an MgO(001) substrate. Pt is a key element in systems showing all optical switching observed in Co/Pt multilayers and granular L1${}_0$-FePt. They investigated the dynamics of its magnetic moment. 
TrXMCD measurements were performed at an X-ray energy of 11.567 keV, at which the PtXMCD spectrum reaches the maximum. Figure \ref{sacla_bl3} (b) shows the results of the measurements with a laser fluence of 32 mJ/cm${}^2$ by varying the delay time. 
This behavior indicates ultrafast demagnetization of the Pt $5d$ magnetic moment, first revealed by the element-specific trXMCD technique developed in this study. The demagnetization time of Pt was quantitatively extracted by fitting the data with a single exponential function, and the decay time was determined as $\tau_{\mbox{Pt}} = 0.$ ps. 

It is known that the Fe magnetic moment dominates the total magnetic moment. It is responsible for about 88\% of the total moment as determined by static XMCD measurements in Ref.~\cite{10.1063/1.4993077}. 
To investigate not only Pt-specific dynamics but also the overall magnetization dynamics of the sample, they performed trMOKE measurements with a laboratory laser at the same excitation fluence as in the trXMCD experiments at SACLA BL3. 
One can see in Figure \ref{sacla_bl3} (b) that Fe demagnetizes much faster than Pt, which was characterized by the different demagnetization times, $\tau_{\mbox{total}} < 0.1$ ps and $\tau_{\mbox{Pt}} = 0.6$ ps.

Fe has a significant partial density of states (pDOS) below $E_F$ that can be excited by the laser pump, and even has a larger pDOS than Pt around $E_F$ because the Pt $5d$ bands lie at lower energies than the Fe $3d$ bands, which provide the main part of the DOS at $E_F$ \cite{Sueda}. This implies that spin-majority electrons of Fe $3d$ states will be excited to Pt states either immediately through optical intersite electron transfer or by initial excitation to unoccupied Fe states followed by hot electron hopping to Pt sites as in superdiffusive transport \cite{Superdiffusive}. 
On the very short time during and immediately after the laser pump ($t=$ 0-0.1 ps), effectively hot majority-spin electrons are transferred from Fe to Pt atoms in their vicinity (Figure \ref{sacla_bl3} (c)). This causes a rapid reduction in Fe magnetization, but a slower decrease in Pt magnetization. 

Recently, Hofherr \textit{et al.} performed trMOKE measurements on FePt using HHG from a laboratory laser \cite{Hoffer}. 
Their measurements probed the Fe M$_{2,3}$ edge ($3p\rightarrow 3d$, $\sim 54$ eV) and the Pt O$_2$ edge ($5p_{3/2} \rightarrow 5d$, $\sim 65$ eV), and they reported similar demagnetization times for both Fe and Pt. In contrast, our results show different time constants for Fe and Pt, suggesting a deviation from Hofherr’s observations. 
Further investigation is needed to clarify the origin of these differences in demagnetization timescales.

\subsection{Element-selective tracking in Co/Pt multilayer thin films}
In trXMCD measurements, circularly polarized X-rays in the hard X-ray region can be generated using diamond phase retarders. 
However, in the soft X-ray or EUV region, implementing transmissive phase retarders is difficult, so undulators capable of producing circular polarization are required. In practice, when such undulators are not available at XFEL facilities, only linearly polarized X-rays can be used. In that case, time-resolved X-ray MOKE (XMOKE) becomes a viable alternative \cite{PhysRevB.89.064423}. As an example, Yamamoto \textit{et al.} performed time-resolved XMOKE results at SACLA BL1, a soft X-ray XFEL beamline \cite{YamamotoAPL}. The sample used was a multilayer film with the structure Pt(1.7~nm)/[Co(0.4~nm)Pt(0.7~nm)]$_3$/Co(0.4~nm), grown on a sapphire substrate. 

\begin{figure}[H]
\centerline{\includegraphics[width=10cm]{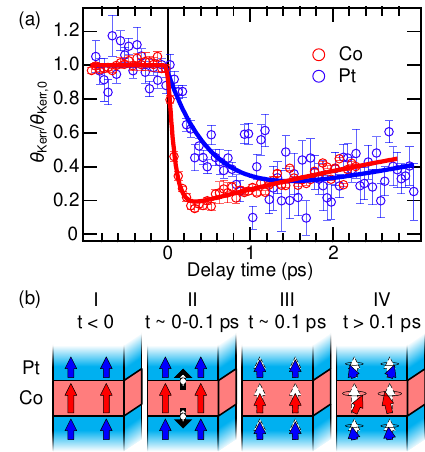}}
\caption{
Photo-induced magnetization dynamics of Co and Pt measured by trXMOKE. (a) Kerr rotation at Co and Pt edges with error bars. (b) Schematics comparing unpumped and pumped states \cite{YamamotoAPL} The element-resolved responses reveal different relaxation pathways within the same multilayer.}
\label{sacla_bl1}
\end{figure}

The XFEL photon energies for the time-resolved XMOKE measurements were determined based on XAS spectra: 60~eV for the Co M$_{2,3}$ edge ($3p$ $\rightarrow$ $3d$) and 72~eV for the Pt N$_{6,7}$ edge ($4f$ $\rightarrow$ $5d$). Figure~\ref{sacla_bl1} shows the results of the time-resolved XMOKE measurements. The vertical axis is normalized to the Kerr rotation angle before laser excitation.

The results show an element-specific demagnetization process induced by laser excitation. The demagnetization timescales were evaluated by fitting the data with exponential decay functions, yielding $t_{\mathrm{Co}} = 80 \pm 60$~fs and $t_{\mathrm{Pt}} = 640 \pm 140$~fs. These findings, obtained using SACLA, reveal a consistent behavior in both L1$_0$-FePt thin films and Co/Pt multilayer systems.

\newpage
\section{Ultrafast antiferromagnetic spin dynamics}
\subsection{Background}
The mainstream research on the laser irradiation of magnetic materials initially focused on ferromagnetic materials. Initially, the primary focus of the study on the laser irradiation of magnetic materials was demagnetization \cite{PhysRevLett.76.4250}, which eliminates the magnetic ordering, and all-optical magnetization reversal \cite{Stanciu,Radu}. Recently, attention has been paid to antiferromagnetic materials, which are expected to undergo ultrafast changes in their magnetic order upon irradiation with ultrashort-pulse lasers. For example, an antiferromagnetic-ferromagnetic transition was realized by the laser irradiation of an iron-rhodium alloy (FeRh), which transitions from antiferromagnetism (low temperature) to ferromagnetism (high temperature) as the temperature increases \cite{PhysRevB.81.104415}. Ferromagnetism was induced by suppressing antiferromagnetism via laser irradiation. 

Antiferromagnets are expected to be controlled more rapidly and with less energy than ferromagnets. How the magnetic ordering evolves is governed by the law of the conservation of angular momentum. When a ferromagnetic material is locally demagnetized, the angular momentum must be transferred from the spin to another system, such as a lattice. On the other hand, antiferromagnetic materials have no such need in principle. Thus, this may enable faster and more energy-efficient control of magnetic ordering. 

\subsection{Laser-induced AM-to-FM phase transition}
\begin{figure}[H]
\centering
\includegraphics[width=8cm]{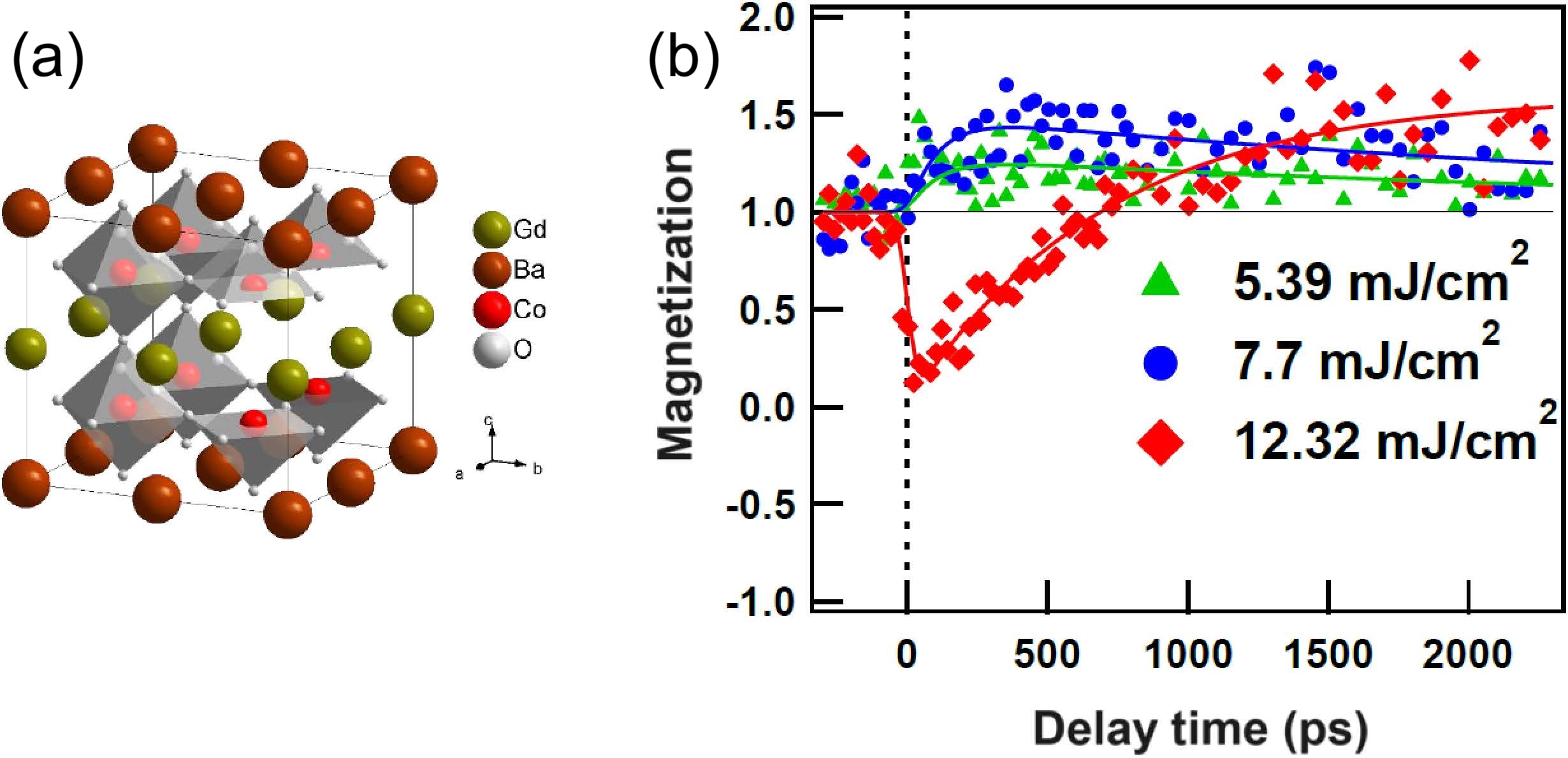}
\caption{
(a) Crystal structure of GBCO. (b) Time-resolved XMCDR signals showing a magnetization increase shortly after laser irradiation (0 ps) in GBCO thin films \cite{Zhang2022}. The enhancement indicates light-induced strengthening of ferromagnetic ordering.}
\label{GBCO1}
\end{figure}

To realize such phenomena of creating ferromagnetic states not only in FeRh alloy but also in oxide thin films, which are more stable and resistant to oxidation in air, Zhang \textit{et al.} \cite{Zhang2022} attempted used laser irradiation to realize ferromagnetism in cobalt oxide, GdBaCo$_2$O$_{5.5}$ (GBCO: Fig.~\ref{GBCO1} (a)), thin films, with a transition from antiferromagnetism to ferromagnetism similar to that for FeRh. GBCO thin films with a thickness of approximately 35 nm were fabricated on SrTiO$_3$ (001) substrates using the PLD method \cite{katayama}. The energy of the X-rays was that of the Co $L_3$ absorption edge ($\sim 776$ eV). The XMCDR was measured using circularly polarized light, as in Ref.~\cite{tsuyama}. The GBCO thin films exhibited AFM-FM transitions with an antiferromagnetic (AFM) phase at low temperatures and a ferromagnetic (FM) phase at high temperatures ($T_{AFM-FM}=230$ K, $T_C=300$ K). Therefore, as observed in FeRh (=$T_{AFM-FM}=375$ K, $T_C=680$ K), it is expected that a transient ferromagnetic state can be realized by laser irradiation at a temperature lower than the AFM-FM transition. Such ferromagnetism was observed using time-resolved XMCDR measurements, as shown in Fig.~\ref{GBCO1} (b). At 150 K, a temporary increase in the XMCDR was observed, exceeding the value when it was not excited. When the pump light fluence was low, the XMCDR increased immediately after laser irradiation, reaching approximately 1.5 times its pre-irradiation value. At a higher fluence of 12.32 $\mathrm{mJ/cm}^2$, the XMCDR showed rapid decay, and then increased beyond the XMCDR value before excitation. At 250 K, which was higher than TAFM FM, an XMCDR value exceeding the pre-excitation value was not observed; instead, only rapid demagnetization and a slow recovery were observed. From these results, it can be concluded that a photo-induced AFM–FM transition was observed in GBCO thin films. Interestingly, the X-ray reflectance value itself showed a behavior change that was simultaneous to the change in the XMCDR. 

\begin{figure}[H]
\centering
\includegraphics[width=12cm]{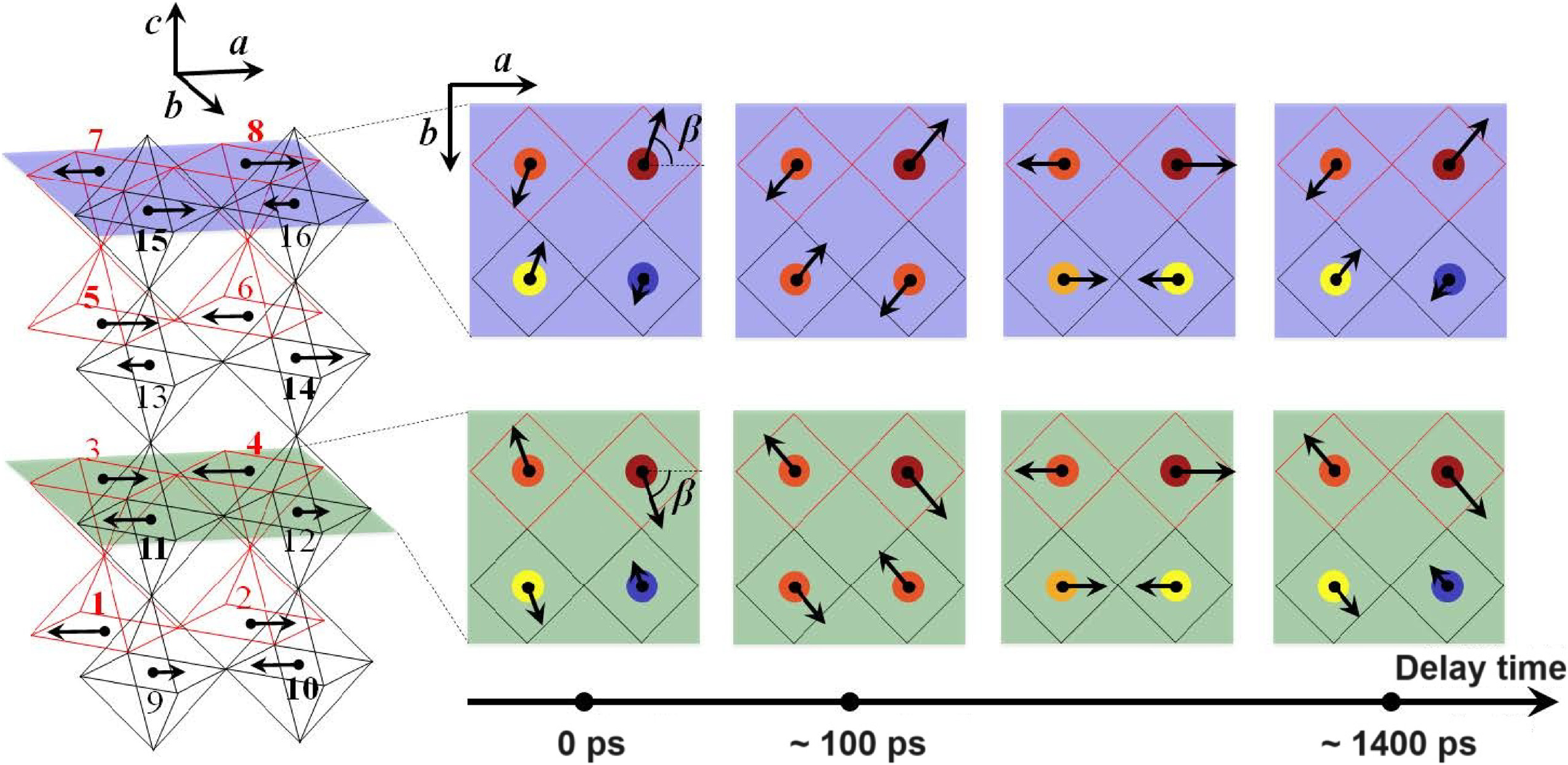}
\caption{
Photoexcited cobalt-spin states of GBCO. Laser irradiation increases in-plane magnetization; HS and LS denote high- and low-spin states \cite{Zhang2022}. This demonstrates ultrafast spin-state switching driven by optical excitation.}
\label{GBCO2}
\end{figure}

It is thought that the above results can be explained by the photo-induced spin dynamics of GBCO, as shown in the schematic diagram of Fig.~\ref{GBCO2}. As evidenced by changes in X-ray reflectance, ultrashort-pulse laser irradiation significantly altered the electronic states of the GBCO thin films, leading to a spin-state transition. At the same time, it caused the AFM-FM transition, as evidenced by an increase in the XMCDR intensity and a decrease in the antiferromagnetic Bragg peak intensity observed in the RSXS. The change in magnetism could be interpreted as a change in the spin-tilt angle or the AFM/FM ratio in the antiferromagnetic phase. Figure \ref{GBCO2} illustrates the change in the spin tilt angle of the antiferromagnetic sublattice. This system was first pumped into the high-spin ferromagnetic state of octahedral $\mathrm{Co}^{3+}$. The increase in the XMCDR was attributed to both the more parallel alignment of the Co magnetic moments and the increase in the individual Co magnetic moments resulting from the spin state transition. After that, the system entered a transient state in which the magnetization exceeded its unexcited state value. Finally, the octahedral $\mathrm{Co}^{3+}$ returned to its original antiferromagnetic state with mainly low-spin states. 

Thus, by using time-resolved X-ray measurements, they succeeded in observing the spin state transition and photo-induced AFM-FM transition at the octahedral $\mathrm{Co}^{3+}$ site in GBCO thin films. In this study, they discovered a new photo-induced phenomenon involving both charge and spin, resulting in a photo-induced AFM-FM transition and associated ultrafast ferromagnetism in oxide thin films. This result showed that the bonding between magnetism and electronic structure in transition-metal oxides is essential not only in the static state but also in the transient and dynamic states.

\subsection{Energy-efficient antiferromagnetism control}
Yamamoto \textit{et al.} observed the ultrafast response of antiferromagnetic materials through the laser irradiation of thin films of perovskite-type iron oxides (La$_{1/3}$Sr$_{2/3}$FeO$_3$ (LSFO) and SrFeO$_{3-\delta}$ (SFO)) with antiferromagnetic and helimagnetic orders, respectively \cite{Yamamoto2022}. Although it was difficult to directly observe the antiferromagnetic order with light, direct measurements of the antiferromagnetic Bragg peaks were possible through RSXS using synchrotron radiation X-rays \cite{,Okamoto,PhysRevB.88.220405,YamamotoLSFOPRB}.

\begin{figure}[H]
\centering
\includegraphics[width=8cm]{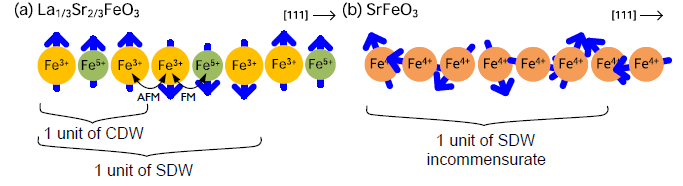}
\caption{Antiferromagnetic orderings of LSFO (a) and SFO (b). LSFO shows commensurate antiferromagnetic peaks, and SFO shows incommensurate antiferromagnetic peaks. FM and AFM indicate ferromagnetic and antiferromagnetic coupling, respectively. CDW and SDW indicate charge and spin density wave, respectively. \cite{Yamamoto2022}}
\label{LSFO}
\end{figure}

Figure \ref{LSFO} shows the magnetic orders of the LSFO and SFO. In the LSFO, the average valence of Fe was 3.67+. As shown in Fig.~\ref{LSFO} (a), a charge density wave with three times the period and a spin density wave with six times the period in the (111) direction existed, i.e., a charge/spin order such as Fe$^{3+}(\uparrow)$-Fe$^{5+}(\uparrow)$-Fe$^{3+}(\uparrow)$-Fe$^{3+}(\downarrow)$-Fe$^{5+}(\downarrow)$-Fe$^{3+}(\downarrow)$. On the other hand, in the SFO, a lattice-incommensurate helimagnetic order existed in the (111) direction, as shown in Fig.~\ref{LSFO} (b), reflecting the stronger ferromagnetic interaction of Fe$^{4+}$. LSFO thin films with a thickness of approximately 40 nm were fabricated using the pulsed laser deposition (PLD) method on SrTiO$_3$(111) substrates \cite{10.1063/1.4958670}. SFO thin films with a thickness of approximately 27 nm were fabricated by oxidizing SrFeO$_{2.5}$ thin films grown epitaxially on SrTiO$_3$(100) substrates using the PLD method. UV/O3 DRY CLEANER UV-1 (Samco Inc.) was used for ozone oxidation. The X-ray energy was the Fe $L_3$ absorption edge ($\sim 710$ eV), and RSXS was measured using horizontally polarized light.

\begin{figure}[H]
\centering
\includegraphics[width=9cm]{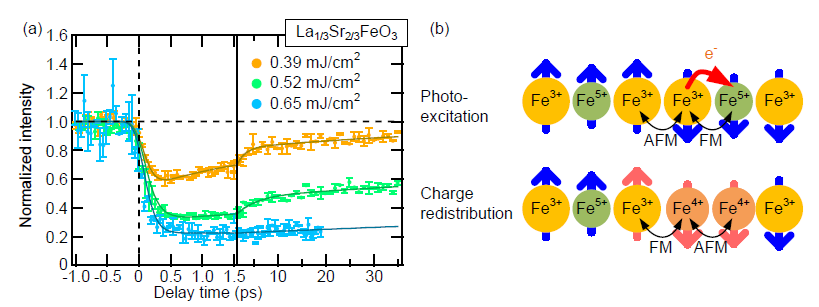}
\caption{(a) Antiferromagnetic Bragg peaks of antiferromagnetic ordering of LSFO thin films observed with trRSXS measurements. The antiferromagnetic peak intensity changed by approximately 0.1 ps after laser irradiation. (b) Schematics of the photo-induced state of Fe spins. Laser irradiation induces a sequential change in charge and magnetism \cite{Yamamoto2022}.}
\label{fig44}
\end{figure}

Figure \ref{fig44} (a) shows the time evolution of the intensity of the antiferromagnetic Bragg peak (1/6 1/6 1/6) after laser excitation of the LSFO thin film. This peak corresponded to the spin density wave with six times the period, as shown in Fig.~\ref{fig44} (a). Because the peak intensity was sufficient, laser slicing was used to obtain a time resolution of approximately 130 fs. From Fig. 4(a), it can be seen that the antiferromagnetic order appears to be decreasing within the time resolution range after laser irradiation (0 s). When the pump light fluence was 0.5 mJ/cm$^2$, the peak intensity decreased to half its initial value. The peak intensity recovered within approximately 30 ps after laser excitation in the low-fluence region, but the recovery slowed at high fluence. From this result, one can quantitatively estimate the time constant of the decrease and recovery of the diffraction intensity, as well as the magnitude of the change.

The time scale of the change in the antiferromagnetic intensity of the LSFO thin film immediately after light irradiation was estimated to be $\tau_{decay} < 0.1$ ps. Changes on this timescale are considered ultrafast magnetization dynamics, driven by a photo-induced change in charge order that breaks magnetic order via strong charge-spin coupling. The demagnetization due to such a photo-induced change in the charge order has also been observed, for example, in the antiferromagnetic peak of NdNiO$_3$ thin films \cite{NNO}. Figure 4(b) shows the process for the optical excitation of the spin of iron in the LSFO thin film. From the previous studies on the optical spectrum, the excitation at 0.15 eV was considered to cause the charge transfer at the Fe$^{3+}(\uparrow)$-Fe$^{3+}(\downarrow)$-Fe$^{5+}(\downarrow)$ site and realize the Fe$^{3+}(\uparrow)$-Fe$^{4+}(\downarrow)$-Fe$^{4+}(\downarrow)$ state. Previous studies \cite{Zhu2018} on the photo-induced transient states of LSFO suggested that this charge transfer state exists in a quasi-stable manner as a result of the change in the Fe–O bond length. This charge transfer modulates the magnetic interaction, thereby altering magnetic order. Specifically, because the Fe$^{3+}(\uparrow)$-Fe$^{4+}(\downarrow)$ spot has ferromagnetic interaction, the original antiferromagnetism is destroyed.
Furthermore, trRSXS measurements were performed on SFO thin films at 35K and 80 K, and changes in the intensity of the antiferromagnetic Bragg peak were observed. A comparison of these materials showed that the LSFO peak intensity decreased by approximately 70\% at a pump light fluence of 0.65 mJ/cm$^2$. Still, for the SFO, it decreased by approximately 60\% at a pump light fluence of 2.8 mJ/cm$^2$. This fluence value corresponded to the energy of light absorbed by the thin film, and it was shown that the antiferromagnetic order of LSFO changed at a lower energy than that of SFO. In the antiferromagnetic phase, a fast and energy-efficient change was expected because the magnetization was canceled out. TrRSXS has revealed the energy efficiency of this phase change in dysprosium metal \cite{Dy}. This study showed that antiferromagnetic changes are induced more energy-efficiently in LSFOs than in SFOs, which exhibit strong ferromagnetic interactions, and that these changes are ultrafast. The antiferromagnetism associated with charge alignment is a characteristic property of transition-metal oxides, and these systems have been successfully observed to exhibit energy-efficient, ultrafast photo-induced phenomena.

\newpage
\section{Ultrafast valence dynamics in strongly correlated systems}
Lanthanides containing intermetallics are also of great interest in the field of ultrafast all-optical magnetization switching (AOS). In that field, fs optical excitation pulses can induce magnetization reversal, e.g., in GdFeCo~\cite{Stanciu}. The excitation is expected to differently affect the magnetization 4$f$ and the 3$d$ metal ions~\cite{PhysRevLett.102.117201}, and it is therefore of more general importance to understand the excited 4$f$ electron system and its interaction with 3$d$ transition metal ions in intermetallics. A direct view of the ultrafast behavior of 4$f$ electrons by probing with fs soft X-ray pulses is therefore of great importance for a better understanding of both AOS and ultrafast electron localization, including valence transitions. 

\begin{figure}[H]
\centerline{\includegraphics[width=7cm]{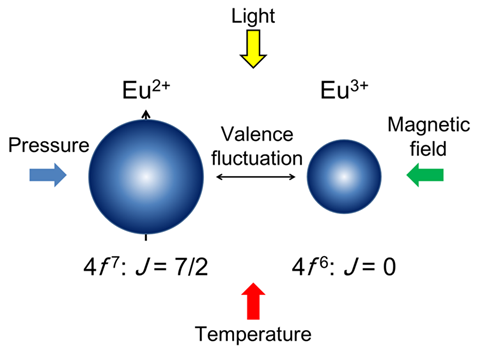}}
\caption{Valence transitions between Eu$^{2+}$ ($4f^7$) and Eu$^{3+}$ ($4f^6$), highlighting the difference in electronic occupation. The energy separation between these states underpins valence conversion under excitation.}
\label{EuEu}
\end{figure}

Rare-earth europium compounds provide an ideal platform for studying ultrafast valence dynamics, owing to the strong instability of the Eu$^{2+}$/Eu$^{3+}$ valence configuration and the tight coupling between 4$f$ electronic states and magnetism. While static studies have demonstrated that the Eu valence can be tuned by pressure, chemical substitution, and temperature, as shown in Fig.~\ref{EuEu}, the microscopic mechanism and timescales of light-driven valence switching remain essentially unexplored. Therefore, Eu-based materials represent a unique opportunity to unravel the interplay among charge, spin, and valence degrees of freedom under nonequilibrium conditions.

The Eu intermetallic compound EuNi$_2$(Si$_{1-x}$Ge$_x$)$_2$ shows the valence transition between Eu$^{2+}$ (4$f^7$) and Eu$^{3+}$ (4$f^6$) as a function of temperature \cite{Eu1,Wada1997,YamamotoEu1,YAMAMOTO2006681,Ichiki,SHIMOKASA}, magnetic field \cite{Wada1997, YHMatsuda1, YHMatsuda2, nakamura}, and physical and chemical pressure \cite{PhysRevB.59.1141}. 
The phase transition temperature reaches the maximal value of $\sim 95$ K at around $x=0.79$. An interesting aspect is the decreasing Eu valence with increasing temperature, 

Yokoyama \textit{et al.} and Mardegan \textit{et al.} conducted trXAS measurements of EuNi$_2$(Si$_{0.21}$Ge$_{0.79}$)$_2$ at SPring-8 (Eu $M$ edge) and SwissFEL (Eu $L$ edge), respectively \cite{YokoyamaEu,PhysRevResearch.3.033211}. Then Yamagami \textit{et al.} probed the sub-picosecond $4f$ electronic dynamics of EuNi$_2$(Si$_{0.21}$Ge$_{0.78}$)$_2$  using $M_5$-edge trXAS, which directly accesses the $4f$ states and enables extraction of the transient electronic $4f$ temperature in the femtosecond range \cite{Yamagami}. This parameter is crucial for understanding ultrafast electron localization and serves as a fundamental microscopic quantity in the study of ultrafast magnetization dynamics of lanthanide intermetallics.
The trXAS experiments on polycrystalline samples were carried out at the Soft X-ray Scattering and Spectroscopy (SSS) beamline of the PAL-XFEL in Korea \cite{10.1063/1.5023557,Kim,Park2}, using an 800-nm Ti: sapphire pump laser with a 50-fs pulse duration and an XFEL probe pulse of $\sim 1130$ eV photon energy (Eu M edge) and 100-fs duration.

\begin{figure}[H]
\centerline{\includegraphics[width=10cm]{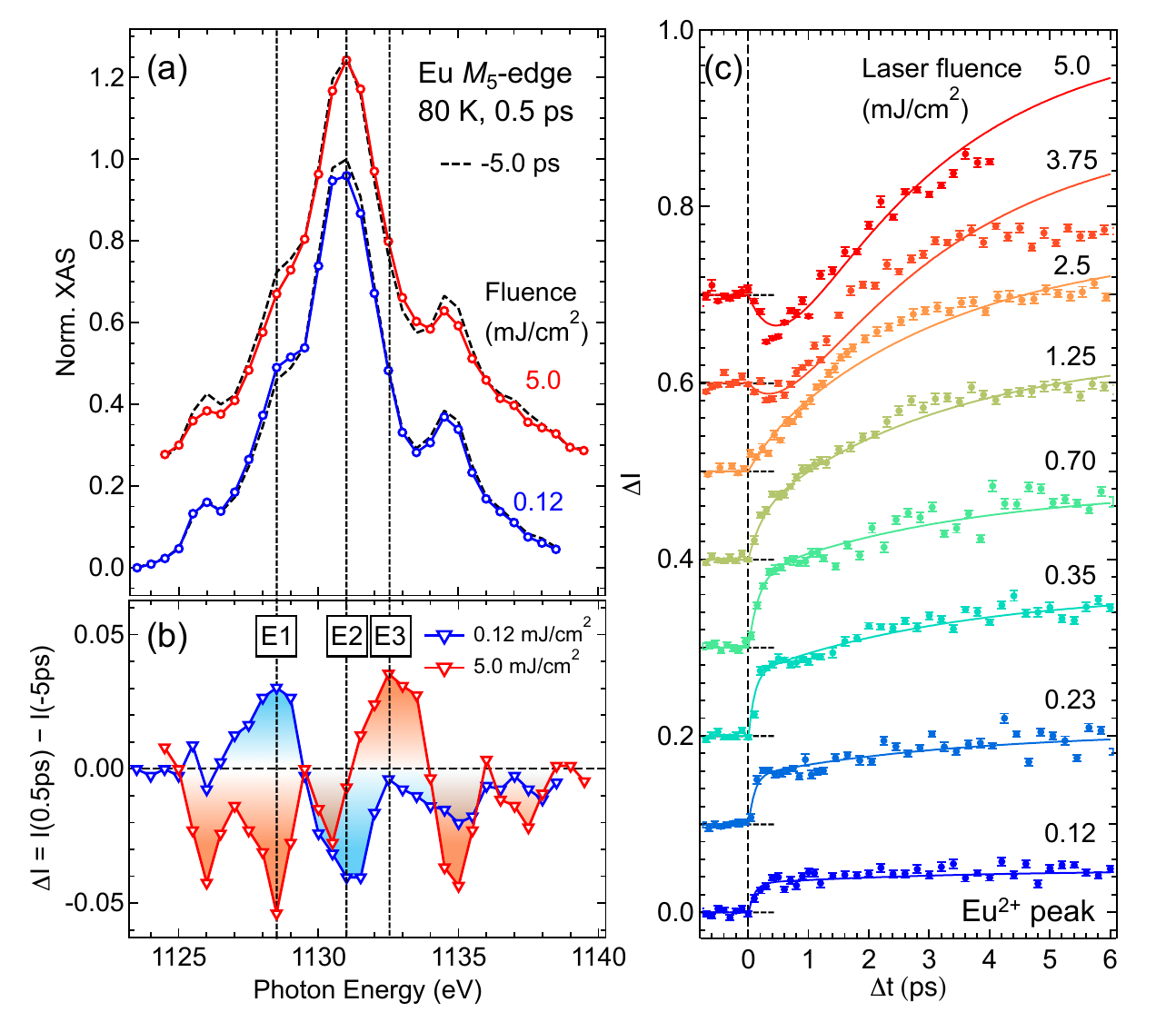}}
\caption{(a) The Eu $M_5$-edge trXAS spectra at 80 K under low (0.12 mJ/cm$^2$, blue) and high fluences (5.0 mJ/cm$^2$, red) with ($\Delta t=0.5$ ps) and without (black, taken at $\Delta t=-5.0$ ps) laser excitation at early times. (b) XAS spectral intensity changes ($\Delta I$) at 0.5 ps. In panels (a) and (b), the vertical dashed lines denote the photon energies, labeled as E1 = 1128.5 eV, E2 = 1131 eV, and E3 = 1132.5 eV. (c) Ultrafast $\Delta$I change as a function of $\Delta t$ for selected fluences at 80 K of the Eu$^{2+}$ peak (E1). \cite{Yamagami}}
\label{Eucompound1}
\end{figure}

Figure~\ref{Eucompound1}(a) shows $M_5$-edge trXAS for the lowest (0.12 mJ/cm$^2$) and highest (5.0 mJ/cm$^2$) fluences at delay time ($\Delta t$) of 0.5 ps directly probing the excited Eu 4$f$ states. The unperturbed XAS data ($\Delta t = -5.0$ ps) are also shown for comparison. The photon-induced change in the XAS at 0.5 ps shown in Fig. Figure~\ref{Eucompound1}(b) is qualitatively different between low and high fluence. There are some sign changes in the pump effect indicating larger and smaller intensities for XAS attributed to Eu$^{2+}$ and Eu$^{3+}$, respectively. A simple spectral analysis from the individual divalent and trivalent spectra results in an effective 4$f$ valency ($v_{4f}$) of $2.67 \pm 0.02$ for 0.12 mJ/cm$^2$ and $2.77 \pm 0.02$ for 5.0 mJ/cm$^2$, suggesting the increase and decrease of the Eu$^{2+}$ population after photoexcitation at low and high fluence, respectively. 
By adjusting the photon energy to the Eu$^{2+}$ peak (E1 = 1128.5 eV), they directly investigated the time evolution of XAS spectral intensity changes [see Fig.~\ref{Eucompound1}(c)]. A stepwise increasing component of the Eu$^{2+}$ peak was observed from the low fluence region (0.12-0.35 mJ/cm$^2$). Furthermore, a prominent decrease in the Eu$^{2+}$ signal is captured at earlier times ($\Delta t$$\sim$0.5 ps) through the high fluence region (3.75-5.0 mJ/cm$^2$).

\begin{figure}[H]
\centerline{\includegraphics[width=14cm]{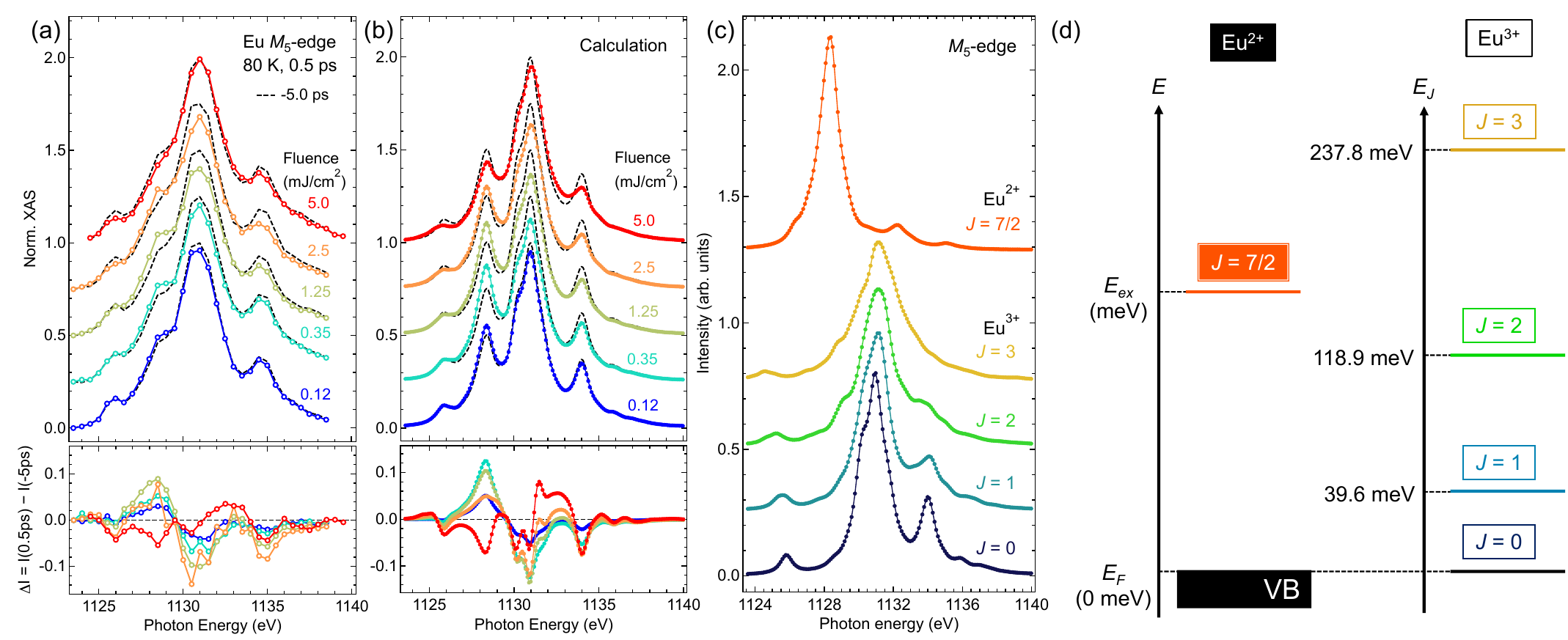}}
\caption{(a)The $M_5$-edge XAS and $\Delta$I spectra for selected fluences at 80 K and $t=0.5$ ps, together with equilibrium data taken at $t=-5.0$ ps. (b) The calculated XAS and their difference from the equilibrium spectrum reproduce the experimental results well. (c) Atomic multiplet calculations for each $J$ state. (d) Energy level of the ground and excited states for Eu$^{2+}$ and Eu$^{3+}$ free ions. \cite{Yamagami}}
\label{Eucompound2}
\end{figure}

Figures~\ref{Eucompound2}(a) and (b) show the experimental and calculated data of the trXAS and their spectral deviations from the equilibrium data for selected laser fluences at $\Delta t = 0.5$ ps and $T = 80$ K, respectively. The experimental data are well described by a fit to the interconfigurational fluctuation (ICF) model~\cite{Wada1997}. The ICF model is a high-temperature model for incoherent thermal charge fluctuations in rare-earth metals. As shown in Fig.~\ref{Eucompound2} (c) and (d), $E_J$ is the energy of the Eu$^{3+}$ atomic multiplet states with $J$, neglecting their crystal field splitting. $E_{ex}$ is the energy to take one electron out of the conduction band and to put it into the 4$f$ levels, converting the Eu valence state from 3+ to 2+. 

In this model, the populations of the Eu$^{2+}$ and Eu$^{3+}$ states, represented as $p_2$ and $p_3$, respectively, are given by the Boltzmann statistic
\begin{equation}
\frac{p_2}{p_3} = \frac{N_{J=7/2}\exp(-\displaystyle\frac{E_{ex}}{k_B T^*})}{\Sigma_{J=0}^{3}N_J\exp(-\displaystyle\frac{E_{J}}{k_B T^*})}
\end{equation}
where $k_B$ is the Boltzmann constant. The degeneracy $N_J$ = 2$J$ + 1 is adopted as the coefficient of each Boltzmann distribution for the $J$= 7/2 state of Eu$^{2+}$ and $J$= 0, 1, 2, and 3 states of Eu$^{3+}$ and $p_2+p_3=1$. $T^{\ast}$ = $\sqrt{T^2+T_{4f}^2}$ is the effective temperature including the temperature in the system ($T$) and the constant value of the broadening of the 4$f$ state ($T_{4f}$). In the equilibrium state, the broadening of the various energy levels of the 4$f$ electronic states, due to quantum-mechanical hybridization (e.g., orbital hybridization and electron-lattice interactions), can be accounted for by introducing an effective temperature. Here, $E_{ex}$ and $T^{\ast}$ are the tuning parameters to reproduce the XAS data, and $v_{4f}$ is given by $3-p_2$. The excited Eu$^{3+}$ spectra for a given electronic temperature T$^*$ were obtained by summing the XAS spectra of each $J$ state multiplied by their population. The ICF model-calculated spectra reproduce the experimental data well.

Looking ahead, clarifying whether ultrafast valence switching in Eu compounds is driven by electronic screening, lattice collapse, or magnetic exchange remains an open question. The combination of HHG and XFEL experiments will be essential to establish a quantitative understanding of valence dynamics in nonequilibrium correlated materials.
\newpage
\section{HHG X-ray measurements}
Time-resolved measurements using X-rays are evolving with synchrotron radiation and XFEL, but the former suffers from limited time resolution, and the latter has more limited beam time. Thus, there are difficulties in making steady progress in future research. In addition, with these X-rays
sources, the energy of the X-rays must be measured by sequentially selecting the resonance energies of the constituent elements of the material, making it impossible to investigate the spin dynamics of different elements simultaneously. Therefore, recent research has used laboratory-based HHG to observe magnetization dynamics via the transverse MOKE in a reflective configuration \cite{HHGMOKE}.

\begin{figure}[H]
\centering
\includegraphics[width=10cm]{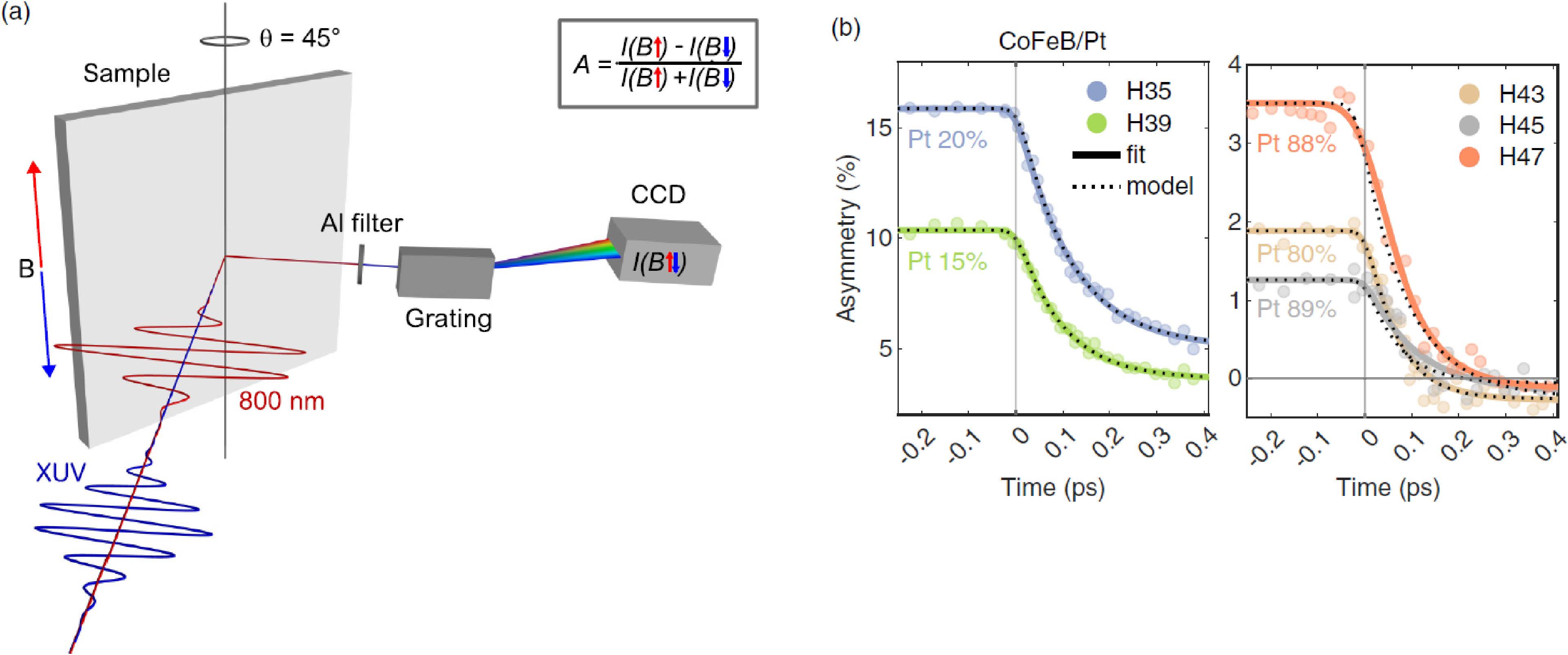}
\caption{
(a) Schematic of time-resolved measurements using ultraviolet HHG. (b) Spin dynamics of a CoFeB/Pt thin film observed via HHG-based measurements \cite{HHGMOKE}. The HHG probe provides a tabletop alternative to synchrotron-based ultrafast magnetization studies.}
\label{HHG22}
\end{figure}

A schematic diagram of the setup in Ref.~\cite{HHGMOKE} is shown in Fig.~\ref{HHG22} (a). An ultrashort-pulse laser with a repetition rate of 3 kHz, a center wavelength of 800 nm, and a pulse width of 25 fs is focused into a gas cell filled with the noble gas neon, generating high-order harmonics. The spectrum of the emitted ultraviolet light consists of discrete peaks with a spectral width of approximately 200 meV. The p-polarized ultraviolet HHG is reflected by a thin ferromagnetic CoFeB sample at 45°, spectrally resolved by a diffraction grating, and detected by a CCD. The magnetization of the thin-film sample is aligned by an external magnetic field perpendicular to the incident polarized light from ultraviolet HHG, resulting in a transverse MOKE arrangement. The magnetic signal is obtained as the difference in HHG reflectance when the magnetic field is switched between positive and negative. The time-resolved experiment was performed using the pump-probe method, employing an 800 nm ultrashort-pulse laser for the pump and ultraviolet HHG for the probe. The upper limit of the time resolution of the experiment was estimated to be approximately 35 fs, which was equivalent to or higher than that of XFEL.

Using this method, the spin dynamics could be observed for each element in the laboratory taking advantage of the fact that, for example, the 35th harmonic corresponded to 53.5 eV at the $M_{2,3}$ absorption edge ($3p\rightarrow 3d$) of Fe, the 39th harmonic corresponded to 60 eV at the $M_{2,3}$ edge ($3p\rightarrow 3d$) of Co, and the 43rd-47th harmonics corresponded to 65–71 eV at the $O_2$ absorption edge ($5p_{3/2}\rightarrow 5d$) and $N_7$ absorption edge ($4d_{7/2}\rightarrow 5d$) of Pt. From the results in Fig. 7(b), the demagnetization time constant was $\tau_{H35} = 115$ fs for the 35th harmonic, $\tau_{H39} = 105$ fs for the 39th harmonic, and $\tau_{H43-H47} = 70$ fs for the 43rd–47th harmonics. These results showed that the demagnetization time constant of Pt was 70 fs, shorter than 115 fs for Fe and 105 fs for Co. Research on dynamics using HHG is currently expanding not only to the region below 100 eV but also to the soft X-ray region above 500 eV. For example, trXAS at the K-edge of nitrogen has already been conducted using HHG \cite{SaitoHHG}.

Compared with XFELs, HHG provides intrinsically synchronized pump–probe operation and high repetition rates of $10^4$–$10^5$ Hz, enabling statistical measurements with high sensitivity. Although the photon flux is lower than XFELs, HHG excels in systematic parameter scans and offers full access to the laboratory without beamtime limitations. The coexistence of HHG and XFEL facilities is therefore not competitive but complementary: HHG enables hypothesis generation and exploratory studies, while XFELs deliver the highest spectral and momentum resolution once promising samples or phenomena are identified.

The tabletop nature of HHG is particularly important for the next generation of ultrafast X-ray researchers, enabling graduate students and young scientists to learn ultrafast spectroscopy without the barrier of access to large-scale facilities.
\newpage
\section{Summary and future prospects}
\begin{figure}[H]
\centering
\includegraphics[width=8cm]{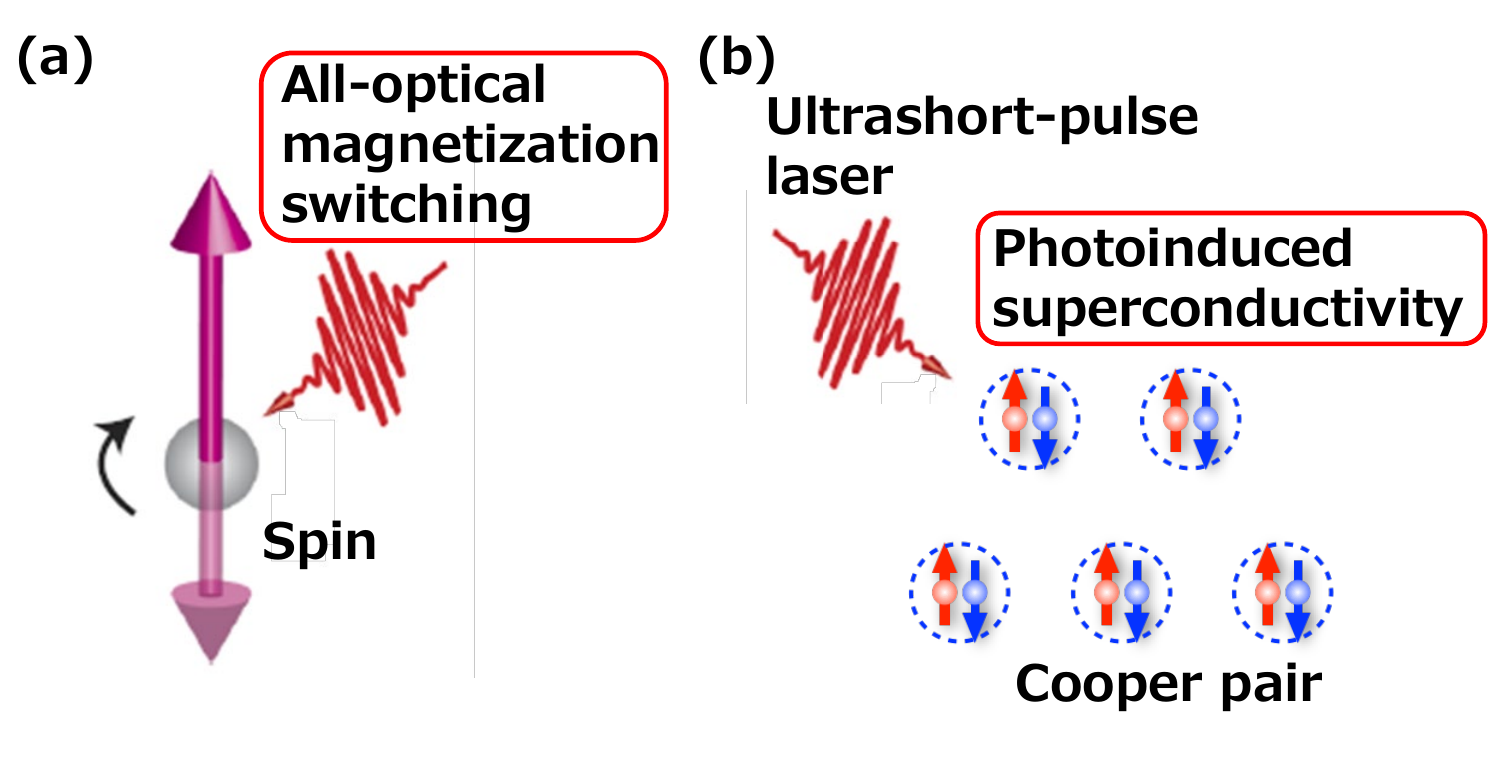}
\caption{
Schematic illustration of (a) all-optical magnetization switching and (b) photo-induced superconductivity, representing ultrafast optical control of electronic phases. These concepts motivate the exploration of materials for energy-efficient opto-spintronic functionality.}
\label{future}
\end{figure}

In recent years, advances in femtosecond time-resolved techniques such as trXAS, trXMCD, and trRSXS have enabled a unified understanding of the coupled dynamics of charge, spin, orbital, and lattice degrees of freedom in complex materials. These developments have opened new avenues for investigating photo-induced phase transitions and element-specific ultrafast magnetism. In this chapter, we highlight representative frontier studies—illustrated in Fig.~\ref{future}—that address key issues such as all-optical magnetization switching and photo-induced superconductivity, as well as emerging approaches combining X-ray and terahertz.

\subsection{All-optical magnetization switching}
Since the discovery of ultrafast demagnetization in Ni in 1996, with characteristic demagnetization times shorter than 1 ps \cite{PhysRevLett.76.4250}, the microscopic origin of laser-induced spin dynamics has been extensively investigated because of its potential for next-generation nonvolatile spintronic devices \cite{RevModPhys.82.2731,Koo}. 
All-optical switching (AOS), in which magnetization is reversed solely by femtosecond laser irradiation without an external magnetic field, represents a particularly promising route toward ultrafast, energy-efficient magnetic control \cite{Stanciu, Hübner}. 
Several types of AOS have been identified, including all-optical helicity-independent switching (AO-HIS) \cite{Ostler,Graves}, all-optical helicity-dependent switching (AO-HDS) \cite{Lambert2014,Koo2}, and magnetization precession associated with laser-induced anisotropy reorientation \cite{10.1063/1.5010915}. 
The relative importance of mechanisms such as the inverse Faraday effect, magnetic circular dichroism, and thermally induced domain-wall motion remains an active topic of debate \cite{RevModPhys.82.2731,AOS1}. 

A key challenge in understanding and optimizing AOS is disentangling the contributions of individual elements in multi-sublattice systems, especially in ferrimagnets and complex oxides. 
To date, most AOS studies have relied on optical or magneto-optical probes that integrate over all constituent elements, making it difficult to separate the ultrafast responses of different atomic species. 
In this respect, XFELs provide a unique opportunity to probe AOS on an element-specific basis by exploiting XMCD or RSXS at core-level absorption edges. 
XFEL-based trXMCD enables direct observation of sublattice-resolved spin and orbital dynamics at femtosecond time resolution, providing insight into how angular momentum is redistributed among distinct atomic sites during optical excitation. 
Such measurements are crucial for clarifying whether laser-induced magnetization reversal originates from spin-lattice coupling, inter-sublattice exchange, or purely electronic processes.

Recent progress has expanded the scope of AOS beyond rare-earth-based alloys. 
In particular, Takahashi \textit{et al.} have reported all-optical helicity-independent switching in the rare-earth-free spinel oxide NiCo$_2$O$_4$, demonstrating that laser-induced magnetization reversal can occur even in transition-metal-only systems \cite{Takahashi2023,Takahashi2025}. 
This finding highlights the versatility of AOS phenomena and the growing relevance of oxide materials for sustainable spintronics. 
Future XFEL studies targeting such systems will enable direct tracking of the transient oxidation states and element-resolved spin dynamics at the Co and Ni sites, thereby revealing the microscopic origin of light-driven magnetization switching in correlated oxides.

Looking forward, the combination of ultrafast optical excitation with element-selective XFEL probing will be indispensable for establishing a unified picture of AOS across metals, ferrimagnets, and oxides. By resolving the interplay among charge, spin, and orbital degrees of freedom on the natural timescales of exchange interactions, XFEL-based approaches will provide a microscopic foundation for the design of rare-earth-free, energy-efficient, and ultrafast magnetic memory materials.

\subsection{Photo-induced superconductivity}
Photo-induced superconductivity refers to a transient state in which a normally non-superconducting material exhibits superconducting-like behavior for a short period following optical excitation.
The phenomenon was first reported in the stripe-ordered cuprate La$_{
1.675}$Eu$_{0.2}$Sr$_{0.125}$CuO$_4$, where the irradiation of a mid-infrared pulse induced a Josephson plasma resonance persisting for several picoseconds, indicative of light-induced superconductivity \cite{SC1}.
Similar signatures have subsequently been observed in other cuprates, such as La$_{2-x}$Ba$_x$CuO$_4$ and YBaCu$_3$O$_y$ \cite{SC2, SC3, SC4}. In a distinct class of materials, the fulleride compound K$_3$C$_{60}$ has also shown transient changes in optical conductivity consistent with superconductivity - namely, the emergence of a gap in the real part and a $1/\omega$ divergence in the imaginary part - lasting for a few picoseconds after mid-infrared excitation \cite{SC5}.
Remarkably, both YBa$_2$Cu$_3$O$_y$ and K$_3$C$_{60}$ display such behavior even near room temperature, which makes these findings particularly intriguing.

While these results strongly suggest the possibility of light-induced superconductivity, they do not constitute definitive proof.
In conventional superconductors, the identification of a superconducting state requires verification of both zero resistivity and the Meissner effect.
Recent experiments have reported encouraging progress on both fronts.
In K$_3$C$_{60}$, the emergence of a long-lived metastable state lasting over 10 ns allowed transport measurements showing behavior approaching zero resistance \cite{SC6}. In cuprates such as YBa$_2$Cu$_3$O$_{6.48}$, transient magneto-optical Faraday rotation measurements revealed a diamagnetic response concomitant with the photo-induced state, suggesting a light-induced Meissner-like effect \cite{SC7}. The authors argued that the magnitude of the diamagnetic signal cannot be explained solely by the generation of high-mobility carriers, implying the emergence of transient superconductivity.

Despite these remarkable advances, the microscopic origin of photo-induced superconductivity remains under debate. For instance, in YBa$_2$Cu$_3$O$_y$, third-harmonic generation measurements have failed to detect the Josephson current expected along the c-axis, suggesting that the light-induced state may not possess actual long-range phase coherence \cite{SC8}.
Moreover, several groups have raised concerns about possible systematic errors in data analysis, arguing that some reported signatures may not directly evidence superconductivity itself \cite{SC9}.
Further cross-verification using complementary probes is therefore crucial to establish the nature of these nonequilibrium states.

In a seminal trXAS work on the cuprate superconductor Bi$_2$Sr$_2$CaCu$_2$O$_{8+\delta}$, Mitrano \textit{et al.} observed an ultrafast renormalization of the on-site Coulomb repulsion $U$ within a few hundred femtoseconds after optical excitation \cite{SC10}. This result revealed that photoexcitation can transiently screen electron–electron interactions, effectively weakening the Mott-like correlations that compete with superconductivity.
Such findings highlight the unique capability of time-resolved X-ray measurements to probe nonequilibrium many-body interactions beyond charge, spin, or lattice dynamics, and suggest that controlling $U$ on ultrafast timescales may provide a microscopic route to photo-induced superconductivity.
Beyond these developments, in the field of resonant inelastic X-ray scattering (RIXS) \cite{deGroot2024,Yamamoto2025,yamamoto20252drixsresonantinelasticxray}, time-resolved RIXS studies will reveal strong electronic correlations dynamically modulated by light and offer solutions to the realization of photo-induced superconductivity. 

Time-resolved X-ray techniques such as trXAS and trRSXS are expected to play a decisive role in this endeavor. They can simultaneously track valence changes, lattice distortions, and the evolution of charge and spin correlations on femtosecond timescales, thereby providing direct insight into whether light-induced superconductivity involves a genuine pairing mechanism or merely a transient enhancement of conductivity. In particular, element-specific femtosecond X-ray scattering at XFELs can visualize the melting or reformation of charge-density-wave and stripe orders that compete with superconductivity. At the same time, high-repetition-rate XFEL and HHG sources will enable statistically robust measurements of transient structural coherence.
The synergy between optical-pump terahertz-probe studies and ultrafast X-ray measurements is expected to elucidate the microscopic pathways that give rise to emergent superconducting-like states, thereby offering a promising frontier for nonequilibrium quantum-material research.

\subsection{Combination of XFEL and THz}
The ongoing development of ultrashort-pulse laser technology has revealed nonequilibrium phase transitions and collective excitations in condensed matter that are inaccessible under thermal equilibrium. 
Among these, recent advances in the terahertz (THz) frequency range have provided new opportunities for selective and coherent control of low-energy excitations. 
THz pulses can resonantly drive elementary modes such as phonons, magnons, and polar lattice oscillations. They can serve as intense, transient electric or magnetic fields confined within a few optical cycles. 
In contrast, XFELs deliver femtosecond, coherent, and element-specific probes that resolve structural and electronic changes at the atomic scale. 
The combination of these two complementary light sources-the THz-pump/X-ray-probe technique—has emerged as a powerful platform for exploring the interplay among lattice, spin, and charge degrees of freedom in the ultrafast regime.

This approach has already enabled the visualization of lattice and superlattice dynamics following resonant excitation of phonons, charge-density waves, or polar vortices \cite{Kuba,10.1063/1.4983153,Kozina2019}. 
The strong-field nature of THz radiation, often exceeding hundreds of kV/cm, enables nonlinear driving of atomic motions or magnetic order parameters, while femtosecond XFEL pulses provide structural snapshots of the resulting symmetry breaking and relaxation processes. 
Recent developments in accelerator-based THz sources synchronized with XFELs have further expanded the accessible field strengths and frequency ranges, facilitating systematic studies of field-induced phase transitions~\cite{Kampfrath2013}. 

\begin{figure}[H]
\centering
\includegraphics[width=8cm]{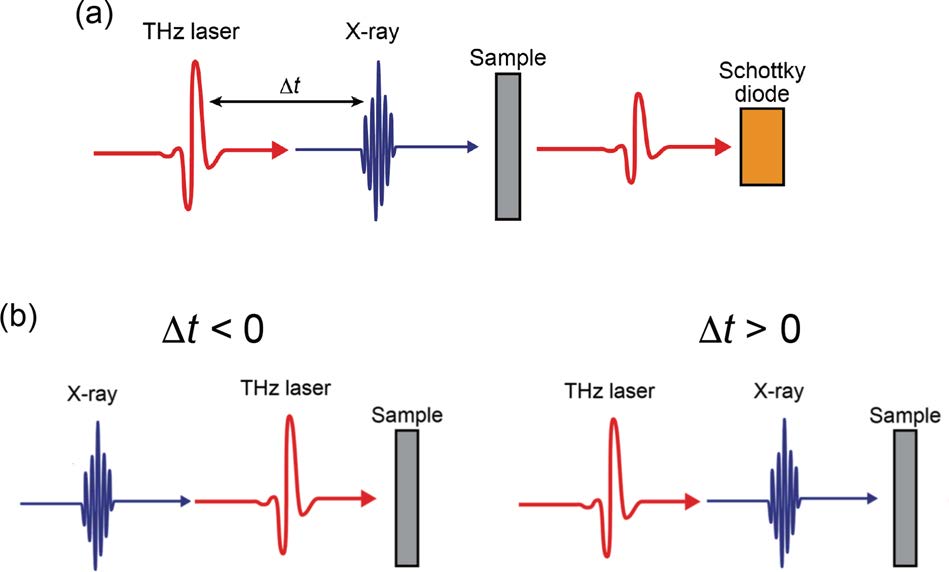}
\caption{
(a) Schematic of the X-ray pump–THz probe experiment for semiconductors.
(b) Schematic of the X-ray and THz pulses reaching the sample. At positive (negative) delay times, the X-ray (THz) pulse reaches the sample first. \cite{KubotaTHz}}
\label{future2}
\end{figure}

A major technical challenge in this field has been the precise determination and stabilization of the temporal overlap between THz and X-ray pulses as shown in Fig.~\ref{future2}. 
This synchronization is essential for achieving reproducible pump-probe measurements with femtosecond accuracy. 
While earlier experiments relied on indirect methods involving near-infrared (NIR) intermediary pulses and electro-optic sampling, Kubota \textit{et al.} recently demonstrated a reliable and straightforward approach for direct THz-X-ray temporal overlap determination with picosecond accuracy, eliminating the need for intermediate NIR pulses \cite{KubotaTHz}. This advancement marks a significant step toward fully synchronized THz-XFEL facilities. Future improvements in shot-to-shot arrival-time diagnostics and active feedback control are expected to push the timing precision to the tens-of-femtosecond level, enabling phase-resolved measurements of coherent phonon, magnon, and polar excitations.

Furthermore, intense THz magnetic fields have already been shown to directly manipulate spin order on sub-picosecond timescales, leading to ultrafast demagnetization in ferromagnetic thin films \cite{Polley}.
These THz-driven spin dynamics demonstrate the capability of strong-field THz pulses to coherently control magnetic order parameters without significant electronic heating.
Combining such THz control with time-resolved X-ray techniques will enable element-specific observation of spin and orbital responses, providing microscopic insight into ultrafast magnetic processes that have so far been accessible only through optical pump-probe studies.
Looking ahead, the integration of high-field THz sources with high-repetition-rate XFEL beamlines will open a new regime in which low-energy collective modes can be both driven and tracked with femtosecond precision.
This integration will not only deepen our understanding of coupled charge-spin-lattice dynamics but also pave the way for the realization of all-optical magnetic switching and photo-induced superconductivity, in which ultrafast control of quantum states is achieved entirely through light-matter interactions.

\section*{Acknowledgments}
The authors would like to thank Y. Hirata, K. Takubo, Y. Yokoyama, R. Takahashi, Y. Kubota, N. Ishii, and K. Takasan for informative discussions, and K. Sakurai for critical reading of the manuscript during its preparation. This work was supported by the JSPS KAKENHI under Grant Nos. 23K25805 and 25H01251, and the MEXT Quantum Leap Flagship Program (MEXT Q-LEAP) under Grant No. JPMXS0118068681. 




\bibliographystyle{aipnum4-1}
\nocite{*} 
\bibliography{sample}

@Article{Ueda2023,
author={Ueda, Hiroki
and Garc{\'i}a-Fern{\'a}ndez, Mirian
and Agrestini, Stefano
and Romao, Carl P.
and van den Brink, Jeroen
and Spaldin, Nicola A.
and Zhou, Ke-Jin
and Staub, Urs},
title={Chiral phonons in quartz probed by X-rays},
journal={Nature},
year={2023},
month={Jun},
day={01},
volume={618},
number={7967},
pages={946-950},
issn={1476-4687},
doi={10.1038/s41586-023-06016-5},
url={https://doi.org/10.1038/s41586-023-06016-5}
}

@Article{XASrev,
author={Chantler, Christopher T.
and Bunker, Grant
and D'Angelo, Paola
and Diaz-Moreno, Sofia},
title={X-ray absorption spectroscopy},
journal={Nature Reviews Methods Primers},
year={2024},
month={Dec},
day={05},
volume={4},
number={1},
pages={89},
issn={2662-8449},
doi={10.1038/s43586-024-00366-8},
url={https://doi.org/10.1038/s43586-024-00366-8}
}

@Article{Vaz2025,
author={Vaz, C. A. F.
and van der Laan, G.
and Cavill, S. A.
and D{\"u}rr, H. A.
and Fraile Rodr{\'i}guez, A.
and Kronast, F.
and Kuch, W.
and Sainctavit, Ph.
and Sch{\"u}tz, G.
and Wende, H.
and Weschke, E.
and Wilhelm, F.},
title={X-ray magnetic circular dichroism},
journal={Nature Reviews Methods Primers},
year={2025},
month={May},
day={02},
volume={5},
number={1},
pages={27},
issn={2662-8449},
doi={10.1038/s43586-025-00397-9},
url={https://doi.org/10.1038/s43586-025-00397-9}
}

@article{RevModPhys.82.2731,
  title = {Ultrafast optical manipulation of magnetic order},
  author = {Kirilyuk, Andrei and Kimel, Alexey V. and Rasing, Theo},
  journal = {Rev. Mod. Phys.},
  volume = {82},
  issue = {3},
  pages = {2731--2784},
  numpages = {0},
  year = {2010},
  month = {Sep},
  publisher = {American Physical Society},
  doi = {10.1103/RevModPhys.82.2731},
  url = {https://link.aps.org/doi/10.1103/RevModPhys.82.2731}
}

@article{PhysRevLett.76.4250,
  title = {Ultrafast Spin Dynamics in Ferromagnetic Nickel},
  author = {Beaurepaire, E. and Merle, J.-C. and Daunois, A. and Bigot, J.-Y.},
  journal = {Phys. Rev. Lett.},
  volume = {76},
  issue = {22},
  pages = {4250--4253},
  numpages = {0},
  year = {1996},
  month = {May},
  publisher = {American Physical Society},
  doi = {10.1103/PhysRevLett.76.4250},
  url = {https://link.aps.org/doi/10.1103/PhysRevLett.76.4250}
}

@Article{Emma2010,
author={Emma, P.
and Akre, R.
and Arthur, J.
and Bionta, R.
and Bostedt, C.
and Bozek, J.
and Brachmann, A.
and Bucksbaum, P.
and Coffee, R.
and Decker, F.-J.
and Ding, Y.
and Dowell, D.
and Edstrom, S.
and Fisher, A.
and Frisch, J.
and Gilevich, S.
and Hastings, J.
and Hays, G.
and Hering, Ph.
and Huang, Z.
and Iverson, R.
and Loos, H.
and Messerschmidt, M.
and Miahnahri, A.
and Moeller, S.
and Nuhn, H.-D.
and Pile, G.
and Ratner, D.
and Rzepiela, J.
and Schultz, D.
and Smith, T.
and Stefan, P.
and Tompkins, H.
and Turner, J.
and Welch, J.
and White, W.
and Wu, J.
and Yocky, G.
and Galayda, J.},
title={First lasing and operation of an {\aa}ngstrom-wavelength free-electron laser},
journal={Nature Photonics},
year={2010},
month={Sep},
day={01},
volume={4},
number={9},
pages={641-647},
issn={1749-4893},
doi={10.1038/nphoton.2010.176},
url={https://doi.org/10.1038/nphoton.2010.176}
}

@Article{Ishikawa2012,
author={Ishikawa, Tetsuya
and Aoyagi, Hideki
and Asaka, Takao
and Asano, Yoshihiro
and Azumi, Noriyoshi
and Bizen, Teruhiko
and Ego, Hiroyasu
and Fukami, Kenji
and Fukui, Toru
and Furukawa, Yukito
and Goto, Shunji
and Hanaki, Hirofumi
and Hara, Toru
and Hasegawa, Teruaki
and Hatsui, Takaki
and Higashiya, Atsushi
and Hirono, Toko
and Hosoda, Naoyasu
and Ishii, Miho
and Inagaki, Takahiro
and Inubushi, Yuichi
and Itoga, Toshiro
and Joti, Yasumasa
and Kago, Masahiro
and Kameshima, Takashi
and Kimura, Hiroaki
and Kirihara, Yoichi
and Kiyomichi, Akio
and Kobayashi, Toshiaki
and Kondo, Chikara
and Kudo, Togo
and Maesaka, Hirokazu
and Mar{\'e}chal, Xavier M.
and Masuda, Takemasa
and Matsubara, Shinichi
and Matsumoto, Takahiro
and Matsushita, Tomohiro
and Matsui, Sakuo
and Nagasono, Mitsuru
and Nariyama, Nobuteru
and Ohashi, Haruhiko
and Ohata, Toru
and Ohshima, Takashi
and Ono, Shun
and Otake, Yuji
and Saji, Choji
and Sakurai, Tatsuyuki
and Sato, Takahiro
and Sawada, Kei
and Seike, Takamitsu
and Shirasawa, Katsutoshi
and Sugimoto, Takashi
and Suzuki, Shinsuke
and Takahashi, Sunao
and Takebe, Hideki
and Takeshita, Kunikazu
and Tamasaku, Kenji
and Tanaka, Hitoshi
and Tanaka, Ryotaro
and Tanaka, Takashi
and Togashi, Tadashi
and Togawa, Kazuaki
and Tokuhisa, Atsushi
and Tomizawa, Hiromitsu
and Tono, Kensuke
and Wu, Shukui
and Yabashi, Makina
and Yamaga, Mitsuhiro
and Yamashita, Akihiro
and Yanagida, Kenichi
and Zhang, Chao
and Shintake, Tsumoru
and Kitamura, Hideo
and Kumagai, Noritaka},
title={A compact X-ray free-electron laser emitting in the sub-{\aa}ngstr{\"o}m region},
journal={Nature Photonics},
year={2012},
month={Aug},
day={01},
volume={6},
number={8},
pages={540-544},
issn={1749-4893},
doi={10.1038/nphoton.2012.141},
url={https://doi.org/10.1038/nphoton.2012.141}
}

@article{Schoenlein2019,
  title = {Recent advances in ultrafast X-ray sources},
  volume = {377},
  ISSN = {1471-2962},
  url = {http://dx.doi.org/10.1098/rsta.2018.0384},
  DOI = {10.1098/rsta.2018.0384},
  number = {2145},
  journal = {Philosophical Transactions of the Royal Society A: Mathematical,  Physical and Engineering Sciences},
  publisher = {The Royal Society},
  author = {Schoenlein,  Robert and Elsaesser,  Thomas and Holldack,  Karsten and Huang,  Zhirong and Kapteyn,  Henry and Murnane,  Margaret and Woerner,  Michael},
  year = {2019},
  month = apr,
  pages = {20180384}
}

@article{Uemura2022,
  title = {Hole Dynamics in Photoexcited Hematite Studied with Femtosecond Oxygen K-edge X-ray Absorption Spectroscopy},
  volume = {13},
  ISSN = {1948-7185},
  url = {http://dx.doi.org/10.1021/acs.jpclett.2c00295},
  DOI = {10.1021/acs.jpclett.2c00295},
  number = {19},
  journal = {The Journal of Physical Chemistry Letters},
  publisher = {American Chemical Society (ACS)},
  author = {Uemura,  Yohei and Ismail,  Ahmed S. M. and Park,  Sang Han and Kwon,  Soonnam and Kim,  Minseok and Elnaggar,  Hebatalla and Frati,  Federica and Wadati,  Hiroki and Hirata,  Yasuyuki and Zhang,  Yujun and Yamagami,  Kohei and Yamamoto,  Susumu and Matsuda,  Iwao and Halisdemir,  Ufuk and Koster,  Gertjan and Milne,  Christopher and Ammann,  Markus and Weckhuysen,  Bert M. and de Groot,  Frank M. F.},
  year = {2022},
  month = may,
  pages = {4207–4214}
}

@article{Shen,
    author = {Shen, Yufan and Kan, Daisuke and Lin, I-Ching and Chu, Ming-Wen and Suzuki, Ikumi and Shimakawa, Yuichi},
    title = {Perpendicular magnetic tunnel junctions based on half-metallic NiCo2O4},
    journal = {Applied Physics Letters},
    volume = {117},
    number = {4},
    pages = {042408},
    year = {2020},
    month = {07},
    issn = {0003-6951},
    doi = {10.1063/5.0017637},
    url = {https://doi.org/10.1063/5.0017637},
}

@article{Tokura,
doi = {10.1088/0034-4885/77/7/076501},
url = {https://doi.org/10.1088/0034-4885/77/7/076501},
year = {2014},
month = {jul},
publisher = {IOP Publishing},
volume = {77},
number = {7},
pages = {076501},
author = {Tokura, Yoshinori and Seki, Shinichiro and Nagaosa, Naoto},
title = {Multiferroics of spin origin},
journal = {Reports on Progress in Physics}
}

@article{PhysRevLett.75.152,
  title = {Experimental Confirmation of the X-Ray Magnetic Circular Dichroism Sum Rules for Iron and Cobalt},
  author = {Chen, C. T. and Idzerda, Y. U. and Lin, H.-J. and Smith, N. V. and Meigs, G. and Chaban, E. and Ho, G. H. and Pellegrin, E. and Sette, F.},
  journal = {Phys. Rev. Lett.},
  volume = {75},
  issue = {1},
  pages = {152--155},
  numpages = {0},
  year = {1995},
  month = {Jul},
  publisher = {American Physical Society},
  doi = {10.1103/PhysRevLett.75.152},
  url = {https://link.aps.org/doi/10.1103/PhysRevLett.75.152}
}

@article{VANDERLAAN2014,
title = {X-ray magnetic circular dichroism—A versatile tool to study magnetism},
journal = {Coordination Chemistry Reviews},
volume = {277-278},
pages = {95-129},
year = {2014},
note = {Following Chemical Structures using Synchrotron Radiation},
issn = {0010-8545},
doi = {https://doi.org/10.1016/j.ccr.2014.03.018},
url = {https://www.sciencedirect.com/science/article/pii/S0010854514000733},
author = {Gerrit {van der Laan} and Adriana I. Figueroa}
}

@article{Fink,
doi = {10.1088/0034-4885/76/5/056502},
url = {https://doi.org/10.1088/0034-4885/76/5/056502},
year = {2013},
month = {apr},
publisher = {IOP Publishing},
volume = {76},
number = {5},
pages = {056502},
author = {Fink, J and Schierle, E and Weschke, E and Geck, J},
title = {Resonant elastic soft x-ray scattering},
journal = {Reports on Progress in Physics},
}

@article{BL07LSU,
author = "Yamamoto, Susumu and Senba, Yasunori and Tanaka, Takashi and Ohashi, Haruhiko and Hirono, Toko and Kimura, Hiroaki and Fujisawa, Masami and Miyawaki, Jun and Harasawa, Ayumi and Seike, Takamitsu and Takahashi, Sunao and Nariyama, Nobuteru and Matsushita, Tomohiro and Takeuchi, Masao and Ohata, Toru and Furukawa, Yukito and Takeshita, Kunikazu and Goto, Shunji and Harada, Yoshihisa and Shin, Shik and Kitamura, Hideo and Kakizaki, Akito and Oshima, Masaharu and Matsuda, Iwao",
title = "{New soft X-ray beamline BL07LSU at SPring-8}",
journal = "Journal of Synchrotron Radiation",
year = "2014",
volume = "21",
number = "2",
pages = "352--365",
month = "Mar",
doi = {10.1107/S1600577513034796},
url = {https://doi.org/10.1107/S1600577513034796},
}

@article{TakuboAPL,
    author = {Takubo, Kou and Yamamoto, Kohei and Hirata, Yasuyuki and Yokoyama, Yuichi and Kubota, Yuya and Yamamoto, Shingo and Yamamoto, Susumu and Matsuda, Iwao and Shin, Shik and Seki, Takeshi and Takanashi, Koki and Wadati, Hiroki},
    title = {Capturing ultrafast magnetic dynamics by time-resolved soft x-ray magnetic circular dichroism},
    journal = {Applied Physics Letters},
    volume = {110},
    number = {16},
    pages = {162401},
    year = {2017},
    month = {04},
    issn = {0003-6951},
    doi = {10.1063/1.4981769},
    url = {https://doi.org/10.1063/1.4981769},
}

@article{Ogawa,
    author = {Ogawa, Manami and Yamamoto, Susumu and Kousa, Yuka and Nakamura, Fumitaka and Yukawa, Ryu and Fukushima, Akiko and Harasawa, Ayumi and Kondoh, Hiroshi and Tanaka, Yoshihito and Kakizaki, Akito and Matsuda, Iwao},
    title = {Development of soft x-ray time-resolved photoemission spectroscopy system with a two-dimensional angle-resolved time-of-flight analyzer at SPring-8 BL07LSU},
    journal = {Review of Scientific Instruments},
    volume = {83},
    number = {2},
    pages = {023109},
    year = {2012},
    month = {02},
    issn = {0034-6748},
    doi = {10.1063/1.3687428},
    url = {https://doi.org/10.1063/1.3687428},
}

@Article{Stamm2007,
author={Stamm, C.
and Kachel, T.
and Pontius, N.
and Mitzner, R.
and Quast, T.
and Holldack, K.
and Khan, S.
and Lupulescu, C.
and Aziz, E. F.
and Wietstruk, M.
and D{\"u}rr, H. A.
and Eberhardt, W.},
title={Femtosecond modification of electron localization and transfer of angular momentum in nickel},
journal={Nature Materials},
year={2007},
month={Oct},
day={01},
volume={6},
number={10},
pages={740-743},
issn={1476-4660},
doi={10.1038/nmat1985},
url={https://doi.org/10.1038/nmat1985}
}

@article{Higley,
    author = {Higley, Daniel J. and Hirsch, Konstantin and Dakovski, Georgi L. and Jal, Emmanuelle and Yuan, Edwin and Liu, Tianmin and Lutman, Alberto A. and MacArthur, James P. and Arenholz, Elke and Chen, Zhao and Coslovich, Giacomo and Denes, Peter and Granitzka, Patrick W. and Hart, Philip and Hoffmann, Matthias C. and Joseph, John and Le Guyader, Loïc and Mitra, Ankush and Moeller, Stefan and Ohldag, Hendrik and Seaberg, Matthew and Shafer, Padraic and Stöhr, Joachim and Tsukamoto, Arata and Nuhn, Heinz-Dieter and Reid, Alex H. and Dürr, Hermann A. and Schlotter, William F.},
    title = {Femtosecond X-ray magnetic circular dichroism absorption spectroscopy at an X-ray free electron laser},
    journal = {Review of Scientific Instruments},
    volume = {87},
    number = {3},
    pages = {033110},
    year = {2016},
    month = {03},
    issn = {0034-6748},
    doi = {10.1063/1.4944410},
    url = {https://doi.org/10.1063/1.4944410},
}

@article{JPSJ.82.021003,
author = {Yamamoto ,Susumu and Matsuda ,Iwao},
title = {Time-Resolved Photoelectron Spectroscopies Using Synchrotron Radiation: Past, Present, and Future},
journal = {Journal of the Physical Society of Japan},
volume = {82},
number = {2},
pages = {021003},
year = {2013},
doi = {10.7566/JPSJ.82.021003},

URL = { 
    
        https://doi.org/10.7566/JPSJ.82.021003
    
    

}
}

@article{Schoenlein,
  title = {Generation of Femtosecond Pulses of Synchrotron Radiation},
  volume = {287},
  ISSN = {1095-9203},
  url = {http://dx.doi.org/10.1126/science.287.5461.2237},
  DOI = {10.1126/science.287.5461.2237},
  number = {5461},
  journal = {Science},
  publisher = {American Association for the Advancement of Science (AAAS)},
  author = {Schoenlein,  R. W. and Chattopadhyay,  S. and Chong,  H. H. W. and Glover,  T. E. and Heimann,  P. A. and Shank,  C. V. and Zholents,  A. A. and Zolotorev,  M. S.},
  year = {2000},
  month = mar,
  pages = {2237–2240}
}

@article{PhysRevLett.97.074801,
  title = {Femtosecond Undulator Radiation from Sliced Electron Bunches},
  author = {Khan, S. and Holldack, K. and Kachel, T. and Mitzner, R. and Quast, T.},
  journal = {Phys. Rev. Lett.},
  volume = {97},
  issue = {7},
  pages = {074801},
  numpages = {4},
  year = {2006},
  month = {Aug},
  publisher = {American Physical Society},
  doi = {10.1103/PhysRevLett.97.074801},
  url = {https://link.aps.org/doi/10.1103/PhysRevLett.97.074801}
}

@article{Hol,
author = "Holldack, Karsten and Bahrdt, Johannes and Balzer, Andreas and Bovensiepen, Uwe and Brzhezinskaya, Maria and Erko, Alexei and Eschenlohr, Andrea and Follath, Rolf and Firsov, Alexander and Frentrup, Winfried and Le Guyader, Lo{\"\i}c and Kachel, Torsten and Kuske, Peter and Mitzner, Rolf and M{\"{u}}ller, Roland and Pontius, Niko and Quast, Torsten and Radu, Ilie and Schmidt, Jan-Simon and Sch{\"{u}}{\ss}ler-Langeheine, Christian and Sperling, Mike and Stamm, Christian and Trabant, Christoph and F{\"{o}}hlisch, Alexander",
title = "{FemtoSpeX: a versatile optical pump{--}soft X-ray probe facility with 100fs X-ray pulses of variable polarization}",
journal = "Journal of Synchrotron Radiation",
year = "2014",
volume = "21",
number = "5",
pages = "1090--1104",
month = "Sep",
doi = {10.1107/S1600577514012247},
url = {https://doi.org/10.1107/S1600577514012247},
}

@article{PhysRevLett.99.174801,
  title = {Spatiotemporal Stability of a Femtosecond Hard--X-Ray Undulator Source Studied by Control of Coherent Optical Phonons},
  author = {Beaud, P. and Johnson, S. L. and Streun, A. and Abela, R. and Abramsohn, D. and Grolimund, D. and Krasniqi, F. and Schmidt, T. and Schlott, V. and Ingold, G.},
  journal = {Phys. Rev. Lett.},
  volume = {99},
  issue = {17},
  pages = {174801},
  numpages = {4},
  year = {2007},
  month = {Oct},
  publisher = {American Physical Society},
  doi = {10.1103/PhysRevLett.99.174801},
  url = {https://link.aps.org/doi/10.1103/PhysRevLett.99.174801}
}

@article{Labat,
author = "Labat, Marie and Brubach, Jean-Blaise and Ciavardini, Alessandra and Couprie, Marie-Emmanuelle and Elkaim, Erik and Fertey, Pierre and Ferte, Tom and Hollander, Philippe and Hubert, Nicolas and Jal, Emmanuelle and Laulh{\'{e}}, Claire and Luning, Jan and Marcouill{\'{e}}, Olivier and Moreno, Thierry and Morin, Paul and Polack, Francois and Prigent, Pascale and Ravy, Sylvain and Ricaud, Jean-Paul and Roy, Pascale and Silly, Mathieu and Sirotti, Fausto and Taleb, Amina and Tordeux, Marie-Agn{\`{e}}s and Nadji, Amor",
title = "{Commissioning of a multi-beamline femtoslicing facility at SOLEIL}",
journal = "Journal of Synchrotron Radiation",
year = "2018",
volume = "25",
number = "2",
pages = "385--398",
month = "Mar",
doi = {10.1107/S1600577518000863},
url = {https://doi.org/10.1107/S1600577518000863},
}

@article{PhysRevA.39.5751,
  title = {Multiple-harmonic generation in rare gases at high laser intensity},
  author = {Li, X. F. and L'Huillier, A. and Ferray, M. and Lompr\'e, L. A. and Mainfray, G.},
  journal = {Phys. Rev. A},
  volume = {39},
  issue = {11},
  pages = {5751--5761},
  numpages = {0},
  year = {1989},
  month = {Jun},
  publisher = {American Physical Society},
  doi = {10.1103/PhysRevA.39.5751},
  url = {https://link.aps.org/doi/10.1103/PhysRevA.39.5751}
}

@article{Paul2001,
  title = {Observation of a Train of Attosecond Pulses from High Harmonic Generation},
  volume = {292},
  ISSN = {1095-9203},
  url = {http://dx.doi.org/10.1126/science.1059413},
  DOI = {10.1126/science.1059413},
  number = {5522},
  journal = {Science},
  publisher = {American Association for the Advancement of Science (AAAS)},
  author = {Paul,  P. M. and Toma,  E. S. and Breger,  P. and Mullot,  G. and Augé,  F. and Balcou,  Ph. and Muller,  H. G. and Agostini,  P.},
  year = {2001},
  month = jun,
  pages = {1689–1692}
}

@Article{Hentschel2001,
author={Hentschel, M.
and Kienberger, R.
and Spielmann, Ch.
and Reider, G. A.
and Milosevic, N.
and Brabec, T.
and Corkum, P.
and Heinzmann, U.
and Drescher, M.
and Krausz, F.},
title={Attosecond metrology},
journal={Nature},
year={2001},
month={Nov},
day={01},
volume={414},
number={6863},
pages={509-513},
issn={1476-4687},
doi={10.1038/35107000},
url={https://doi.org/10.1038/35107000}
}

@article{RSXSWater,
    author = {Jarecki, Jasmin and Hennecke, Martin and Sidiropoulos, Themistoklis and Schnuerer, Matthias and Eisebitt, Stefan and Schick, Daniel},
    title = {Ultrafast energy-dispersive soft-x-ray diffraction in the water window with a laser-driven source},
    journal = {Structural Dynamics},
    volume = {11},
    number = {5},
    pages = {054303},
    year = {2024},
    month = {10},
    issn = {2329-7778},
    doi = {10.1063/4.0000270},
    url = {https://doi.org/10.1063/4.0000270},
}

@article{CHEN2025,
title = {Ultrafast demagnetization in ferromagnetic materials: Origins and progress},
journal = {Physics Reports},
volume = {1102},
pages = {1-63},
year = {2025},
note = {Ultrafast demagnetization in ferromagnetic materials: Origins and progress},
issn = {0370-1573},
doi = {https://doi.org/10.1016/j.physrep.2024.10.008},
url = {https://www.sciencedirect.com/science/article/pii/S037015732400379X},
author = {Xiaowen Chen and Roman Adam and Daniel E. Bürgler and Fangzhou Wang and Zhenyan Lu and Lining Pan and Sarah Heidtfeld and Christian Greb and Meihong Liu and Qingfang Liu and Jianbo Wang and Claus M. Schneider and Derang Cao}
}

@article{feld,
title = {Electron and lattice dynamics following optical excitation of metals},
journal = {Chemical Physics},
volume = {251},
number = {1},
pages = {237-258},
year = {2000},
issn = {0301-0104},
doi = {https://doi.org/10.1016/S0301-0104(99)00330-4},
url = {https://www.sciencedirect.com/science/article/pii/S0301010499003304},
author = {J. Hohlfeld and S.-S. Wellershoff and J. Güdde and U. Conrad and V. Jähnke and E. Matthias}
}

@article{Lambert2014,
  title = {All-optical control of ferromagnetic thin films and nanostructures},
  volume = {345},
  ISSN = {1095-9203},
  url = {http://dx.doi.org/10.1126/science.1253493},
  DOI = {10.1126/science.1253493},
  number = {6202},
  journal = {Science},
  publisher = {American Association for the Advancement of Science (AAAS)},
  author = {Lambert,  C-H. and Mangin,  S. and Varaprasad,  B. S. D. Ch. S. and Takahashi,  Y. K. and Hehn,  M. and Cinchetti,  M. and Malinowski,  G. and Hono,  K. and Fainman,  Y. and Aeschlimann,  M. and Fullerton,  E. E.},
  year = {2014},
  month = sep,
  pages = {1337–1340}
}

@article{SuzukiMCD,
author = "Suzuki, Motohiro and Inubushi, Yuichi and Yabashi, Makina and Ishikawa, Tetsuya",
title = "{Polarization control of an X-ray free-electron laser with a diamond phase retarder}",
journal = "Journal of Synchrotron Radiation",
year = "2014",
volume = "21",
number = "3",
pages = "466--472",
month = "May",
doi = {10.1107/S1600577514004780},
url = {https://doi.org/10.1107/S1600577514004780},
}

@article{KubotaMCD,
author = "Kubota, Y. and Suzuki, M. and Katayama, T. and Yamamoto, K. and Tono, K. and Inubushi, Y. and Seki, T. and Takanashi, K. and Wadati, H. and Yabashi, M.",
title = "{Polarization control with an X-ray phase retarder for high-time-resolution pump{--}probe experiments at SACLA}",
journal = "Journal of Synchrotron Radiation",
year = "2019",
volume = "26",
number = "4",
pages = "1139--1143",
month = "Jul",
doi = {10.1107/S1600577519006222},
url = {https://doi.org/10.1107/S1600577519006222},
}

@article{Yamamotonjp,
doi = {10.1088/1367-2630/ab5ac2},
url = {https://doi.org/10.1088/1367-2630/ab5ac2},
year = {2019},
month = {dec},
publisher = {IOP Publishing},
volume = {21},
number = {12},
pages = {123010},
author = {Yamamoto, Kohei and Kubota, Yuya and Suzuki, Motohiro and Hirata, Yasuyuki and Carva, Karel and Berritta, Marco and Takubo, Kou and Uemura, Yohei and Fukaya, Ryo and Tanaka, Kenta and Nishimura, Wataru and Ohkochi, Takuo and Katayama, Tetsuo and Togashi, Tadashi and Tamasaku, Kenji and Yabashi, Makina and Tanaka, Yoshihito and Seki, Takeshi and Takanashi, Koki and Oppeneer, Peter M and Wadati, Hiroki},
title = {Ultrafast demagnetization of Pt magnetic moment in L10-FePt probed by magnetic circular dichroism at a hard x-ray free electron laser},
journal = {New Journal of Physics}
}

@article{10.1063/1.4993077,
    author = {Ikeda, K. and Seki, T. and Shibata, G. and Kadono, T. and Ishigami, K. and Takahashi, Y. and Horio, M. and Sakamoto, S. and Nonaka, Y. and Sakamaki, M. and Amemiya, K. and Kawamura, N. and Suzuki, M. and Takanashi, K. and Fujimori, A.},
    title = {Magnetic anisotropy of L1-ordered FePt thin films studied by Fe and Pt L2,3-edges x-ray magnetic circular dichroism},
    journal = {Applied Physics Letters},
    volume = {111},
    number = {14},
    pages = {142402},
    year = {2017},
    month = {10},
    issn = {0003-6951},
    doi = {10.1063/1.4993077},
    url = {https://doi.org/10.1063/1.4993077},
}

@article{Hoffer,
  title = {Induced versus intrinsic magnetic moments in ultrafast magnetization dynamics},
  author = {Hofherr, M. and Moretti, S. and Shim, J. and H\"auser, S. and Safonova, N. Y. and Stiehl, M. and Ali, A. and Sakshath, S. and Kim, J. W. and Kim, D. H. and Kim, H. J. and Hong, J. I. and Kapteyn, H. C. and Murnane, M. M. and Cinchetti, M. and Steil, D. and Mathias, S. and Stadtm\"uller, B. and Albrecht, M. and Kim, D. E. and Nowak, U. and Aeschlimann, M.},
  journal = {Phys. Rev. B},
  volume = {98},
  issue = {17},
  pages = {174419},
  numpages = {7},
  year = {2018},
  month = {Nov},
  publisher = {American Physical Society},
  doi = {10.1103/PhysRevB.98.174419},
  url = {https://link.aps.org/doi/10.1103/PhysRevB.98.174419}
}

@article{PhysRevB.89.064423,
  title = {Observation of a giant Kerr rotation in a ferromagnetic transition metal by $M$-edge resonant magneto-optic Kerr effect},
  author = {Yamamoto, Sh. and Taguchi, M. and Fujisawa, M. and Hobara, R. and Yamamoto, S. and Yaji, K. and Nakamura, T. and Fujikawa, K. and Yukawa, R. and Togashi, T. and Yabashi, M. and Tsunoda, M. and Shin, S. and Matsuda, I.},
  journal = {Phys. Rev. B},
  volume = {89},
  issue = {6},
  pages = {064423},
  numpages = {6},
  year = {2014},
  month = {Feb},
  publisher = {American Physical Society},
  doi = {10.1103/PhysRevB.89.064423},
  url = {https://link.aps.org/doi/10.1103/PhysRevB.89.064423}
}

@article{YamamotoAPL,
    author = {Yamamoto, Kohei and Moussaoui, Souliman El and Hirata, Yasuyuki and Yamamoto, Susumu and Kubota, Yuya and Owada, Shigeki and Yabashi, Makina and Seki, Takeshi and Takanashi, Koki and Matsuda, Iwao and Wadati, Hiroki},
    title = {Element-selectively tracking ultrafast demagnetization process in Co/Pt multilayer thin films by the resonant magneto-optical Kerr effect},
    journal = {Applied Physics Letters},
    volume = {116},
    number = {17},
    pages = {172406},
    year = {2020},
    month = {05},
    issn = {0003-6951},
    doi = {10.1063/5.0005393},
    url = {https://doi.org/10.1063/5.0005393},
}

@article{Stanciu,
  title = {All-Optical Magnetic Recording with Circularly Polarized Light},
  author = {Stanciu, C. D. and Hansteen, F. and Kimel, A. V. and Kirilyuk, A. and Tsukamoto, A. and Itoh, A. and Rasing, Th.},
  journal = {Phys. Rev. Lett.},
  volume = {99},
  issue = {4},
  pages = {047601},
  numpages = {4},
  year = {2007},
  month = {Jul},
  publisher = {American Physical Society},
  doi = {10.1103/PhysRevLett.99.047601},
  url = {https://link.aps.org/doi/10.1103/PhysRevLett.99.047601}
}

@Article{Radu,
author={Radu, I.
and Vahaplar, K.
and Stamm, C.
and Kachel, T.
and Pontius, N.
and D{\"u}rr, H. A.
and Ostler, T. A.
and Barker, J.
and Evans, R. F. L.
and Chantrell, R. W.
and Tsukamoto, A.
and Itoh, A.
and Kirilyuk, A.
and Rasing, Th.
and Kimel, A. V.},
title={Transient ferromagnetic-like state mediating ultrafast reversal of antiferromagnetically coupled spins},
journal={Nature},
year={2011},
month={Apr},
day={01},
volume={472},
number={7342},
pages={205-208},
issn={1476-4687},
doi={10.1038/nature09901},
url={https://doi.org/10.1038/nature09901}
}

@article{PhysRevB.81.104415,
  title = {Laser-induced generation and quenching of magnetization on FeRh studied with time-resolved x-ray magnetic circular dichroism},
  author = {Radu, I. and Stamm, C. and Pontius, N. and Kachel, T. and Ramm, P. and Thiele, J.-U. and D\"urr, H. A. and Back, C. H.},
  journal = {Phys. Rev. B},
  volume = {81},
  issue = {10},
  pages = {104415},
  numpages = {5},
  year = {2010},
  month = {Mar},
  publisher = {American Physical Society},
  doi = {10.1103/PhysRevB.81.104415},
  url = {https://link.aps.org/doi/10.1103/PhysRevB.81.104415}
}

@Article{Zhang2022,
author={Zhang, Yujun
and Katayama, Tsukasa
and Chikamatsu, Akira
and Sch{\"u}{\ss}ler-Langeheine, Christian
and Pontius, Niko
and Hirata, Yasuyuki
and Takubo, Kou
and Yamagami, Kohei
and Ikeda, Keisuke
and Yamamoto, Kohei
and Hasegawa, Tetsuya
and Wadati, Hiroki},
title={Photo-induced antiferromagnetic-ferromagnetic and spin-state transition in a double-perovskite cobalt oxide thin film},
journal={Communications Physics},
year={2022},
month={Mar},
day={07},
volume={5},
number={1},
pages={50},
issn={2399-3650},
doi={10.1038/s42005-022-00823-4},
url={https://doi.org/10.1038/s42005-022-00823-4}
}

@article{katayama,
author = {Katayama, Tsukasa and Chikamatsu, Akira and Zhang, Yujun and Yasui, Shintaro and Wadati, Hiroki and Hasegawa, Tetsuya},
title = {Ionic Order Engineering in Double-Perovskite Cobaltite},
journal = {Chemistry of Materials},
volume = {33},
number = {14},
pages = {5675-5680},
year = {2021},
doi = {10.1021/acs.chemmater.1c01228},

URL = { 
    
        https://doi.org/10.1021/acs.chemmater.1c01228
    
    

},
}

@article{tsuyama,
  title = {Photoinduced Demagnetization and Insulator-to-Metal Transition in Ferromagnetic Insulating ${\mathrm{BaFeO}}_{3}$ Thin Films},
  author = {Tsuyama, T. and Chakraverty, S. and Macke, S. and Pontius, N. and Sch\"u\ss{}ler-Langeheine, C. and Hwang, H. Y. and Tokura, Y. and Wadati, H.},
  journal = {Phys. Rev. Lett.},
  volume = {116},
  issue = {25},
  pages = {256402},
  numpages = {5},
  year = {2016},
  month = {Jun},
  publisher = {American Physical Society},
  doi = {10.1103/PhysRevLett.116.256402},
  url = {https://link.aps.org/doi/10.1103/PhysRevLett.116.256402}
}

@article{Yamamoto2022,
doi = {10.1088/1367-2630/ac5f31},
url = {https://doi.org/10.1088/1367-2630/ac5f31},
year = {2022},
month = {apr},
publisher = {IOP Publishing},
volume = {24},
number = {4},
pages = {043012},
author = {Yamamoto, Kohei and Tsuyama, Tomoyuki and Ito, Suguru and Takubo, Kou and Matsuda, Iwao and Pontius, Niko and Schüßler-Langeheine, Christian and Minohara, Makoto and Kumigashira, Hiroshi and Yamasaki, Yuichi and Nakao, Hironori and Murakami, Youichi and Katase, Takayoshi and Kamiya, Toshio and Wadati, Hiroki},
title = {Photoinduced transient states of antiferromagnetic orderings in La$_{1/3}$Sr$_{2/3}$FeO$_{3}$ and SrFeO$_{3-\delta}$ thin films observed through time-resolved resonant soft x-ray scattering},
journal = {New Journal of Physics}
}

@article{Okamoto,
  title = {Quasi-two-dimensional $d$-spin and $p$-hole ordering in the three-dimensional perovskite ${\text{La}}_{1/3}{\text{Sr}}_{2/3}{\text{FeO}}_{3}$},
  author = {Okamoto, J. and Huang, D. J. and Chao, K. S. and Huang, S. W. and Hsu, C.-H. and Fujimori, A. and Masuno, A. and Terashima, T. and Takano, M. and Chen, C. T.},
  journal = {Phys. Rev. B},
  volume = {82},
  issue = {13},
  pages = {132402},
  numpages = {4},
  year = {2010},
  month = {Oct},
  publisher = {American Physical Society},
  doi = {10.1103/PhysRevB.82.132402},
  url = {https://link.aps.org/doi/10.1103/PhysRevB.82.132402}
}

@article{PhysRevB.88.220405,
  title = {Multiple helimagnetic phases and topological Hall effect in epitaxial thin films of pristine and Co-doped SrFeO${}_{3}$},
  author = {Chakraverty, S. and Matsuda, T. and Wadati, H. and Okamoto, J. and Yamasaki, Y. and Nakao, H. and Murakami, Y. and Ishiwata, S. and Kawasaki, M. and Taguchi, Y. and Tokura, Y. and Hwang, H. Y.},
  journal = {Phys. Rev. B},
  volume = {88},
  issue = {22},
  pages = {220405},
  numpages = {5},
  year = {2013},
  month = {Dec},
  publisher = {American Physical Society},
  doi = {10.1103/PhysRevB.88.220405},
  url = {https://link.aps.org/doi/10.1103/PhysRevB.88.220405}
}

@article{YamamotoLSFOPRB,
  title = {Thickness dependence and dimensionality effects on charge and magnetic orderings in ${\mathrm{La}}_{1/3}{\mathrm{Sr}}_{2/3}{\mathrm{FeO}}_{3}$ thin films},
  author = {Yamamoto, K. and Hirata, Y. and Horio, M. and Yokoyama, Y. and Takubo, K. and Minohara, M. and Kumigashira, H. and Yamasaki, Y. and Nakao, H. and Murakami, Y. and Fujimori, A. and Wadati, H.},
  journal = {Phys. Rev. B},
  volume = {97},
  issue = {7},
  pages = {075134},
  numpages = {6},
  year = {2018},
  month = {Feb},
  publisher = {American Physical Society},
  doi = {10.1103/PhysRevB.97.075134},
  url = {https://link.aps.org/doi/10.1103/PhysRevB.97.075134}
}

@article{10.1063/1.4958670,
    author = {Minohara, Makoto and Kitamura, Miho and Wadati, Hiroki and Nakao, Hironori and Kumai, Reiji and Murakami, Youichi and Kumigashira, Hiroshi},
    title = {Thickness-dependent physical properties of La1/3Sr2/3FeO3 thin films grown on SrTiO3 (001) and (111) substrates},
    journal = {Journal of Applied Physics},
    volume = {120},
    number = {2},
    pages = {025303},
    year = {2016},
    month = {07},
    issn = {0021-8979},
    doi = {10.1063/1.4958670},
    url = {https://doi.org/10.1063/1.4958670},
}

@article{NNO,
  title = {Photoinduced melting of magnetic order in the correlated electron insulator NdNiO${}_{3}$},
  author = {Caviglia, A. D. and F\"orst, M. and Scherwitzl, R. and Khanna, V. and Bromberger, H. and Mankowsky, R. and Singla, R. and Chuang, Y.-D. and Lee, W. S. and Krupin, O. and Schlotter, W. F. and Turner, J. J. and Dakovski, G. L. and Minitti, M. P. and Robinson, J. and Scagnoli, V. and Wilkins, S. B. and Cavill, S. A. and Gibert, M. and Gariglio, S. and Zubko, P. and Triscone, J.-M. and Hill, J. P. and Dhesi, S. S. and Cavalleri, A.},
  journal = {Phys. Rev. B},
  volume = {88},
  issue = {22},
  pages = {220401},
  numpages = {5},
  year = {2013},
  month = {Dec},
  publisher = {American Physical Society},
  doi = {10.1103/PhysRevB.88.220401},
  url = {https://link.aps.org/doi/10.1103/PhysRevB.88.220401}
}

@Article{Zhu2018,
author={Zhu, Yi
and Hoffman, Jason
and Rowland, Clare E.
and Park, Hyowon
and Walko, Donald A.
and Freeland, John W.
and Ryan, Philip J.
and Schaller, Richard D.
and Bhattacharya, Anand
and Wen, Haidan},
title={Unconventional slowing down of electronic recovery in photoexcited charge-ordered La1/3Sr2/3FeO3},
journal={Nature Communications},
year={2018},
month={May},
day={04},
volume={9},
number={1},
pages={1799},
issn={2041-1723},
doi={10.1038/s41467-018-04199-4},
url={https://doi.org/10.1038/s41467-018-04199-4}
}

@article{Dy,
  title = {Ultrafast and Energy-Efficient Quenching of Spin Order: Antiferromagnetism Beats Ferromagnetism},
  author = {Thielemann-K\"uhn, Nele and Schick, Daniel and Pontius, Niko and Trabant, Christoph and Mitzner, Rolf and Holldack, Karsten and Zabel, Hartmut and F\"ohlisch, Alexander and Sch\"u\ss{}ler-Langeheine, Christian},
  journal = {Phys. Rev. Lett.},
  volume = {119},
  issue = {19},
  pages = {197202},
  numpages = {6},
  year = {2017},
  month = {Nov},
  publisher = {American Physical Society},
  doi = {10.1103/PhysRevLett.119.197202},
  url = {https://link.aps.org/doi/10.1103/PhysRevLett.119.197202}
}

@article{PhysRevLett.102.117201,
  title = {Laser-Induced Magnetization Dynamics of Lanthanide-Doped Permalloy Thin Films},
  author = {Radu, I. and Woltersdorf, G. and Kiessling, M. and Melnikov, A. and Bovensiepen, U. and Thiele, J.-U. and Back, C. H.},
  journal = {Phys. Rev. Lett.},
  volume = {102},
  issue = {11},
  pages = {117201},
  numpages = {4},
  year = {2009},
  month = {Mar},
  publisher = {American Physical Society},
  doi = {10.1103/PhysRevLett.102.117201},
  url = {https://link.aps.org/doi/10.1103/PhysRevLett.102.117201}
}

@article{Eu1,
  title = {Critical evaluation of Eu valences from ${\mathit{L}}_{\mathrm{III}}$-edge x-ray-absorption and M\"ossbauer spectroscopy of ${\mathrm{EuNi}}_{2}$${\mathrm{Si}}_{2\mathrm{\ensuremath{-}}\mathit{x}}$${\mathrm{Ge}}_{\mathit{x}}$},
  author = {Wortmann, G. and Nowik, I. and Perscheid, B. and Kaindl, G. and Felner, I.},
  journal = {Phys. Rev. B},
  volume = {43},
  issue = {7},
  pages = {5261--5268},
  numpages = {0},
  year = {1991},
  month = {Mar},
  publisher = {American Physical Society},
  doi = {10.1103/PhysRevB.43.5261},
  url = {https://link.aps.org/doi/10.1103/PhysRevB.43.5261}
}

@article{Wada1997,
doi = {10.1088/0953-8984/9/37/021},
url = {https://doi.org/10.1088/0953-8984/9/37/021},
year = {1997},
month = {sep},
publisher = {},
volume = {9},
number = {37},
pages = {7913},
author = {H Wada and A Nakamura and A Mitsuda and M Shiga and T Tanaka and H Mitamura and T Goto},
title = {Temperature- and field-induced valence transitions of EuNi$_2$(Si$_{1-x}$Ge$_x$)$_2$},
journal = {Journal of Physics: Condensed Matter}
}

@article{YamamotoEu1,
  title = {Temperature-dependent Eu $3d\text{\ensuremath{-}}4f$ x-ray absorption and resonant photoemission study of the valence transition in $\mathrm{Eu}{\mathrm{Ni}}_{2}{({\mathrm{Si}}_{0.2}{\mathrm{Ge}}_{0.8})}_{2}$},
  author = {Yamamoto, K. and Horiba, K. and Taguchi, M. and Matsunami, M. and Kamakura, N. and Chainani, A. and Takata, Y. and Mimura, K. and Shiga, M. and Wada, H. and Senba, Y. and Ohashi, H. and Shin, S.},
  journal = {Phys. Rev. B},
  volume = {72},
  issue = {16},
  pages = {161101},
  numpages = {4},
  year = {2005},
  month = {Oct},
  publisher = {American Physical Society},
  doi = {10.1103/PhysRevB.72.161101},
  url = {https://link.aps.org/doi/10.1103/PhysRevB.72.161101}
}

@article{YAMAMOTO2006681,
title = {Temperature dependent X-ray absorption spectroscopy of the valence transition in EuNi2(Si0.20Ge0.80)2},
journal = {Physica B: Condensed Matter},
volume = {378-380},
pages = {681-682},
year = {2006},
note = {Proceedings of the International Conference on Strongly Correlated Electron Systems},
issn = {0921-4526},
doi = {https://doi.org/10.1016/j.physb.2006.01.507},
url = {https://www.sciencedirect.com/science/article/pii/S0921452606003188},
author = {Kazuya Yamamoto and Koji Horiba and Munetaka Taguchi and Masaharu Matsunami and Nozomu Kamakura and Yasutaka Takata and Ashish Chainani and Kojiro Mimura and Masayuki Shiga and Hirofumi Wada and Yasunori Senba and Haruhiko Ohashi and Shik Shin}
}

@article{Ichiki,
  title = {Hard x-ray photoemission study of the temperature-induced valence transition system ${\mathrm{EuNi}}_{2}{({\mathrm{Si}}_{1\ensuremath{-}x}{\mathrm{Ge}}_{x})}_{2}$},
  author = {Ichiki, Katsuya and Mimura, Kojiro and Anzai, Hiroaki and Uozumi, Takayuki and Sato, Hitoshi and Utsumi, Yuki and Ueda, Shigenori and Mitsuda, Akihiro and Wada, Hirofumi and Taguchi, Yukihiro and Shimada, Kenya and Namatame, Hirofumi and Taniguchi, Masaki},
  journal = {Phys. Rev. B},
  volume = {96},
  issue = {4},
  pages = {045106},
  numpages = {7},
  year = {2017},
  month = {Jul},
  publisher = {American Physical Society},
  doi = {10.1103/PhysRevB.96.045106},
  url = {https://link.aps.org/doi/10.1103/PhysRevB.96.045106}
}

@article{SHIMOKASA,
title = {Temperature-induced valence transition in EuNi2(Si1–xGex)2 investigated by high-energy resolution fluorescence detection X-ray absorption spectroscopy},
journal = {Radiation Physics and Chemistry},
volume = {175},
pages = {108150},
year = {2020},
note = {17th International Conference on X-ray Absorption Fine Structure - XAFS2018},
issn = {0969-806X},
doi = {https://doi.org/10.1016/j.radphyschem.2019.02.009},
url = {https://www.sciencedirect.com/science/article/pii/S0969806X1830954X},
author = {Ryohei Shimokasa and Naomi Kawamura and Takayuki Matsumoto and Koki Kawakami and Taku Kawabata and Gen Isumi and Takayuki Uozumi and Akihiro Mitsuda and Hirofumi Wada and Masaichiro Mizumaki and Kojiro Mimura}
}

@article{YHMatsuda1,
author = {H. Matsuda ,Yasuhiro and Inami ,Toshiya and Ohwada ,Kenji and Murata ,Yuto and Nojiri ,Hiroyuki and Murakami ,Youichi and Mitsuda ,Akihiro and Wada ,Hirofumi and Miyazaki ,Hiroshi and Harada ,Isao},
title = {High-Magnetic-Field X-ray Absorption Spectroscopy of Field-Induced Valence Transition in EuNi2(Si1-xGex)2},
journal = {Journal of the Physical Society of Japan},
volume = {77},
number = {5},
pages = {054713},
year = {2008},
doi = {10.1143/JPSJ.77.054713},

URL = { 
    
        https://doi.org/10.1143/JPSJ.77.054713
    
    

}
}

@article{YHMatsuda2,
  title = {X-Ray Magnetic Circular Dichroism of a Valence Fluctuating State in Eu at High Magnetic Fields},
  author = {Matsuda, Y. H. and Ouyang, Z. W. and Nojiri, H. and Inami, T. and Ohwada, K. and Suzuki, M. and Kawamura, N. and Mitsuda, A. and Wada, H.},
  journal = {Phys. Rev. Lett.},
  volume = {103},
  issue = {4},
  pages = {046402},
  numpages = {4},
  year = {2009},
  month = {Jul},
  publisher = {American Physical Society},
  doi = {10.1103/PhysRevLett.103.046402},
  url = {https://link.aps.org/doi/10.1103/PhysRevLett.103.046402}
}

@article{nakamura,
author = {Nakamura ,Tetsuya and Hirono ,Toko and Kinoshita ,Toyohiko and Narumi ,Yasuo and Hayashi ,Misaki and Nojiri ,Hiroyuki and Mitsuda ,Akihiro and Wada ,Hirofumi and Kodama ,Kenji and Kindo ,Koichi and Kotani ,Akio},
title = {Soft-X-ray Magnetic Circular Dichroism under Pulsed High Magnetic Fields at Eu M4,5 Edges of Mixed Valence Compound EuNi2(Si0.18Ge0.82)2},
journal = {Journal of the Physical Society of Japan},
volume = {81},
number = {10},
pages = {103705},
year = {2012},
doi = {10.1143/JPSJ.81.103705},

URL = { 
    
        https://doi.org/10.1143/JPSJ.81.103705
    
    

}
}

@article{PhysRevB.59.1141,
  title = {Pressure effect on the valence transition of ${\mathrm{EuNi}}_{2}({\mathrm{Ge}}_{1\ensuremath{-}x}{\mathrm{Si}}_{x}{)}_{2}$},
  author = {Wada, H. and Hundley, M. F. and Movshovich, R. and Thompson, J. D.},
  journal = {Phys. Rev. B},
  volume = {59},
  issue = {2},
  pages = {1141--1144},
  numpages = {0},
  year = {1999},
  month = {Jan},
  publisher = {American Physical Society},
  doi = {10.1103/PhysRevB.59.1141},
  url = {https://link.aps.org/doi/10.1103/PhysRevB.59.1141}
}

@article{YokoyamaEu,
  title = {Photoinduced valence dynamics in ${\mathrm{EuNi}}_{2}({\mathrm{Si}}_{0.21}{\mathrm{Ge}}_{0.79}{)}_{2}$ studied via time-resolved x-ray absorption spectroscopy},
  author = {Yokoyama, Y. and Kawakami, K. and Hirata, Y. and Takubo, K. and Yamamoto, K. and Abe, K. and Mitsuda, A. and Wada, H. and Uozumi, T. and Yamamoto, S. and Matsuda, I. and Kimura, S. and Mimura, K. and Wadati, H.},
  journal = {Phys. Rev. B},
  volume = {100},
  issue = {11},
  pages = {115123},
  numpages = {6},
  year = {2019},
  month = {Sep},
  publisher = {American Physical Society},
  doi = {10.1103/PhysRevB.100.115123},
  url = {https://link.aps.org/doi/10.1103/PhysRevB.100.115123}
}

@article{PhysRevResearch.3.033211,
  title = {Ultrafast electron localization in the $\mathrm{Eu}{\mathrm{Ni}}_{2}{({\mathrm{Si}}_{0.21}{\mathrm{Ge}}_{0.79})}_{2}$ correlated metal},
  author = {Mardegan, Jose R. L. and Zerdane, Serhane and Mancini, Giulia and Esposito, Vincent and Rouxel, J\'er\'emy R. and Mankowsky, Roman and Svetina, Cristian and Gurung, Namrata and Parchenko, Sergii and Porer, Michael and Burganov, Bulat and Deng, Yunpei and Beaud, Paul and Ingold, Gerhard and Pedrini, Bill and Arrell, Christopher and Erny, Christian and Dax, Andreas and Lemke, Henrik and Decker, Martin and Ortiz, Nazaret and Milne, Chris and Smolentsev, Grigory and Maurel, Laura and Johnson, Steven L. and Mitsuda, Akihiro and Wada, Hirofumi and Yokoyama, Yuichi and Wadati, Hiroki and Staub, Urs},
  journal = {Phys. Rev. Res.},
  volume = {3},
  issue = {3},
  pages = {033211},
  numpages = {12},
  year = {2021},
  month = {Sep},
  publisher = {American Physical Society},
  doi = {10.1103/PhysRevResearch.3.033211},
  url = {https://link.aps.org/doi/10.1103/PhysRevResearch.3.033211}
}

@article{Yamagami,
  title = {$4f$ electron temperature driven ultrafast electron localization},
  author = {Yamagami, Kohei and Ueda, Hiroki and Staub, Urs and Zhang, Yujun and Yamamoto, Kohei and Park, Sang Han and Kwon, Soonnam and Mitsuda, Akihiro and Wada, Hirofumi and Uozumi, Takayuki and Mimura, Kojiro and Wadati, Hiroki},
  journal = {Phys. Rev. Res.},
  volume = {6},
  issue = {2},
  pages = {023099},
  numpages = {9},
  year = {2024},
  month = {Apr},
  publisher = {American Physical Society},
  doi = {10.1103/PhysRevResearch.6.023099},
  url = {https://link.aps.org/doi/10.1103/PhysRevResearch.6.023099}
}

@article{10.1063/1.5023557,
    author = {Park, Sang Han and Kim, Minseok and Min, Changi-Ki and Eom, Intae and Nam, Inhyuk and Lee, Heung-Soo and Kang, Heung-Sik and Kim, Hyeong-Do and Jang, Ho Young and Kim, Seonghan and Hwang, Sun-min and Park, Gi-Soo and Park, Jaehun and Koo, Tae-Yeong and Kwon, Soonnam},
    title = {PAL-XFEL soft X-ray scientific instruments and X-ray optics: First commissioning results},
    journal = {Review of Scientific Instruments},
    volume = {89},
    number = {5},
    pages = {055105},
    year = {2018},
    month = {05},
    issn = {0034-6748},
    doi = {10.1063/1.5023557},
    url = {https://doi.org/10.1063/1.5023557},
}

@article{Kim,
author = "Kim, Minseok and Min, Chang-Ki and Eom, Intae",
title = "{Laser systems for time-resolved experiments at the Pohang Accelerator Laboratory X-ray Free-Electron Laser beamlines}",
journal = "Journal of Synchrotron Radiation",
year = "2019",
volume = "26",
number = "3",
pages = "868--873",
month = "May",
doi = {10.1107/S1600577519003515},
url = {https://doi.org/10.1107/S1600577519003515},
}

@article{Park2,
author = "Park, Sang Han and Yoon, Jungbum and Kim, Changsoo and Hwang, Chanyong and Kim, Dong-Hyun and Lee, Sang-Hyuk and Kwon, Soonnam",
title = "{Scientific instruments for soft X-ray photon-in/photon-out spectroscopy on the PAL-XFEL}",
journal = "Journal of Synchrotron Radiation",
year = "2019",
volume = "26",
number = "4",
pages = "1031--1036",
month = "Jul",
doi = {10.1107/S1600577519004272},
url = {https://doi.org/10.1107/S1600577519004272},
}

@article{HHGMOKE,
  title = {Ultrafast behavior of induced and intrinsic magnetic moments in CoFeB/Pt bilayers probed by element-specific measurements in the extreme ultraviolet spectral range},
  author = {von Korff Schmising, Clemens and Jana, Somnath and Yao, Kelvin and Hennecke, Martin and Scheid, Philippe and Sharma, Sangeeta and Viret, Michel and Chauleau, Jean-Yves and Schick, Daniel and Eisebitt, Stefan},
  journal = {Phys. Rev. Res.},
  volume = {5},
  issue = {1},
  pages = {013147},
  numpages = {8},
  year = {2023},
  month = {Feb},
  publisher = {American Physical Society},
  doi = {10.1103/PhysRevResearch.5.013147},
  url = {https://link.aps.org/doi/10.1103/PhysRevResearch.5.013147}
}

@article{SaitoHHG,
  title = {Attosecond electronic dynamics of core-excited states of ${\mathrm{N}}_{2}\mathrm{O}$ in the soft x-ray region},
  author = {Saito, Nariyuki and Douguet, Nicolas and Sannohe, Hiroki and Ishii, Nobuhisa and Kanai, Teruto and Wu, Yi and Chew, Andrew and Han, Seunghwoi and Schneider, Barry I. and Olsen, Jeppe and Argenti, Luca and Chang, Zenghu and Itatani, Jiro},
  journal = {Phys. Rev. Res.},
  volume = {3},
  issue = {4},
  pages = {043222},
  numpages = {8},
  year = {2021},
  month = {Dec},
  publisher = {American Physical Society},
  doi = {10.1103/PhysRevResearch.3.043222},
  url = {https://link.aps.org/doi/10.1103/PhysRevResearch.3.043222}
}

@Article{Koo,
author={Koopmans, B.
and Malinowski, G.
and Dalla Longa, F.
and Steiauf, D.
and F{\"a}hnle, M.
and Roth, T.
and Cinchetti, M.
and Aeschlimann, M.},
title={Explaining the paradoxical diversity of ultrafast laser-induced demagnetization},
journal={Nature Materials},
year={2010},
month={Mar},
day={01},
volume={9},
number={3},
pages={259-265},
issn={1476-4660},
doi={10.1038/nmat2593},
url={https://doi.org/10.1038/nmat2593}
}

@article{Hübner,
doi = {10.1088/1361-648X/ad5bae},
url = {https://doi.org/10.1088/1361-648X/ad5bae},
year = {2024},
month = {jul},
publisher = {IOP Publishing},
volume = {36},
number = {40},
pages = {403001},
author = {Hübner, Wolfgang and Lefkidis, Georgios and Zhang, G P},
title = {All-optical spin switching on an ultrafast time scale},
journal = {Journal of Physics: Condensed Matter}
}

@Article{Ostler,
author={Ostler, T. A.
and Barker, J.
and Evans, R. F. L.
and Chantrell, R. W.
and Atxitia, U.
and Chubykalo-Fesenko, O.
and El Moussaoui, S.
and Le Guyader, L.
and Mengotti, E.
and Heyderman, L. J.
and Nolting, F.
and Tsukamoto, A.
and Itoh, A.
and Afanasiev, D.
and Ivanov, B. A.
and Kalashnikova, A. M.
and Vahaplar, K.
and Mentink, J.
and Kirilyuk, A.
and Rasing, Th.
and Kimel, A. V.},
title={Ultrafast heating as a sufficient stimulus for magnetization reversal in a ferrimagnet},
journal={Nature Communications},
year={2012},
month={Feb},
day={07},
volume={3},
number={1},
pages={666},
issn={2041-1723},
doi={10.1038/ncomms1666},
url={https://doi.org/10.1038/ncomms1666}
}

@Article{Graves,
author={Graves, C. E.
and Reid, A. H.
and Wang, T.
and Wu, B.
and de Jong, S.
and Vahaplar, K.
and Radu, I.
and Bernstein, D. P.
and Messerschmidt, M.
and M{\"u}ller, L.
and Coffee, R.
and Bionta, M.
and Epp, S. W.
and Hartmann, R.
and Kimmel, N.
and Hauser, G.
and Hartmann, A.
and Holl, P.
and Gorke, H.
and Mentink, J. H.
and Tsukamoto, A.
and Fognini, A.
and Turner, J. J.
and Schlotter, W. F.
and Rolles, D.
and Soltau, H.
and Str{\"u}der, L.
and Acremann, Y.
and Kimel, A. V.
and Kirilyuk, A.
and Rasing, Th.
and St{\"o}hr, J.
and Scherz, A. O.
and D{\"u}rr, H. A.},
title={Nanoscale spin reversal by non-local angular momentum transfer following ultrafast laser excitation in ferrimagnetic GdFeCo},
journal={Nature Materials},
year={2013},
month={Apr},
day={01},
volume={12},
number={4},
pages={293-298},
issn={1476-4660},
doi={10.1038/nmat3597},
url={https://doi.org/10.1038/nmat3597}
}

@article{Koo2,
    author = {Cornelissen, T. D. and Córdoba, R. and Koopmans, B.},
    title = {Microscopic model for all optical switching in ferromagnets},
    journal = {Applied Physics Letters},
    volume = {108},
    number = {14},
    pages = {142405},
    year = {2016},
    month = {04},
    issn = {0003-6951},
    doi = {10.1063/1.4945660},
    url = {https://doi.org/10.1063/1.4945660},
}

@article{10.1063/1.5010915,
    author = {Vomir, M. and Albrecht, M. and Bigot, J.-Y.},
    title = {Single shot all optical switching of intrinsic micron size magnetic domains of a Pt/Co/Pt ferromagnetic stack},
    journal = {Applied Physics Letters},
    volume = {111},
    number = {24},
    pages = {242404},
    year = {2017},
    month = {12},
    issn = {0003-6951},
    doi = {10.1063/1.5010915},
    url = {https://doi.org/10.1063/1.5010915},
}

@article{AOS1,
  title = {Manipulation of ferromagnets via the spin-selective optical Stark effect},
  author = {Qaiumzadeh, Alireza and Bauer, Gerrit E. W. and Brataas, Arne},
  journal = {Phys. Rev. B},
  volume = {88},
  issue = {6},
  pages = {064416},
  numpages = {5},
  year = {2013},
  month = {Aug},
  publisher = {American Physical Society},
  doi = {10.1103/PhysRevB.88.064416},
  url = {https://link.aps.org/doi/10.1103/PhysRevB.88.064416}
}

@article{Takahashi2023,
  title = {Optically Induced Magnetization Switching in NiCo2O4 Thin Films Using Ultrafast Lasers},
  volume = {5},
  ISSN = {2637-6113},
  url = {http://dx.doi.org/10.1021/acsaelm.2c01233},
  DOI = {10.1021/acsaelm.2c01233},
  number = {2},
  journal = {ACS Applied Electronic Materials},
  publisher = {American Chemical Society (ACS)},
  author = {Takahashi,  Ryunosuke and Ohkochi,  Takuo and Kan,  Daisuke and Shimakawa,  Yuichi and Wadati,  Hiroki},
  year = {2023},
  month = jan,
  pages = {748–753}
}

@article{Takahashi2025,
    author = {Takahashi, Ryunosuke and Guen, Yann Le and Nakata, Suguru and Igarashi, Junta and Hohlfeld, Julius and Malinowski, Grégory and Xie, Lingling and Kan, Daisuke and Shimakawa, Yuichi and Mangin, Stéphane and Wadati, Hiroki},
    title = {All-optical helicity-dependent switching in NiCo2O4 thin films},
    journal = {Applied Physics Letters},
    volume = {126},
    number = {21},
    pages = {212405},
    year = {2025},
    month = {05},
    issn = {0003-6951},
    doi = {10.1063/5.0253785},
    url = {https://doi.org/10.1063/5.0253785},
}

@article{SC1,
  title = {Light-Induced Superconductivity in a Stripe-Ordered Cuprate},
  volume = {331},
  ISSN = {1095-9203},
  url = {http://dx.doi.org/10.1126/science.1197294},
  DOI = {10.1126/science.1197294},
  number = {6014},
  journal = {Science},
  publisher = {American Association for the Advancement of Science (AAAS)},
  author = {Fausti,  D. and Tobey,  R. I. and Dean,  N. and Kaiser,  S. and Dienst,  A. and Hoffmann,  M. C. and Pyon,  S. and Takayama,  T. and Takagi,  H. and Cavalleri,  A.},
  year = {2011},
  month = jan,
  pages = {189–191}
}

@article{SC2,
  title = {Optically induced superconductivity in striped ${\mathrm{La}}_{2\ensuremath{-}x}{\mathrm{Ba}}_{x}{\mathrm{CuO}}_{4}$ by polarization-selective excitation in the near infrared},
  author = {Nicoletti, D. and Casandruc, E. and Laplace, Y. and Khanna, V. and Hunt, C. R. and Kaiser, S. and Dhesi, S. S. and Gu, G. D. and Hill, J. P. and Cavalleri, A.},
  journal = {Phys. Rev. B},
  volume = {90},
  issue = {10},
  pages = {100503},
  numpages = {6},
  year = {2014},
  month = {Sep},
  publisher = {American Physical Society},
  doi = {10.1103/PhysRevB.90.100503},
  url = {https://link.aps.org/doi/10.1103/PhysRevB.90.100503}
}

@article{SC3,
  title = {Optically induced coherent transport far above ${T}_{c}$ in underdoped ${\mathrm{YBa}}_{2}{\mathrm{Cu}}_{3}{\mathrm{O}}_{6+\ensuremath{\delta}}$},
  author = {Kaiser, S. and Hunt, C. R. and Nicoletti, D. and Hu, W. and Gierz, I. and Liu, H. Y. and Le Tacon, M. and Loew, T. and Haug, D. and Keimer, B. and Cavalleri, A.},
  journal = {Phys. Rev. B},
  volume = {89},
  issue = {18},
  pages = {184516},
  numpages = {9},
  year = {2014},
  month = {May},
  publisher = {American Physical Society},
  doi = {10.1103/PhysRevB.89.184516},
  url = {https://link.aps.org/doi/10.1103/PhysRevB.89.184516}
}

@Article{SC4,
author={Hu, W.
and Kaiser, S.
and Nicoletti, D.
and Hunt, C. R.
and Gierz, I.
and Hoffmann, M. C.
and Le Tacon, M.
and Loew, T.
and Keimer, B.
and Cavalleri, A.},
title={Optically enhanced coherent transport in YBa2Cu3O6.5 by ultrafast redistribution of interlayer coupling},
journal={Nature Materials},
year={2014},
month={Jul},
day={01},
volume={13},
number={7},
pages={705-711},
issn={1476-4660},
doi={10.1038/nmat3963},
url={https://doi.org/10.1038/nmat3963}
}

@Article{SC5,
author={Mitrano, M.
and Cantaluppi, A.
and Nicoletti, D.
and Kaiser, S.
and Perucchi, A.
and Lupi, S.
and Di Pietro, P.
and Pontiroli, D.
and Ricc{\`o}, M.
and Clark, S. R.
and Jaksch, D.
and Cavalleri, A.},
title={Possible light-induced superconductivity in K3C60 at high temperature},
journal={Nature},
year={2016},
month={Feb},
day={01},
volume={530},
number={7591},
pages={461-464},
issn={1476-4687},
doi={10.1038/nature16522},
url={https://doi.org/10.1038/nature16522}
}

@Article{SC6,
author={Budden, M.
and Gebert, T.
and Buzzi, M.
and Jotzu, G.
and Wang, E.
and Matsuyama, T.
and Meier, G.
and Laplace, Y.
and Pontiroli, D.
and Ricc{\`o}, M.
and Schlawin, F.
and Jaksch, D.
and Cavalleri, A.},
title={Evidence for metastable photo-induced superconductivity in K3C60},
journal={Nature Physics},
year={2021},
month={May},
day={01},
volume={17},
number={5},
pages={611-618},
issn={1745-2481},
doi={10.1038/s41567-020-01148-1},
url={https://doi.org/10.1038/s41567-020-01148-1}
}

@Article{SC7,
author={Fava, S.
and De Vecchi, G.
and Jotzu, G.
and Buzzi, M.
and Gebert, T.
and Liu, Y.
and Keimer, B.
and Cavalleri, A.},
title={Magnetic field expulsion in optically driven YBa2Cu3O6.48},
journal={Nature},
year={2024},
month={Aug},
day={01},
volume={632},
number={8023},
pages={75-80},
issn={1476-4687},
doi={10.1038/s41586-024-07635-2},
url={https://doi.org/10.1038/s41586-024-07635-2}
}

@article{SC8,
  title = {Near-infrared light-induced superconducting-like state in underdoped $\mathrm{Y}{\mathrm{Ba}}_{2}{\mathrm{Cu}}_{3}{\mathrm{O}}_{y}$ studied by $c$-axis terahertz third-harmonic generation},
  author = {Katsumi, Kota and Nishida, Morihiko and Kaiser, Stefan and Miyasaka, Shigeki and Tajima, Setsuko and Shimano, Ryo},
  journal = {Phys. Rev. B},
  volume = {107},
  issue = {21},
  pages = {214506},
  numpages = {18},
  year = {2023},
  month = {Jun},
  publisher = {American Physical Society},
  doi = {10.1103/PhysRevB.107.214506},
  url = {https://link.aps.org/doi/10.1103/PhysRevB.107.214506}
}

@article{SC9,
  title = {Optical Saturation Produces Spurious Evidence for Photoinduced Superconductivity in ${\mathrm{K}}_{3}{\mathrm{C}}_{60}$},
  author = {Dodge, J. Steven and Lopez, Leya and Sahota, Derek G.},
  journal = {Phys. Rev. Lett.},
  volume = {130},
  issue = {14},
  pages = {146002},
  numpages = {6},
  year = {2023},
  month = {Apr},
  publisher = {American Physical Society},
  doi = {10.1103/PhysRevLett.130.146002},
  url = {https://link.aps.org/doi/10.1103/PhysRevLett.130.146002}
}

@article{SC10,
  title = {Ultrafast Renormalization of the On-Site Coulomb Repulsion in a Cuprate Superconductor},
  author = {Baykusheva, Denitsa R. and Jang, Hoyoung and Husain, Ali A. and Lee, Sangjun and TenHuisen, Sophia F. R. and Zhou, Preston and Park, Sunwook and Kim, Hoon and Kim, Jin-Kwang and Kim, Hyeong-Do and Kim, Minseok and Park, Sang-Youn and Abbamonte, Peter and Kim, B. J. and Gu, G. D. and Wang, Yao and Mitrano, Matteo},
  journal = {Phys. Rev. X},
  volume = {12},
  issue = {1},
  pages = {011013},
  numpages = {14},
  year = {2022},
  month = {Jan},
  publisher = {American Physical Society},
  doi = {10.1103/PhysRevX.12.011013},
  url = {https://link.aps.org/doi/10.1103/PhysRevX.12.011013}
}

@article{Yamamoto2025,
doi = {10.1088/1742-6596/3010/1/012115},
url = {https://doi.org/10.1088/1742-6596/3010/1/012115},
year = {2025},
month = {may},
publisher = {IOP Publishing},
volume = {3010},
number = {1},
pages = {012115},
author = {Yamamoto, Kohei and Ugalino, Ralph and Fujii, Kentaro and Ohtsubo, Yoshiyuki and Iwasawa, Hideaki and Kitamura, Miho and Imazono, Takashi and Inami, Nobuhito and Nakatani, Takeshi and Inaba, Kento and Agui, Akane and Takeuchi, Tomoyuki and Kimura, Hiroaki and Takahasi, Masamitu and Horiba, Koji and Miyawaki, Jun},
title = {Status of RIXS Beamline BL02U at NanoTerasu},
journal = {Journal of Physics: Conference Series}
}

@article{Kuba,
  title = {Large-Amplitude Spin Dynamics Driven by a THz Pulse in Resonance with an Electromagnon},
  volume = {343},
  ISSN = {1095-9203},
  url = {http://dx.doi.org/10.1126/science.1242862},
  DOI = {10.1126/science.1242862},
  number = {6177},
  journal = {Science},
  publisher = {American Association for the Advancement of Science (AAAS)},
  author = {Kubacka,  T. and Johnson,  J. A. and Hoffmann,  M. C. and Vicario,  C. and de Jong,  S. and Beaud,  P. and Gr\"{u}bel,  S. and Huang,  S.-W. and Huber,  L. and Patthey,  L. and Chuang,  Y.-D. and Turner,  J. J. and Dakovski,  G. L. and Lee,  W.-S. and Minitti,  M. P. and Schlotter,  W. and Moore,  R. G. and Hauri,  C. P. and Koohpayeh,  S. M. and Scagnoli,  V. and Ingold,  G. and Johnson,  S. L. and Staub,  U.},
  year = {2014},
  month = mar,
  pages = {1333–1336}
}

@article{10.1063/1.4983153,
    author = {Kozina, M. and van Driel, T. and Chollet, M. and Sato, T. and Glownia, J. M. and Wandel, S. and Radovic, M. and Staub, U. and Hoffmann, M. C.},
    title = {Ultrafast X-ray diffraction probe of terahertz field-driven soft mode dynamics in SrTiO3},
    journal = {Structural Dynamics},
    volume = {4},
    number = {5},
    pages = {054301},
    year = {2017},
    month = {05},
    issn = {2329-7778},
    doi = {10.1063/1.4983153},
    url = {https://doi.org/10.1063/1.4983153},
}

@Article{Kozina2019,
author={Kozina, M.
and Fechner, M.
and Marsik, P.
and van Driel, T.
and Glownia, J. M.
and Bernhard, C.
and Radovic, M.
and Zhu, D.
and Bonetti, S.
and Staub, U.
and Hoffmann, M. C.},
title={Terahertz-driven phonon upconversion in SrTiO3},
journal={Nature Physics},
year={2019},
month={Apr},
day={01},
volume={15},
number={4},
pages={387-392},
issn={1745-2481},
doi={10.1038/s41567-018-0408-1},
url={https://doi.org/10.1038/s41567-018-0408-1}
}

@Article{Kampfrath2013,
author={Kampfrath, Tobias
and Tanaka, Koichiro
and Nelson, Keith A.},
title={Resonant and nonresonant control over matter and light by intense terahertz transients},
journal={Nature Photonics},
year={2013},
month={Sep},
day={01},
volume={7},
number={9},
pages={680-690},
issn={1749-4893},
doi={10.1038/nphoton.2013.184},
url={https://doi.org/10.1038/nphoton.2013.184}
}

@article{KubotaTHz,
    author = {Kubota, Yuya and Suzuki, Takeshi and Owada, Shigeki and Tamasaku, Kenji and Osawa, Hitoshi and Togashi, Tadashi and Okazaki, Kozo and Yabashi, Makina},
    title = {A simple method to find temporal overlap between THz and x-ray pulses using x-ray-induced carrier dynamics in semiconductors},
    journal = {Applied Physics Letters},
    volume = {126},
    number = {5},
    pages = {052101},
    year = {2025},
    month = {02},
    issn = {0003-6951},
    doi = {10.1063/5.0242393},
    url = {https://doi.org/10.1063/5.0242393},
}

@article{Polley,
doi = {10.1088/1361-6463/aaa863},
url = {https://doi.org/10.1088/1361-6463/aaa863},
year = {2018},
month = {feb},
publisher = {IOP Publishing},
volume = {51},
number = {8},
pages = {084001},
author = {Polley, Debanjan and Pancaldi, Matteo and Hudl, Matthias and Vavassori, Paolo and Urazhdin, Sergei and Bonetti, Stefano},
title = {THz-driven demagnetization with perpendicular magnetic anisotropy: towards ultrafast ballistic switching},
journal = {Journal of Physics D: Applied Physics}
}

@book{Atto,
  title = {Soft X-Rays and Extreme Ultraviolet Radiation: Principles and Applications},
  ISBN = {9781139164429},
  url = {http://dx.doi.org/10.1017/CBO9781139164429},
  DOI = {10.1017/cbo9781139164429},
  publisher = {Cambridge University Press},
  author = {Attwood,  David},
  year = {2007},
  month = aug 
}

@book{Satokatsu,
  title = {Light and magnetism},
  ISBN = {9784254136289},
  publisher = {Asakura},
  author = {Sato,  Katsuaki},
  year = {2001},
}

@book{store,
  title = {Magnetism},
  ISBN = {9783540302827},
  url = {http://dx.doi.org/10.1007/978-3-540-30283-4},
  DOI = {10.1007/978-3-540-30283-4},
  publisher = {Springer Berlin Heidelberg},
  year = {2006},
  author = {St\"{o}hr,  Joachim and Siegmann, H. C.},
}

@Article{deGroot2024,
author={de Groot, Frank M. F.
and Haverkort, Maurits W.
and Elnaggar, Hebatalla
and Juhin, Am{\'e}lie
and Zhou, Ke-Jin
and Glatzel, Pieter},
title={Resonant inelastic X-ray scattering},
journal={Nature Reviews Methods Primers},
year={2024},
month={Jul},
day={04},
volume={4},
number={1},
pages={45},
issn={2662-8449},
doi={10.1038/s43586-024-00322-6},
url={https://doi.org/10.1038/s43586-024-00322-6}
}

@Article{Chakraborty2025,
author={Chakraborty, Swaroop
and Britto, Sylvia
and Gomez-Gonzalez, Miguel
and Buzanich, Ana Guilherme
and Mikulska, Iuliia},
title={Synchrotrons, neutron sources, and XFELs guiding the future of safe and sustainable nanomaterials},
journal={Cell Reports Physical Science},
year={2025},
month={Sep},
day={17},
publisher={Elsevier},
volume={6},
number={9},
issn={2666-3864},
doi={10.1016/j.xcrp.2025.102806},
url={https://doi.org/10.1016/j.xcrp.2025.102806}
}

@article{Sueda,
    author = {Ueda, S. and Mizuguchi, M. and Miura, Y. and Kang, J. G. and Shirai, M. and Takanashi, K.},
    title = {Electronic structure and magnetic anisotropy of L1-FePt thin film studied by hard x-ray photoemission spectroscopy and first-principles calculations},
    journal = {Applied Physics Letters},
    volume = {109},
    number = {4},
    pages = {042404},
    year = {2016},
    month = {07},
    issn = {0003-6951},
    doi = {10.1063/1.4959957},
    url = {https://doi.org/10.1063/1.4959957},
}

@article{Superdiffusive,
  title = {Superdiffusive Spin Transport as a Mechanism of Ultrafast Demagnetization},
  author = {Battiato, M. and Carva, K. and Oppeneer, P. M.},
  journal = {Phys. Rev. Lett.},
  volume = {105},
  issue = {2},
  pages = {027203},
  numpages = {4},
  year = {2010},
  month = {Jul},
  publisher = {American Physical Society},
  doi = {10.1103/PhysRevLett.105.027203},
  url = {https://link.aps.org/doi/10.1103/PhysRevLett.105.027203}
}

@phdthesis{phdthesis, author = {Y. Kubota}, title = {Magnetism of buried layers studied by soft X-ray resonant
magneto-optical effect using polarization modulation}, school = {University of Tokyo}, year = 2016, }

@misc{yamamoto20252drixsresonantinelasticxray,
      title={2D-RIXS: Resonant inelastic x-ray scattering microscopy with high energy and spatial resolutions}, 
      author={Kohei Yamamoto and Hakuto Suzuki and Jun Miyawaki},
      year={2025},
      eprint={2511.17966},
      archivePrefix={arXiv},
      primaryClass={cond-mat.mtrl-sci},
      url={https://arxiv.org/abs/2511.17966}, 
}
\end{document}